\documentclass[12pt]{article}

\usepackage{scicite}
\usepackage{graphicx}
\usepackage{etoolbox}
\usepackage{hyperref}
\usepackage{times}
\usepackage{bm}

\usepackage{color}
\usepackage{amssymb}
\usepackage{amsmath}
\usepackage{braket}
\usepackage[table]{xcolor}

\newenvironment{sciabstract}{
\begin{quote} \bf}
{\end{quote}}

\usepackage{graphicx}

\graphicspath{ {figures/} }

\makeatletter
\let\saved@includegraphics\includegraphics
\AtBeginDocument{\let\includegraphics\saved@includegraphics}
\renewenvironment*{figure}{\@float{figure}}{\end@float}
\makeatother

\DeclareFontFamily{OMS}{oasy}{\skewchar\font48 }
\DeclareFontShape{OMS}{oasy}{m}{n}{
         <-5.5> oasy5     <5.5-6.5> oasy6
      <6.5-7.5> oasy7     <7.5-8.5> oasy8
      <8.5-9.5> oasy9     <9.5->  oasy10
      }{}
\DeclareFontShape{OMS}{oasy}{b}{n}{
       <-6> oabsy5
      <6-8> oabsy7
      <8->  oabsy10
      }{}
\DeclareSymbolFont{oasy}{OMS}{oasy}{m}{n}
\SetSymbolFont{oasy}{bold}{OMS}{oasy}{b}{n}

\DeclareMathSymbol{\smallleftarrow}     {\mathrel}{oasy}{"20}
\DeclareMathSymbol{\smallrightarrow}    {\mathrel}{oasy}{"21}
\DeclareMathSymbol{\smallleftrightarrow}{\mathrel}{oasy}{"24}

\newcommand{\tensor}[1]{\overset{\scriptscriptstyle\smallleftrightarrow}{#1}}

\begin{document}

\title{Imaging stress and magnetism at high pressures using a nanoscale quantum sensor}

\date{}

\author{
\hspace{-13mm}
S.~Hsieh,$^{1,2,*}$ 
P.~Bhattacharyya,$^{1,2,*}$ 
C. Zu,$^{1,*}$ 
T. Mittiga,$^1$ 
T. J. Smart,$^3$ 
F. Machado,$^1$\\
\hspace{-13mm}
B. Kobrin,$^{1,2}$ 
T. O. H{\"o}hn,$^{1,4}$ 
N. Z. Rui,$^{1}$ 
M. Kamrani,$^5$ 
S. Chatterjee,$^1$ 
S. Choi,$^1$\\
\hspace{-13mm}
M. Zaletel,$^1$ 
V. V. Struzhkin,$^6$ 
J. E. Moore,$^{1,2}$ 
V. I. Levitas,$^{5,7}$ 
R. Jeanloz,$^3$ 
N. Y. Yao$^{1,2,\dag}$\\
\\
\normalsize{\hspace{-13mm}$^{1}$Department of Physics, University of California, Berkeley, CA 94720, USA}\\
\normalsize{\hspace{-13mm}$^{2}$Materials Science Division, Lawrence Berkeley National Laboratory, Berkeley, CA 94720, USA}\\
\normalsize{\hspace{-13mm}$^{3}$Department of Earth and Planetary Science, University of California, Berkeley, CA 94720, USA}\\
\normalsize{\hspace{-13mm}$^{4}$Fakult{\"a}t f{\"u}r Physik, Ludwig-Maximilians-Universit{\"a}t M{\"u}nchen, 80799 Munich, Germany}\\
\normalsize{\hspace{-13mm}$^{5}$Department of Aerospace Engineering, Iowa State University, Ames, IA 50011, USA}\\
\normalsize{\hspace{-13mm}$^{6}$Geophysical Laboratory, Carnegie Institution of Washington, Washington, DC 20015, USA}\\
\normalsize{\hspace{-13mm}$^{7}$Departments of Mechanical Engineering and Material Science and Engineering,}\\
\normalsize{\hspace{-13mm}Iowa State University, Ames, IA 50011, USA}\\
\normalsize{\hspace{-13mm}$^\dag$To whom correspondence should be addressed; E-mail:  norman.yao@berkeley.edu}
}   

\baselineskip24pt
\maketitle

\vspace{5mm}
\begin{sciabstract}

Pressure alters the physical, chemical and electronic properties of matter. 
The development of the diamond anvil cell (DAC) enables tabletop experiments to investigate a diverse landscape of high-pressure phenomena ranging from the properties of planetary interiors to transitions between quantum mechanical phases. 
In this work, we introduce and utilize a novel nanoscale sensing platform, which integrates nitrogen-vacancy (NV) color centers directly into the culet (tip) of diamond anvils.
We demonstrate the versatility of this platform by performing diffraction-limited imaging ($\sim$$600$~nm) of both stress fields and magnetism, up to pressures $\sim$$30$~GPa and for temperatures ranging from $25-340$~K.
For the former, we quantify all six (normal and shear) stress components with accuracy $\lesssim 0.01$~GPa, offering 
unique new capabilities for characterizing the strength and effective viscosity of solids and fluids under pressure. 
For the latter, we demonstrate vector magnetic field imaging with dipole accuracy $\lesssim 10^{-11}$~emu, enabling us to 
measure the pressure-driven $\alpha\leftrightarrow\epsilon$  phase transition in iron as well as the complex pressure-temperature phase diagram of gadolinium.
In addition to DC vector magnetometry, we highlight a complementary NV-sensing modality using $T_1$ \emph{noise} spectroscopy; crucially, this demonstrates our ability to characterize phase transitions even in the absence of \emph{static} magnetic signatures. 
By integrating an atomic-scale sensor directly into DACs, our platform enables the \emph{in situ} imaging of elastic, electric and magnetic phenomena at high pressures.
\end{sciabstract}

A tremendous amount of recent attention has focused on the development of hybrid quantum sensing devices, in which sensors are directly integrated into existing toolsets ranging from biological imaging to materials spectroscopy \cite{Kucsko:2013bio2,maletinsky2012,cai:2014hybrid1,dovzhenko2018magnetostatic}. 
In this work, we demonstrate the versatility of a novel platform based upon quantum spin defects combined with static high pressure technologies \cite{Jayaraman:1983,Mao:2018fd}.
In particular, we instrument diamond anvil cells with a layer of nitrogen-vacancy (NV) centers directly at the culet, enabling the pursuit of two complementary objectives in high pressure science: first, to understand the strength and failure of materials under pressure (e.g.~the brittle-ductile transition) and second, to discover and characterize new phases of matter (e.g.~high temperature superconductors) \cite{wigner1935possibility,horii1986brittle,Gilioli_Ehm_2014,Drozdov:2015conventional}.
Achieving these goals hinges upon the sensitive \emph{in situ} imaging of signals within the high pressure chamber.
In the former case, measuring the \textit{local} stress environment permits the direct observation of inhomogeneities in plastic flow and the formation of line defects.
In the latter case, the ability to spatially resolve field distributions can provide a direct image of complex order parameters and textured phenomena such as magnetic domains.
Unfortunately, the enormous stress gradients generated near the sample limit the utility of most conventional tabletop spectroscopy techniques; as a result, one is often restricted to measuring bulk properties 
averaged over the entire DAC geometry. 

\begin{figure}[ht]
     \hspace{-15mm}
     \includegraphics[width=1.2\textwidth]{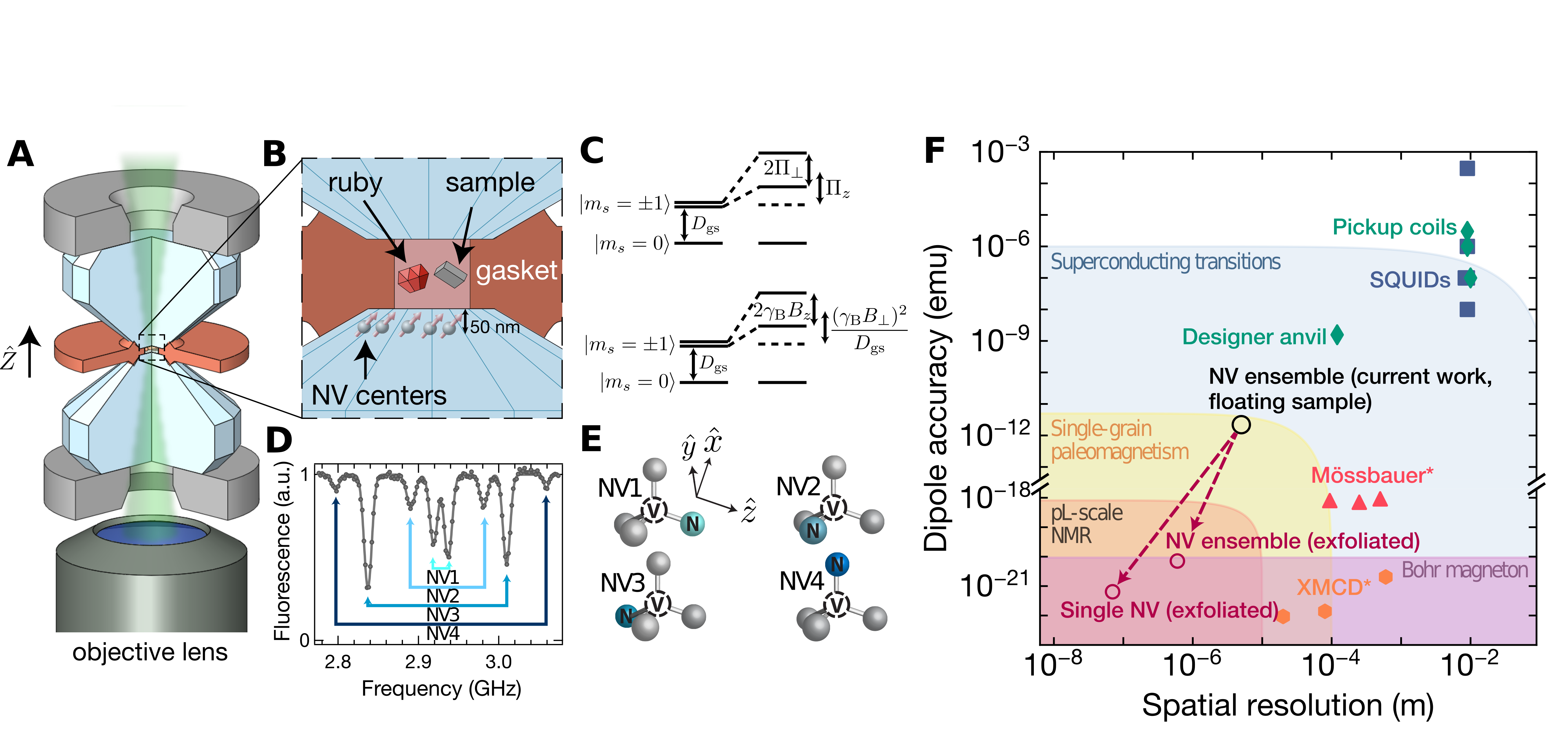}
    \caption{\small \textbf{NV centers integrated into a diamond anvil cell.} (\textbf{A}) Schematic of the DAC geometry.
    Two opposing anvils are compressed by a nonmagnetic steel cell and cubic boron nitride backing plates (gray).
    NV centers are initialized and read out using a 532~nm laser focused to a diffraction-limited spot ($\sim$600 nm) which is scanned across the culet surface.
    (\textbf{B}) The DAC sample chamber is defined by the gasket-anvil assembly; it is loaded with the sample of interest, a pressure-transmitting medium, and a single ruby microsphere (pressure calibration).
    A  $\sim$50~nm layer of NV centers is embedded into the diamond anvil directly below the sample chamber.
    (\textbf{C}) Stress (top) both shifts and splits the $\ket{m_s=\pm1}$ sublevels at first order;
    in particular, the shifting is characterized by  $\Pi_{z} =  \alpha_1 (\sigma_{xx} + \sigma_{yy} ) + \beta_1 \sigma_{zz}$, and the splitting is characterized by $\Pi_\perp^2 = \left [  \alpha_2 (\sigma_{yy} - \sigma_{xx} ) +\beta_2 (2 \sigma_{xz})   \right ]^2 
+\left [  \alpha_2 (2\sigma_{xy}) + \beta_2 (2\sigma_{yz}) \right ]^2$.
    An axial magnetic field (bottom) splits the $\ket{m_s=\pm1}$ sublevels at first order, but a transverse magnetic field leads to shifts only at second order.
    (\textbf{D}) ODMR spectrum from an NV center ensemble under an applied magnetic field.
    (\textbf{E}) Each pair of resonances in (D) corresponds to one of the four NV crystallographic orientations.
    (\textbf{F}) Comparison of high pressure magnetometry techniques. The system characterized in this work is shown here assuming a sample suspended in a pressure medium $5~\mu$m away from the culet (black open circle).
    We project that by exfoliating a sample directly onto the culet surface and using $5$~nm implanted NV centers, the distance from the sample can be significantly reduced, thus improving dipole accuracy (open red circles).
    Inductive methods (pickup coils [green diamonds] and SQUIDs [blue squares]) integrate the magnetization of a sample over their area \cite{supp}. %\red{\cite{Feng:2014coil2,Mito:2003coil3,Endo:1998squid1,alireza:2009squid2,Mito:2001squid3,Giriat:2010squid4,Takeda:2002squid6,Marizy:2017squid7}}.
    In contrast, high energy photon scattering techniques (x-ray magnetic circular dichroism [orange hexagons], and M{\"o}ssbauer spectroscopy [pink triangles]) probe atomic scale magnetism \cite{supp}. %\red{\cite{Pasternak:1991mossbauer1,Pasternak:2001mossbauer2,Kantor:2004mossbauer3,Mathon:2004it,Watanabe:2011hk,Ishimatsu:2007eu,Chen:2018}}.
    Note that the length scale for these methods is shown here as the spot size of the excitation beam.}\label{fig:1}
\end{figure}

Our approach to these challenges is to utilize an ensemble of NV centers  ($\sim$$1$ ppm density) implanted $\sim$$50$ nm from the surface of the diamond anvil culet (Fig.~\ref{fig:1}A,B).
Each NV center represents an atomic-scale defect (i.e.~a substitutional nitrogen impurity adjacent to a vacancy) inside the diamond lattice and exhibits an $S=1$ electronic spin ground state \cite{doherty2013nitrogen}.
In the absence of external fields, the $|m_\textrm{s}=\pm 1\rangle$ spin sublevels  are degenerate  and separated by $D_\textrm{gs} = (2\pi)\times 2.87~ \mathrm{GHz}$ from the $|m_\textrm{s}= 0\rangle$ state. 
Crucially, both the nature and energy of these spin states are sensitive to local changes in stress, temperature, magnetic and electric fields (Fig.~\ref{fig:1}C) \cite{ovartchaiyapong2014dynamic,acosta2010temperature,maze2008nanoscale,dolde2011electric}.
These spin states can be optically initialized and read out, as well as coherently manipulated via microwave fields.
Their energy levels can be probed by performing optically detected magnetic resonance (ODMR) spectroscopy where one measures a change in the NV's fluorescence intensity when an applied microwave field is on resonance between two NV spin sublevels
(Fig.~\ref{fig:1}D), thus enabling sensing of a variety of external signals over a wide range of environmental conditions \cite{Kucsko:2013bio2,Casola:2018nvmagnetometry1,mittiga2018imaging}.%\red{\cite{dolde:2014electrometry,Grazioso:2013ba}}.

Here, we focus on the sensing of stress and magnetic fields, wherein the NV is governed by the Hamiltonian, $H= H_0 + H_\textrm{B} + H_\textrm{S}$, with $H_0 =  D_{\textrm{gs}} S_z^2$ (zero-field splitting), $H_\textrm{B} = \gamma_\textrm{B} \vec{B} \cdot  \vec{S}$ (Zeeman splitting), and 
$H_\textrm{S} = \left [ \alpha_1 (\sigma_{xx} + \sigma_{yy} ) + \beta_1 \sigma_{zz}   \right ]  S_z^2 +   \left [  \alpha_2 (\sigma_{yy} - \sigma_{xx} ) +\beta_2 (2 \sigma_{xz})   \right ] (S_y^2-S_x^2) 
+\left [  \alpha_2 (2\sigma_{xy}) + \beta_2 (2\sigma_{yz}) \right ] (S_xS_y+S_yS_x)$ capturing the NV's response to the local diamond stress tensor, $\tensor{\sigma}$ %$\overset{\text{\tiny$\leftrightarrow$}}{\sigma}$
(Fig.~\ref{fig:1}C).
Note that in the above, $\gamma_\textrm{B} \approx (2\pi)\times2.8$ MHz/G is the gyromagnetic ratio, $\{ \alpha_{1,2},  \beta_{1,2} \}$ are the stress susceptibility coefficients \cite{Barson:2017ba},
$\hat{z}$ is the NV orientation axis, and $\hat{x}$ is defined such that the $xz$-plane contains one of the carbon-vacancy bonds (Fig.~\ref{fig:1}E).
In general, the resulting ODMR spectra exhibit eight resonances arising from the four possible crystallographic orientations of the NV (Fig.~\ref{fig:1}D).
By extracting the energy shifting and splitting of the spin sublevels for each NV orientation group, one obtains an overconstrained set of equations enabling the reconstruction of either the (six component) local stress tensor or the (three component) vector magnetic field \cite{supp}.

In our experiments, we utilize a miniature DAC (Fig.~\ref{fig:1}A,B) consisting of two opposing anvils compressing either a beryllium copper or rhenium gasket \cite{Sterer:1998hl}. 
The sample chamber defined by the gasket and diamond-anvil culets is filled with a pressure-transmitting medium (either a 16:3:1 methanol/ethanol/water solution or cesium iodide) 
to provide a quasi-hydrostatic environment.
Microwave excitation is applied via a 4~$\mu$m thick platinum foil compressed between the gasket and anvil pavilion facets, while scanning confocal microscopy (with a transverse diffraction-limited spot size $\sim$$600$~nm,  containing $\sim$$10^3$ NVs) allows us to obtain two-dimensional ODMR maps across the culet. 

\begin{figure}[ht]
     \centering
     \includegraphics[width=\textwidth]{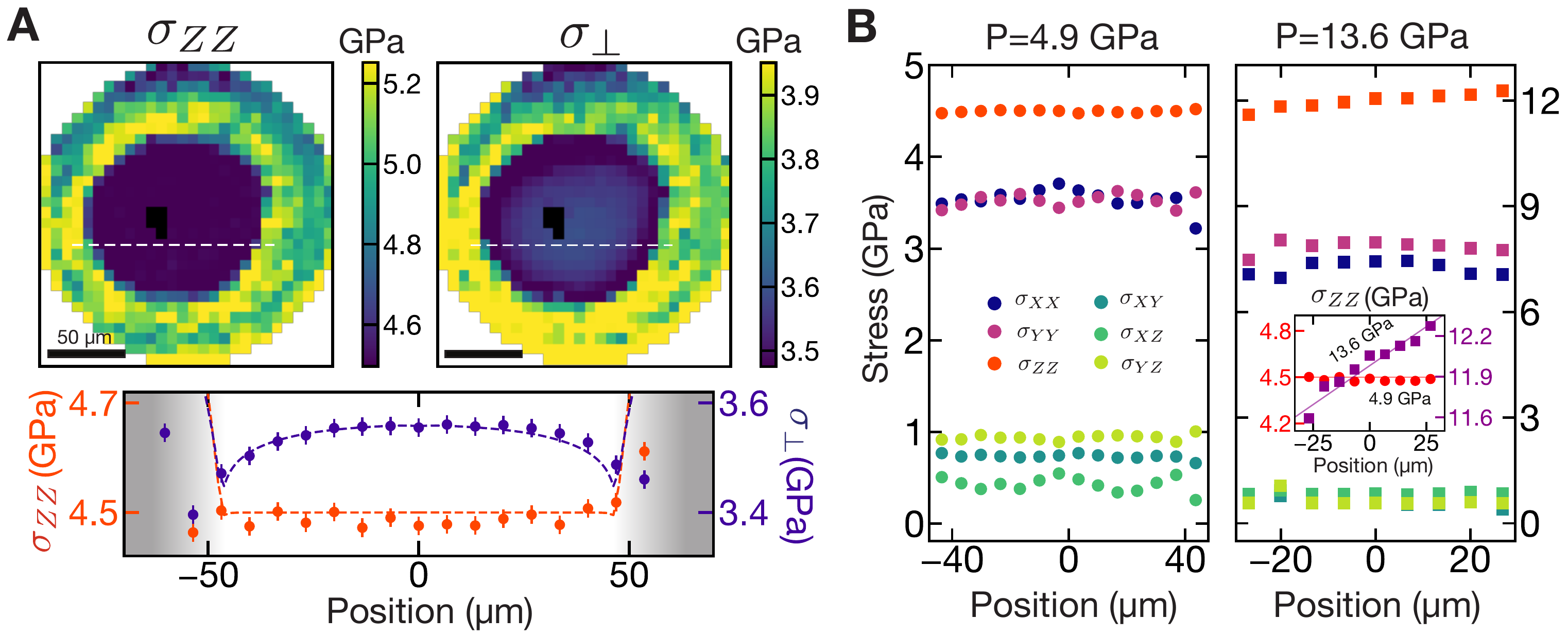}
      \caption{\small \textbf{Full tensorial reconstruction of the stresses in a (111)-cut diamond anvil.} (\textbf{A}) Spatially resolved maps of the loading  stress (left)  and mean lateral stress (right), $\sigma_\perp = \frac 1 2 (\sigma_{XX}+\sigma_{YY})$, across the culet surface. In the inner region, where the culet surface contacts the pressure-transmitting medium (16:3:1 methanol/ethanol/water), the loading stress is spatially uniform, while the lateral stress is concentrated towards the center; this qualitative difference is highlighted by a linecut of the two stresses 
      (below), and reconstructed by finite element analysis (orange and purple dashed lines). The black pixels indicate where the NV spectrum was obfuscated by the ruby microsphere. (\textbf{B}) Comparison of all stress tensor components in the fluid-contact region at $P=$~4.9~GPa and $P=$~13.6~GPa.
    At $P=$~13.6~GPa, the pressure-transmitting medium has entered its glassy phase and we observe a spatial gradient in the loading stress $\sigma_{ZZ}$ (inset).}\label{fig:2}
 \end{figure}

We begin by probing the stress tensor across the culet surface using two different cuts of diamond (i.e.~(111)-cut and (110)-cut culet). %\blue{@Satcher and Bryce: confirm parentheses}
For a generic stress environment, the intrinsic degeneracy associated with the four NV orientations is not sufficiently lifted, implying that individual resonances cannot be resolved. 
In order to resolve these  resonances  while preserving the stress contribution, we sequentially tune a well-controlled external magnetic field to be perpendicular to each of the different NV orientations  \cite{supp}.
For each perpendicular field choice, three of the four NV orientations exhibit a strong Zeeman splitting proportional to the projection of the external magnetic field along their symmetry axes. 
Crucially, this enables one to resolve the stress information encoded in the remaining NV orientation, while the other three groups of NVs are spectroscopically split away.
Using this method, we obtain sufficient information to extract the full stress tensor, as depicted in Fig.~\ref{fig:2}.
A number of intriguing features are observed at the interface between the culet and the sample chamber, which provide insight into both elastic (reversible) and plastic (irreversible) deformations. 

At low pressures ($P=4.9$~GPa), the normal stress along the loading axis, $\sigma_{ZZ}$,  is spatially uniform (Fig.~\ref{fig:2}A), while all shear stresses, $\{\sigma_{XY}$, $\sigma_{XZ}$, $\sigma_{YZ}\}$, are minimal (Fig.~\ref{fig:2}B)\cite{footnote1}.
These observations are in agreement with conventional stress continuity predictions for the interface between a solid and an ideal fluid \cite{falkovich_2018}.
Moreover, $\sigma_{ZZ}$ is consistent with the independently measured pressure inside the sample chamber (via ruby fluorescence), demonstrating the NV's potential as a built-in pressure scale \cite{DewaeleHighPressure}.
In contrast to the uniformity of $\sigma_{ZZ}$, the field profile for the mean lateral stress, $\sigma_\perp\equiv\frac{1}{2}(\sigma_{XX}+\sigma_{YY})$, exhibits a concentration of forces toward the center of the culet 
(Fig.~\ref{fig:2}A). 
Using the measured $\sigma_{ZZ}$ as a boundary condition, we perform finite element simulations to reproduce this spatial pattern\cite{supp}. %\red{which shows an excellent agreement with the measured data} 

Upon increasing pressure ($P=13.6$~GPa), a pronounced spatial gradient in $\sigma_{ZZ}$ emerges (Fig.~\ref{fig:2}B inset). 
This qualitatively distinct feature is consistent with the solidification of the pressure-transmitting medium  into its glassy phase above $P_g \approx 10.5$ GPa \cite{Klotz:2009kc}.
Crucially, this 
 demonstrates our ability to characterize the effective viscosity of solids and liquids under pressure.
To characterize the sensitivity of our system, we perform ODMR spectroscopy with a static applied magnetic field and pressure under varying integration times and extract the frequency uncertainty from a Gaussian fit. We observe a stress sensitivity of $\{0.023,0.030,0.027\}$ GPa/$\sqrt{\textrm{Hz}}$ for hydrostatic, average normal, and average shear stresses, respectively. This is consistent with the theoretically derived stress sensitivity, $\eta_S \sim\frac{\Delta\nu}{\xi C \sqrt{Nt}} = \{0.017, 0.022, 0.020\}$ GPa/$\sqrt{\textrm{Hz}}$, respectively, where $N$ is the number of NV centers, $\Delta\nu$ is the linewidth, $\xi$ is the relevant stress susceptibility, $t$ is the integration time, and $C$ is an overall factor accounting for measurement infidelity \cite{supp}.
In combination with diffraction-limited imaging resolution, this sensitivity opens the door to measuring and ultimately controlling the full stress tensor distribution across a sample.

Having characterized the stress environment, we now utilize the NV centers as an \emph{in situ} magnetometer to detect phase transitions inside the high-pressure chamber. 
Analogous to the case of stress, we observe a magnetic sensitivity of $12 ~\mathrm{\mu}$T/$\sqrt{\textrm{Hz}}$, in agreement with the theoretically estimated value, $\eta_B \sim\frac{\delta\nu}{C \gamma_\textrm{B}\sqrt{Nt}}=8.8~\mathrm{\mu}$T/$\sqrt{\textrm{Hz}}$.
Assuming a point dipole located a distance $d \sim 5~\mu$m from the NV layer, this corresponds to an experimentally measured magnetic moment sensitivity: $7.5\times 10^{-12}$~emu/$\sqrt{\textrm{Hz}}$ (Fig.~\ref{fig:1}F).

\begin{figure}[p]
    \centering
    \includegraphics[width=0.65\textwidth]{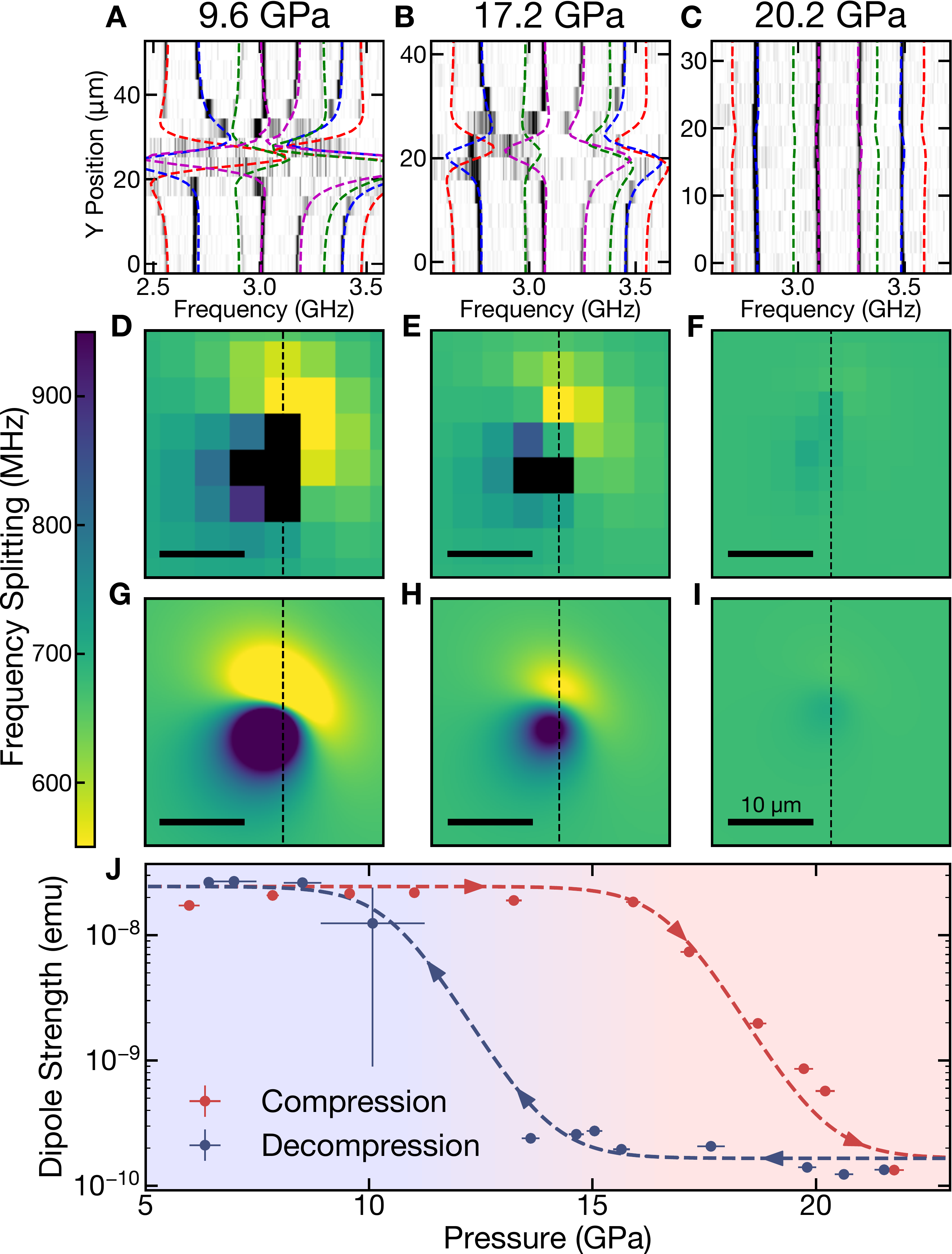}
    \caption{\small
        \textbf{Imaging iron's $\alpha \leftrightarrow \epsilon$ phase transition.}
        Applying an external magnetic field (${\bf B}_\textrm{ext}$$\sim$$180$~G) induces a dipole moment in the polycrystalline iron pellet which generates a spatially varying magnetic field across the culet of the diamond anvil.
        By mapping the ODMR spectra across the culet surface, we reconstruct the local magnetic field which characterizes the iron pellet's magnetization.
        (\textbf{A}-\textbf{C}) Comparison between the measured ODMR spectra (dark regions correspond to resonances) 
        and the theoretical resonance positions (different colors correspond to different NV crystallographic orientations)
        across vertical spatial cuts at pressures 9.6~GPa, 17.2~GPa and 20.2~GPa, respectively (16:3:1 methanol/ethanol/water solution). 
        (\textbf{D}-\textbf{F}) Map of the measured energy difference of a particular NV crystallographic orientation (blue lines in (A-C)). 
        Black pixels correspond to ODMR spectra where the splitting could not be accurately extracted owing to large magnetic field gradients \cite{supp}.
        (\textbf{G}-\textbf{I}) Theoretical reconstruction of the energy differences shown in (D-F).
        Data depicted in (A-C) are taken along the thin black dashed lines.
        (\textbf{J}) Measured dipole moment of the iron pellet as a function of applied pressure at room temperature, for both compression (red) and decompression (blue).
        Based on the hysteresis observed ($\sim$$6$~GPa), we find the critical pressure $P_c = 13.6\pm 3.6$~GPa, in excellent agreement with previous studies \cite{taylor1991hysteresis}.
      }
      \label{fig:3}
  \end{figure}

Sensitivity in hand, we begin by directly measuring the magnetization of iron as it undergoes the pressure-driven $\alpha\leftrightarrow \epsilon$ phase transition from body-centered cubic (bcc) to hexagonal close-packed (hcp) crystal structures \cite{taylor1991hysteresis}; crucially, this structural phase transition is accompanied by the depletion of the %\blue{[removed permanent]} %\red{\cite{Jeanloz:1990fe,Stixrude:1994gh}} 
magnetic moment, and it is this change in the iron's magnetic behavior that we image.
Our sample chamber is loaded with a $\sim$$10~\mathrm{\mu m}$ polycrystalline iron pellet as well as a ruby microsphere (pressure scale), and we apply an external magnetic field ${\bf B}_\textrm{ext}$$\sim$$180$~G. 
As before, by performing a confocal scan across the culet, we acquire a two-dimensional magnetic resonance map (Fig.~\ref{fig:3}).
At low pressures (Fig.~\ref{fig:3}A), near the iron pellet, we observe significant shifts in the eight NV resonances, owing to the presence of a ferromagnetic field from the iron pellet.
As one increases pressure (Fig.~\ref{fig:3}B), these shifts begin to diminish, signaling a reduction in the magnetic susceptibility.
Finally, at the highest pressures ($P\sim 22$~GPa, Fig.~\ref{fig:3}C), the magnetic field from the pellet has reduced by over two orders of magnitude.

To quantify this phase transition, we reconstruct the full vector magnetic field produced by the iron sample from the aforementioned two-dimensional NV magnetic resonance maps (Fig.~\ref{fig:3}D-F). 
We then compare this information with the expected field distribution at the NV layer inside the culet, assuming the iron pellet generates a dipole field \cite{supp}. 
This enables us to extract an effective dipole moment as a function of applied pressure (Fig.~\ref{fig:3}G). 
In order to identify the critical pressure, we fit the transition using a logistic function \cite{supp}. %\red{\cite{aoki2000physics}}.
This procedure yields the transition at $P= 16.7\pm 0.7~\mathrm{GPa}$ (Fig.~\ref{fig:3}J).

In addition to changes in the magnetic behavior, another key signature of this first order transition is the presence of hysteresis.
We investigate this by slowly decompressing the diamond anvil cell and monitoring the dipole moment; the decompression transition occurs at $P=10.5\pm 0.7$~GPa (Fig.~\ref{fig:3}J), suggesting a hysteresis width of approximately $\sim$$6$ GPa, consistent with a combination of intrinsic hysteresis and finite shear stresses in the methanol/ethanol/water pressure-transmitting medium \cite{taylor1991hysteresis}.
Taking the average of the forward and backward hysteresis pressures, we find a critical pressure of $P_\textrm{c} = 13.6\pm3.6$~GPa, in excellent agreement with independent measurements by M\"{o}ssbauer spectroscopy, where $P_\textrm{c} \approx 12~$GPa (Fig.~\ref{fig:3}J) \cite{taylor1991hysteresis}.

\begin{figure}[ht]
  \centering %\hspace{-15mm}
      \includegraphics[width=0.75\textwidth]{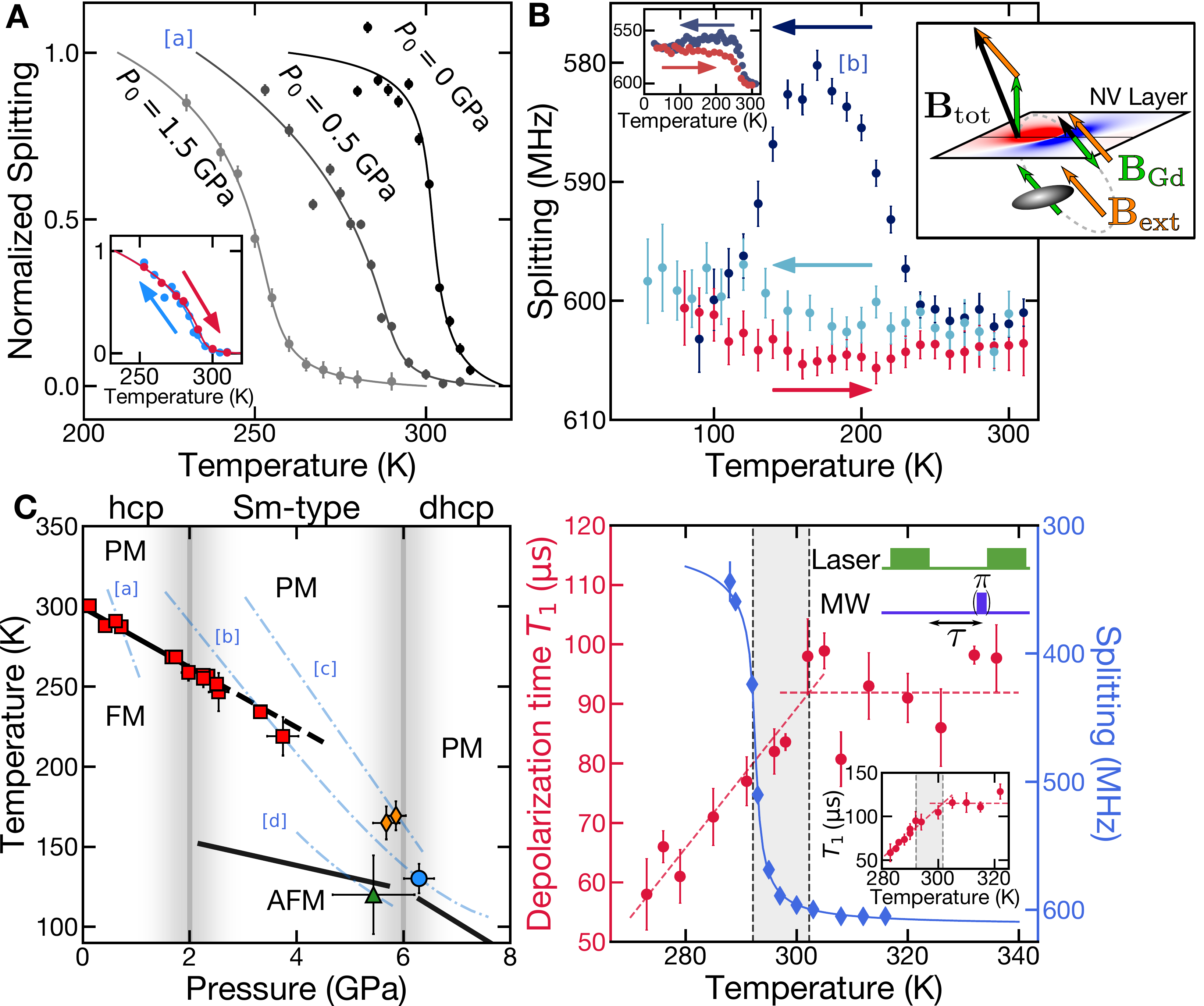}
\caption{\small %\scriptsize 
    \textbf{Magnetic $P$-$T$ phase diagram of gadolinium.}
    A $\sim 30~\mathrm{\mu m}\times30~\mathrm{\mu m}\times25~\mathrm{\mu m}$ polycrystalline Gd foil is loaded into a beryllium copper gasket with a cesium iodide pressure medium. 
    An external magnetic field, ${\bf B}_\textrm{ext}$$\sim$$120$~G,  induces a dipole field, ${\bf B}_\textrm{Gd}$, detected by the splitting of the NVs (right inset, (B)). 
        (\textbf{A}) The FM Curie temperature $T_\textrm{C}$ decreases with increasing pressure up to $\sim$$4$~GPa. 
        NV splittings for three $P$-$T$ paths, labeled by their initial pressure $P_0$, are shown. 
        The $P$-$T$ path for run [a] ($P_0 = 0.5$~GPa) is shown in (C).
        (\textbf{Inset A}) depicts the cool-down (blue) and heat-up (red) of a single $P$-$T$ cycle, which shows negligible hysteresis. 
         (\textbf{B}) 
         If a $P$-$T$ path starting in hcp is taken into the dhcp phase (at pressures $\gtrsim 6$~GPa) \cite{JAYARAMAN1978707}, %\red{\cite{PhysRevB.71.184416,IWAMOTO2003667,PhysRev.139.A682,doi:10.1080/08957959.2014.977277}},
         the FM signal is lost and not reversible.
         Such a $P$-$T$ path [b], is shown in {\bf C}.
         On cool-down (dark blue), we observe the aforementioned Curie transition, followed by the loss of FM signal at $6.3$ GPa, $130$ K.
         But upon heat-up (red) and second cool-down (light blue), the FM signal is not recovered.
         ({\bf Left Inset}) When the pressure does not go beyond $\sim 6$~GPa, the FM signal is recoverable \cite{supp}.
          (\textbf{C}) Magnetic $P$-$T$ phase diagram of Gd.
          At low pressures, we observe the linear decrease of $T_\mathrm{C}$ (black line) with slope $-18.7\pm0.2$~K/GPa, in agreement with previous measurements \cite{JAYARAMAN1978707}.
          This linear regime extends into the Sm-type phase (black dashed line) due to the slow dynamics of the hcp $\to$ Sm-type transition \cite{JAYARAMAN1978707}. %\red{\cite{PhysRev.139.A682, IWAMOTO2003667,PhysRevB.71.184416}}.
          When starting in the Sm-type phase, we no longer observe a FM signal, but rather a small change in the magnetic field at either the transition from Sm-type to dhcp (orange diamonds) or from PM to AFM (green triangle), depending on the $P$-$T$ path.
          The bottom two phase boundaries (black lines) are taken from Ref.~\cite{doi:10.1080/08957959.2014.977277}.
        (\textbf{D}) At ambient pressure, we observe a Curie temperature, $T_\textrm{C} = 292.2\pm0.1$~K, via 
         DC magnetometry (blue data).
        Using nanodiamonds drop-cast onto a Gd foil (and no applied external magnetic field), we find that the depolarization time ($T_1$) of the NVs is qualitatively different in the two phases (red data). 
        $T_1$ is measured using the pulse sequence shown in the top right inset. 
        ({\bf Bottom inset}) The $T_1$ measurement on another nanodiamond exhibits nearly identical behavior.
    }
  \label{fig:4}
\end{figure}

Next, we demonstrate the integration of our platform into a cryogenic system, enabling us to make spatially resolved \emph{in situ} measurements across the pressure-temperature ($P$-$T$) phase diagram of materials. 
Specifically, we investigate the  magnetic $P$-$T$ phase diagram of the rare-earth element gadolinium (Gd) up to pressures $P \approx 8~$GPa and between temperatures $T=25-340$~K. %\cite{Dankov:1998em} -- dont think we need to cite Dankov. Their measurements are at ambient pressure
Owing to an interplay between localized 4f electrons and mobile conduction electrons, Gd represents an interesting playground for studying metallic magnetism; in particular, the itinerant electrons mediate RKKY-type interactions between the local moments, which in turn induce spin-polarization of the itinerant electrons \cite{oroszlany2015}.  %\red{\cite{Santos2004}}.
Moreover, much like its other rare-earth cousins, Gd exhibits a series of pressure-driven structural phase transitions from hexagonal close-packed (hcp) to samarium-type (Sm-type) to double hexagonal close-packed (dhcp) (Fig.~\ref{fig:4})\cite{JAYARAMAN1978707}. %Replace all references contained within this review
The interplay between these different structural phases, various types of magnetic ordering and metastable transition dynamics leads to a complex magnetic $P$-$T$ phase diagram that remains the object of study to this day \cite{JAYARAMAN1978707, doi:10.1080/08957959.2014.977277, oroszlany2015}. %Removed McWhan, Jayaraman 1964

In analogy to our measurements of iron, we monitor the magnetic ordering of a Gd flake via the NV's ODMR spectra at two different locations inside the culet: close to and far away from the sample (the latter to be used as a control) \cite{supp}.
Due to thermal contraction of the DAC (which induces a change in pressure), each experimental run traces a distinct non-isobaric path through the $P$-$T$ phase diagram (Fig.~\ref{fig:4}C, blue curves).
In addition to these DC magnetometry measurements, we also operate the NV sensors in a complementary mode, i.e.~as a noise spectrometer.

We begin by characterizing Gd's well-known ferromagnetic Curie transition at ambient pressure, which induces a sharp jump in the splitting of the NV resonances at $T_\textrm{C} = 292.2\pm0.1$~K (Fig.~4D).
As depicted in Fig.~4A, upon increasing pressure, this transition shifts to lower temperatures, and consonant with its second order nature \cite{PhysRevB.38.2862}, we observe no hysteresis; this motivates us to fit the data and extract $T_\textrm{C}$ by solving a regularized Landau free-energy equation \cite{supp}. 
Combining all of the low pressure data (Fig.~4C, red squares), we find a linear decrease in the Curie temperature at a rate: $dT_\textrm{C} / dP = -18.7\pm0.2 $~K/GPa, consistent with prior studies via both DC conductivity and AC-magnetic susceptibility \cite{JAYARAMAN1978707}. %\red{\cite{PhysRev.139.A682,IWAMOTO2003667, PhysRevB.71.184416, PhysRevB.40.9541}}.
Surprisingly, this linear decrease extends well into the Sm-type phase.
Upon increasing pressure above $\sim 6$~GPa (path [b] in Fig.~4C), we observe the loss of ferromagnetic (FM) signal (Fig.~4B), indicating a first order structural transition into the paramagnetic (PM) dhcp phase
\cite{JAYARAMAN1978707}. %\red{\cite{AKELLA1988573,PhysRevB.71.184416,IWAMOTO2003667,PhysRev.139.A682,doi:10.1080/08957959.2014.977277}}\blue{[See previous blue comment--keep that one]}.
In stark contrast to the previous Curie transition, there is
% espy
no revival of a ferromagnetic signal even after heating up ($\sim$$315$ K) and significantly reducing the pressure ($< 0.1$ GPa).

A few remarks are in order.
The linear decrease of $T_\textrm{C}$ well beyond the $\sim$$2$~GPa structural transition between hcp and Sm-type is consistent with the ``sluggish" equilibration between these two phases at low temperatures
\cite{JAYARAMAN1978707}. %\red{\cite{PhysRev.139.A682}}.
The metastable dynamics of this transition are strongly pressure and temperature dependent, suggesting that different starting points (in the $P$-$T$ phase diagram) can exhibit dramatically different behaviors \cite{JAYARAMAN1978707}.
To highlight this, we probe two \textit{different} transitions out of the paramagnetic Sm-type phase by tailoring specific paths in the $P$-$T$ phase diagram. 
By taking a shallow path in $P$-$T$ space, we observe a small change in the local magnetic field across the \textit{structural} transition into the PM dhcp phase at $\sim$$6$~GPa (Fig.~\ref{fig:4}C path [c], orange diamonds).
%
%We note that this transition is fundamentally distinct from the aforementioned FM hcp to PM dhcp transition since both the initial and final states are paramagnetic, albeit with different suscepitilities. 
%
By taking a steeper path in $P$-$T$ space, one can also investigate the \textit{magnetic} transition into the antiferromagnetic (AFM) Sm-type phase at  $\sim$$150$~K  (Fig.~\ref{fig:4}C path [d], green triangle).
In general, these two transitions are extremely challenging to probe via DC magnetometry since their signals arise only from small differences in the susceptibilities between the various phases \cite{supp}.

To this end, we demonstrate a complementary NV sensing modality based upon
% based on??
noise spectroscopy, which can probe phase transitions even in the absence of a direct magnetic signal \cite{chatterjee2018diagnosing}. 
Specifically, returning to Gd's ferromagnetic Curie transition, we %use nanodiamonds at ambient pressure to 
monitor the NV's depolarization time, $T_1$, as one crosses the phase transition (Fig.~\ref{fig:4}D). 
Normally, the NV's $T_1$ time is limited by spin-phonon interactions and \emph{increases} dramatically as one decreases temperature. 
Here, we observe a strikingly disparate behavior. 
In particular, using nanodiamonds drop-cast on a Gd foil at ambient pressure, we find that the NV $T_1$ is nearly temperature independent in the paramagnetic phase, before exhibiting a kink and subsequent \emph{decrease} as one enters the ferromagnetic phase (Fig.~\ref{fig:4}D). %\blue{[Emphasize AC magnetometry]}
We note two intriguing observations: 
first, one possible microscopic explanation for this behavior is that $T_1$ is dominated by Johnson-Nyquist noise from the thermal fluctuations of charge carriers inside Gd
\cite{Kolkowitz1129,footnote4}. 
Gapless critical spin fluctuations or magnons in the ordered phase, while expected, are less likely to cause this signal \cite{supp}.
Second, we observe that the Curie temperature, as identified via $T_1$-noise spectroscopy, is $\sim$$10$~K higher than that observed via DC magnetometry (Fig.~\ref{fig:4}D).
Similar behavior has previously been reported for the surface of Gd  \cite{oroszlany2015,PhysRevLett.71.444}, %\red{\cite{PhysRevLett.54.1555}} \blue{[FM: Not sure which one is best to keep]}, 
suggesting %the possibility
that our noise spectroscopy %could be 
could be more sensitive to surface physics.

In summary, we have developed a hybrid platform that integrates quantum sensors into diamond anvil cells. 
For the first time, the full stress tensor can be mapped across the sample and gasket, as a function of pressure.  
This provides essential information for investigating mechanical phenomena, such as pressure-dependent yield strength, viscous flow of fluids and plastic deformation of solids, and may ultimately allow control of the deviatoric- as well as normal-stress conditions in high pressure experiments.
Crucially, such information is challenging to obtain via either numerical finite-element simulations or more conventional experimental methods \cite{feng2016large}. %\red{\cite{hanfland1985raman,novikov1994numerical}}.
In the case of magnetometry, the high sensitivity and close proximity of our sensor enables one to probe signals that are beyond the capabilities of existing techniques (Fig.~\ref{fig:1}F); these include for example, nuclear magnetic resonance (NMR) at picoliter volumes \cite{kehayias2017solution} and single grain remnant magnetism  \cite{glenn2017micrometer}, as well as phenomena that exhibit spatial textures such as magnetic skyrmions \cite{dovzhenko2018magnetostatic} and superconducting vortices \cite{thiel2016quantitative}. 

While our work utilizes NV centers, the techniques developed here can be readily extended to other atomic defects.
For instance, recent developments on all-optical control of silicon-vacancy centers in diamond may allow for microwave-free stress imaging with improved sensitivities \cite{Meesala2018}.
In addition, one can consider defects in other anvil substrates beyond diamond; indeed, recent studies have shown that moissanite (6H silicon carbide) hosts optically active defects that show promise as local sensors \cite{falk2013}. 
In contrast to millimeter-scale diamond anvils, moissanite anvils can be  manufactured at the centimeter-scale or larger, and therefore support larger sample volumes that ameliorate the technical requirements of many experiments. %\cite{xu2004large}. 
Finally, the suite of sensing capabilities previously demonstrated for NV centers (i.e.~electric, thermal, gryroscopic precession etc.) can now straightforwardly be extended to high pressure environments, opening up an enormous new range of experiments for quantitatively characterizing materials at such extreme conditions which can test, extend and validate first-principles theory.

\section*{Acknowledgements}
We gratefully acknowledge fruitful discussions with Z. Geballe, G. Samudrala, R. Zieve, J. Jeffries, E. Zepeda-Alarcon, M. Kunz, I. Kim, J. Choi, K. de Greve, P. Maurer, S. Lewin, and D.-H. Lee. We are especially grateful to M. Doherty and M. Barson for sharing their raw data on stress susceptibilities.
We thank C. Laumann for introducing us to the idea of integrating NV centers into diamond anvil cells.
We thank D. Budker, J. Analytis, A. Jarmola, M. Eremets, R. Birgeneau, F. Hellman, R. Ramesh for careful readings of the manuscript. 

\section*{Funding}
This work was supported as part of the Center for Novel
Pathways to Quantum Coherence in Materials, an Energy
Frontier Research Center funded by the U.S. Department of
Energy, Office of Science, Basic Energy Sciences under
Award No. DE-AC02-05CH11231. 
SH acknowledges support from the National Science Foundation Graduate Research Fellowship under Grant No.~DGE-1752814.
VIL and MK acknowledge support from Army Research Office (Grant W911NF-17-1-0225).

\section*{Author Contributions} 
All authors contributed extensively to all aspects of this work. 

\section*{Competing interests} 
The authors declare no competing financial interests. 

\section*{Data and materials availability} 
The data presented in this study are available from the corresponding author on request.

\bibliography{references.bib}

\begin{thebibliography}{10}

\bibitem{Kucsko:2013bio2}
G.~Kucsko, {\it et~al.\/}, {\it Nature\/} {\bf 500}, 54 EP  (2013).

\bibitem{maletinsky2012}
P.~Maletinsky, {\it et~al.\/}, {\it Nature nanotechnology\/} {\bf 7}, 320
  (2012).

\bibitem{cai:2014hybrid1}
J.~Cai, F.~Jelezko, M.~B. Plenio, {\it Nature Communications\/} {\bf 5}, 4065
  EP  (2014). Article.

\bibitem{dovzhenko2018magnetostatic}
Y.~Dovzhenko, {\it et~al.\/}, {\it Nature communications\/} {\bf 9}, 2712
  (2018).

\bibitem{Jayaraman:1983}
A.~Jayaraman, {\it Rev. Mod. Phys.\/} {\bf 55}, 65 (1983).

\bibitem{Mao:2018fd}
H.-k. Mao, X.-J. Chen, Y.~Ding, B.~Li, L.~Wang, {\it Reviews of Modern
  Physics\/} {\bf 90}, 404 (2018).

\bibitem{wigner1935possibility}
E.~Wigner, H.~{\'a}. Huntington, {\it The Journal of Chemical Physics\/} {\bf
  3}, 764 (1935).

\bibitem{horii1986brittle}
H.~Horii, S.~Nemat-Nasser, {\it Philosophical Transactions of the Royal Society
  of London. Series A, Mathematical and Physical Sciences\/} {\bf 319}, 337
  (1986).

\bibitem{Gilioli_Ehm_2014}
E.~Gilioli, L.~Ehm, {\it IUCrJ\/} {\bf 1}, 590–603 (2014).

\bibitem{Drozdov:2015conventional}
A.~P. Drozdov, M.~I. Eremets, I.~A. Troyan, V.~Ksenofontov, S.~I. Shylin, {\it
  Nature\/} {\bf 525}, 73 EP  (2015).

\bibitem{doherty2013nitrogen}
M.~W. Doherty, {\it et~al.\/}, {\it Physics Reports\/} {\bf 528}, 1 (2013).

\bibitem{ovartchaiyapong2014dynamic}
P.~Ovartchaiyapong, K.~W. Lee, B.~A. Myers, A.~C.~B. Jayich, {\it Nature
  communications\/} {\bf 5}, 4429 (2014).

\bibitem{acosta2010temperature}
V.~Acosta, {\it et~al.\/}, {\it Physical review letters\/} {\bf 104}, 070801
  (2010).

\bibitem{maze2008nanoscale}
J.~Maze, {\it et~al.\/}, {\it Nature\/} {\bf 455}, 644 (2008).

\bibitem{dolde2011electric}
F.~Dolde, {\it et~al.\/}, {\it Nature Physics\/} {\bf 7}, 459 (2011).

\bibitem{Casola:2018nvmagnetometry1}
F.~Casola, T.~van~der Sar, A.~Yacoby, {\it Nature Reviews Materials\/} {\bf 3},
  17088 EP  (2018). Review Article.

\bibitem{mittiga2018imaging}
T.~Mittiga, {\it et~al.\/}, {\it Physical Review Letters\/} {\bf 121}, 246402
  (2018).

\bibitem{Barson:2017ba}
M.~S.~J. Barson, {\it et~al.\/}, {\it Nano Letters\/} {\bf 17}, 1496 (2017).

\bibitem{supp}
See Supplementary Material for additional details.

\bibitem{Sterer:1998hl}
E.~Sterer, M.~P. Pasternak, R.~D. Taylor, {\it Review of Scientific
  Instruments\/} {\bf 61}, 1117 (1998).

\bibitem{footnote1}
We note that $\{\hat{X},\hat{Y},\hat{Z}\}$ corresponds to the lab frame while
  $\{\hat{x},\hat{y},\hat{z}\}$ corresponds to the NV frame (Fig.~1).

\bibitem{falkovich_2018}
G.~Falkovich, {\it Fluid Mechanics\/} (Cambridge University Press, 2018),
  second edn.

\bibitem{DewaeleHighPressure}
A.~Dewaele, P.~Loubeyre, M.~Mezouar, {\it PHYSICAL REVIEW B\/} {\bf 70} (2004).

\bibitem{Klotz:2009kc}
S.~Klotz, J.-C. Chervin, P.~Munsch, G.~Le~Marchand, {\it Journal of Physics D:
  Applied Physics\/} {\bf 42}, 075413 (2009).

\bibitem{taylor1991hysteresis}
R.~Taylor, M.~Pasternak, R.~Jeanloz, {\it Journal of Applied Physics\/} {\bf
  69}, 6126 (1991).

\bibitem{oroszlany2015}
L.~Oroszl{\'a}ny, A.~De{\'a}k, E.~Simon, S.~Khmelevskyi, L.~Szunyogh, {\it
  Physical review letters\/} {\bf 115}, 096402 (2015).

\bibitem{JAYARAMAN1978707}
A.~Jayaraman, {\it Metals\/} (Elsevier, 1978), vol.~1 of {\it Handbook on the
  Physics and Chemistry of Rare Earths\/}, pp. 707 -- 747.

\bibitem{doi:10.1080/08957959.2014.977277}
G.~K. Samudrala, G.~M. Tsoi, S.~T. Weir, Y.~K. Vohra, {\it High Pressure
  Research\/} {\bf 34}, 385 (2014).

\bibitem{PhysRevB.38.2862}
P.~Hargraves, R.~A. Dunlap, D.~J.~W. Geldart, S.~P. Ritcey, {\it Phys. Rev.
  B\/} {\bf 38}, 2862 (1988).

\bibitem{chatterjee2018diagnosing}
S.~Chatterjee, J.~F. Rodriguez-Nieva, E.~Demler, {\it arXiv preprint
  arXiv:1810.04183\/}  (2018).

\bibitem{Kolkowitz1129}
S.~Kolkowitz, {\it et~al.\/}, {\it Science\/} {\bf 347}, 1129 (2015).

\bibitem{footnote4}
As opposed to isolated NV samples, where T$_1$ is limited by spin-phonon
  interactions.

\bibitem{PhysRevLett.71.444}
H.~Tang, {\it et~al.\/}, {\it Phys. Rev. Lett.\/} {\bf 71}, 444 (1993).

\bibitem{feng2016large}
B.~Feng, V.~I. Levitas, R.~J. Hemley, {\it International Journal of
  Plasticity\/} {\bf 84}, 33 (2016).

\bibitem{kehayias2017solution}
P.~Kehayias, {\it et~al.\/}, {\it Nature communications\/} {\bf 8}, 188 (2017).

\bibitem{glenn2017micrometer}
D.~R. Glenn, {\it et~al.\/}, {\it Geochemistry, Geophysics, Geosystems\/} {\bf
  18}, 3254 (2017).

\bibitem{thiel2016quantitative}
L.~Thiel, {\it et~al.\/}, {\it Nature nanotechnology\/} {\bf 11}, 677 (2016).

\bibitem{Meesala2018}
S.~Meesala, {\it et~al.\/}, {\it Phys. Rev. B\/} {\bf 97}, 205444 (2018).

\bibitem{falk2013}
A.~L. Falk, {\it et~al.\/}, {\it Nature Communications\/} {\bf 4}, 1819 (2013).

\end{thebibliography}


\begin{thebibliography}{10}

\bibitem{doherty2013nitrogen}
M.~W. Doherty, {\it et~al.\/}, {\it Physics Reports\/} pp. 1--45 (2013).

\bibitem{Dreau:2011ec}
A.~Dr{\'e}au, {\it et~al.\/}, {\it Physical Review B\/} {\bf 84}, 195204
  (2011).

\bibitem{Feng:2014coil2}
Y.~Feng, D.~M. Silevitch, T.~F. Rosenbaum, {\it Review of Scientific
  Instruments\/} {\bf 85}, 033901 (2014).

\bibitem{Mito:2003coil3}
M.~Mito, {\it et~al.\/}, {\it Phys. Rev. B\/} {\bf 67}, 024427 (2003).

\bibitem{Jackson:2003da}
D.~Jackson, {\it et~al.\/}, {\it Review of scientific instruments\/} {\bf 74},
  2467 (2003).

\bibitem{Alireza:2009squid2}
P.~L. Alireza, G.~G. Lonzarich, {\it Review of Scientific Instruments\/} {\bf
  80}, 023906 (2009).

\bibitem{Mito:2001squid3}
M.~Mito, {\it et~al.\/}, {\it Japanese Journal of Applied Physics\/} {\bf 40},
  6641 (2001).

\bibitem{Giriat:2010squid4}
G.~Giriat, W.~Wang, J.~P. Attfield, A.~D. Huxley, K.~V. Kamenev, {\it Review of
  Scientific Instruments\/} {\bf 81}, 073905 (2010).

\bibitem{Takeda:2002squid6}
K.~Takeda, M.~Mito, {\it Journal of the Physical Society of Japan\/} {\bf 71},
  729 (2002).

\bibitem{Marizy:2017squid7}
A.~Marizy, B.~Guigue, F.~Occelli, B.~Leridon, P.~Loubeyre, {\it High Pressure
  Research\/} {\bf 37}, 465 (2017).

\bibitem{Pasternak:1991mossbauer1}
M.~P. Pasternak, R.~D. Taylor, R.~Jeanloz, {\it Journal of Applied Physics\/}
  {\bf 70}, 5956 (1991).

\bibitem{Pasternak:2001mossbauer2}
M.~P. Pasternak, {\it et~al.\/}, {\it Phys. Rev. B\/} {\bf 65}, 035106 (2001).

\bibitem{Kantor:2004mossbauer3}
A.~P. Kantor, {\it et~al.\/}, {\it Phys. Rev. Lett.\/} {\bf 93}, 215502 (2004).

\bibitem{Mathon:2004it}
O.~Mathon, {\it et~al.\/}, {\it Journal of Synchrotron Radiation\/} {\bf 11},
  423 (2004).

\bibitem{Ishimatsu:2007eu}
N.~Ishimatsu, {\it et~al.\/}, {\it Physical Review B\/} {\bf 75}, 180402
  (2007).

\bibitem{Watanabe:2011hk}
S.~Watanabe, {\it et~al.\/}, {\it Journal of the Physical Society of Japan\/}
  {\bf 80}, 093705 (2011).

\bibitem{Chen:2018}
K.~Chen, {\it et~al.\/}, {\it Phys. Rev. B\/} {\bf 97}, 235153 (2018).

\bibitem{rittweger2009resolution}
E.~Rittweger, K.~Y. Han, S.~E. Irvine, C.~Eggeling, S.~W. Hell, {\it Nature
  Photonics\/} {\bf 3}, 144 (2009).

\bibitem{mittiga2018imaging}
T.~Mittiga, {\it et~al.\/}, {\it Physical Review Letters\/} {\bf 121}, 246402
  (2018).

\bibitem{Barson:2017ba}
M.~S.~J. Barson, {\it et~al.\/}, {\it Nano Letters\/} {\bf 17}, 1496 (2017).

\bibitem{MezouarRubyPressure}
A.~Dewaele, P.~Loubeyre, M.~Mezouar, {\it Physical Review B\/} {\bf 70} (2004).

\bibitem{oroszlany2015}
L.~Oroszl{\'a}ny, A.~De{\'a}k, E.~Simon, S.~Khmelevskyi, L.~Szunyogh, {\it
  Physical review letters\/} {\bf 115}, 096402 (2015).

\bibitem{JAYARAMAN1978707}
A.~Jayaraman, {\it Metals\/} (Elsevier, 1978), vol.~1 of {\it Handbook on the
  Physics and Chemistry of Rare Earths\/}, pp. 707 -- 747.

\bibitem{PhysRevB.71.184416}
D.~D. Jackson, V.~Malba, S.~T. Weir, P.~A. Baker, Y.~K. Vohra, {\it Phys. Rev.
  B\/} {\bf 71}, 184416 (2005).

\bibitem{PhysRev.139.A682}
D.~B. McWhan, A.~L. Stevens, {\it Phys. Rev.\/} {\bf 139}, A682 (1965).

\bibitem{IWAMOTO2003667}
T.~Iwamoto, M.~Mito, M.~Hidaka, T.~Kawae, K.~Takeda, {\it Physica B: Condensed
  Matter\/} {\bf 329-333}, 667  (2003). Proceedings of the 23rd International
  Conference on Low Temperature Physics.

\bibitem{doi:10.1080/08957959.2014.977277}
G.~K. Samudrala, G.~M. Tsoi, S.~T. Weir, Y.~K. Vohra, {\it High Pressure
  Research\/} {\bf 34}, 385 (2014).

\bibitem{AKELLA1988573}
J.~Akella, G.~S. Smith, A.~P. Jephcoat, {\it Journal of Physics and Chemistry
  of Solids\/} {\bf 49}, 573  (1988).

\bibitem{Agarwal2017}
K.~Agarwal, {\it et~al.\/}, {\it Phys. Rev. B\/} {\bf 95}, 155107 (2017).

\bibitem{Nigh63}
H.~E. Nigh, S.~Legvold, F.~H. Spedding, {\it Phys. Rev.\/} {\bf 132}, 1092
  (1963).

\bibitem{Jacobsson89}
P.~Jacobsson, B.~Sundqvist, {\it Phys. Rev. B\/} {\bf 40}, 9541 (1989).

\bibitem{Colvin60}
R.~V. Colvin, S.~Legvold, F.~H. Spedding, {\it Phys. Rev.\/} {\bf 120}, 741
  (1960).

\bibitem{Bodryakov99}
V.~Y. Bodryakov, A.~A. Povzner, O.~G. Zelyukova, {\it Physics of the Solid
  State\/} {\bf 41}, 1138 (1999).

\bibitem{FisherLanger}
M.~E. Fisher, J.~S. Langer, {\it Phys. Rev. Lett.\/} {\bf 20}, 665 (1968).

\bibitem{KogutRMP}
J.~B. Kogut, {\it Rev. Mod. Phys.\/} {\bf 51}, 659 (1979).

\bibitem{CRD2018}
S.~{Chatterjee}, J.~F. {Rodriguez-Nieva}, E.~{Demler}, {\it ArXiv e-prints\/}
  p. arXiv:1810.04183 (2018).

\end{thebibliography}

\end{document}

% Double-space the manuscript.

\author{
\hspace{-13mm}
S.~Hsieh,$^{1,2,*}$ 
P.~Bhattacharyya,$^{1,2,*}$ %\protect\CoAuthorMark 
C. Zu,$^{1,*}$ %\protect\CoAuthorMark 
T. Mittiga,$^1$ 
T. J. Smart,$^3$ 
F. Machado,$^1$\\
\hspace{-13mm}
B. Kobrin,$^{1,2}$ 
T. O. H{\"o}hn,$^{1,4}$ 
N. Z. Rui,$^{1}$ 
M. Kamrani,$^5$ 
S. Chatterjee,$^1$ 
S. Choi,$^1$\\
\hspace{-13mm}
M. Zaletel,$^1$ 
V. V. Struzhkin,$^6$ 
J. E. Moore,$^{1,2}$ 
V. I. Levitas,$^{5,7}$ 
R. Jeanloz,$^3$ 
N. Y. Yao$^{1,2,\dag}$\\
\\
\normalsize{\hspace{-13mm}$^{1}$Department of Physics, University of California, Berkeley, CA 94720, USA}\\
\normalsize{\hspace{-13mm}$^{2}$Materials Science Division, Lawrence Berkeley National Laboratory, Berkeley, CA 94720, USA}\\
\normalsize{\hspace{-13mm}$^{3}$Department of Earth and Planetary Science, University of California, Berkeley, CA 94720, USA}\\
\normalsize{\hspace{-13mm}$^{4}$Fakult{\"a}t f{\"u}r Physik, Ludwig-Maximilians-Universit{\"a}t M{\"u}nchen, 80799 Munich, Germany}\\
\normalsize{\hspace{-13mm}$^{5}$Department of Aerospace Engineering, Iowa State University, Ames, IA 50011, USA}\\
\normalsize{\hspace{-13mm}$^{6}$Geophysical Laboratory, Carnegie Institution of Washington, Washington, DC 20015, USA}\\
\normalsize{\hspace{-13mm}$^{7}$Departments of Mechanical Engineering and Material Science and Engineering,}\\
\normalsize{\hspace{-13mm}Iowa State University, Ames, IA 50011, USA}\\
\normalsize{\hspace{-13mm}$^\dag$To whom correspondence should be addressed; E-mail:  norman.yao@berkeley.edu}
}

\baselineskip24pt

% Make the title.

\maketitle 
\tableofcontents

% Place your abstract within the special {sciabstract} environment.

\section{NV center in diamond} % Satcher
% What is NV
% Optical initialization
% Coherent driving by MW
% optical read-out
% ESR to detect external field
The nitrogen-vacancy (NV) center is an atomic defect in diamond in which two adjacent carbon atoms are replaced by a nitrogen atom and a lattice vacancy. 
When negatively charged (by accepting a electron), the ground state of the NV center consists of two unpaired electrons in a spin triplet configuration, resulting in a room temperature zero-field splitting $D_\textrm{gs} = (2\pi)\times2.87~ \mathrm{GHz}$ between $|m_\textrm{s}=0\rangle$ and $|m_\textrm{s}=\pm 1\rangle$ sublevels. 
The NV can be optically initialized into its $|m_\textrm{s}=0\rangle$ sublevel using a laser excitation at wavelength $\lambda=532~\mathrm{nm}$. After initialization, a resonant microwave field is delivered to coherently address the transitions between $|m_\textrm{s}=0\rangle$ and $|m_\textrm{s}=\pm 1\rangle$. At the end, the spin state can be optically read-out via the same laser excitation due to spin-dependent fluorescence spectroscopy \cite{doherty2013nitrogen}.

The presence of externals signals affects the energy levels of the NV, and, in general, lifts the degeneracy of the $| m_{\mathrm{s}} = \pm 1\rangle$ states.
Using optically detected magnetic resonance (ODMR) to characterize the change in the energy levels one can directly measure such external signals.
More specifically, combining the information from the four possible crystallographic orientation of the NV centers, enables the reconstructuction of a signal's vector (e.g. magnetic field) or tensorial (e.g. stress) information.
%In the presence of any external signals, the degeneracy between $|m_\textrm{s}=\pm 1\rangle$ can be lifted, allowing one to probe the change of transition frequencies from $|m_\textrm{s}=0\rangle$ to $|m_\textrm{s}=\pm 1\rangle$ via optically detected magnetic resonance (ODMR), thus probing the external signals relative to the nitrogen-vacancy axis. 
%By measuring four possible crystallographic orientations of the NV centers, one can therefore fully reconstruct vector (e.g. magnetic field) or tensorial (e.g. stress) information of the signals.
\section{Experimental details} 
% Satcher
\subsection{Diamond anvil cell and sample preparation}
% sample prep
All diamond anvils used in this work are synthetic type-Ib ([N]~$\lesssim 200$~ppm) single crystal diamonds cut into a 16-sided standard design with dimensions 0.2~mm diameter culet, 2.75~mm diameter girdle, and 2 mm height (Almax-easyLab and Syntek Co., Ltd.). For stress measurement, both anvils with (111)-cut-culet and (110)-cut-culet are used, while for magentic measurement on iron and gadolinium, (110)-cut-culet anvil is used. We perform $^{12}$C$^+$ ion implantation (CuttingEdge Ions, 30~keV energy, $5\times10^{12}$~cm$^{-2}$) to generate a $\sim$50~nm layer of vacancies near the culet surface. After implantation, the diamonds are annealed in vacuum ($<10^{-6}$~Torr) using a home-built furnace with the following recipe: 12 hours ramp to 400$^\circ$C, dwell for 8 hours, 12 hours ramp to 800$^\circ$C, dwell for 8 hours, 12 hours ramp to 1200$^\circ$C, dwell for 2 hours. During annealing, the vacancies become mobile, and probabilistically form NV centers with intrinsic nitrogen defects. After annealing, the NV concentration is estimated to be around 1~ppm as measured by flourescence intensity. The NV centers are photostable after many iterations of compression and decompression up to 27~GPa, with spin-echo coherence time $T_2\approx 1~\mu s$, mainly limited by nitrogen spin bath. 

The miniature diamond anvil cell body is made of nonmagnetic Vascomax with cubic boron nitride backing plates (Technodiamant). Nonmagnetic gaskets (rhenium or beryllium copper) and pressure media (cesium iodide, methanol/ethanol/water) are used for all experiments.
\subsection{Experimental setup}
% pra- should we consider renaming this subsection to something like "Confocal setup"?
% optical setup
We address NV ensembles integrated inside the DAC using a home-built confocal microscope. A 100~mW 532~nm diode-pumped solid-state laser (Coherent Compass), controlled by an acousto-optic modulator (AOM, Gooch $\&$ Housego AOMO 3110-120) in a double-pass configuration, is used for both NV spin initialization and detection. The laser beam is focused through the light port of the DAC to the NV layer using a long working distance objective lens (Mitutoyo 378-804-3, NA 0.42, for stress and iron measurements; Olympus LCPLFLN-LCD 20X, NA 0.45, for gadolinium measurement in cryogenic environment), with a diffraction-limit spot size $\approx 600 \textrm{ nm}$. The NV fluorescence is collected using the same objective lens, spectrally separated from the laser using a dichroic mirror, further filtered using a $633 \textrm{ nm}$ long-pass filter, and then detected by a fiber coupled single photon counting module (SPCM, Excelitas SPCM-AQRH-64FC). A data aquisition card (National Instruments USB-6343) is used for fluorescence counting and subsequent data processing. The lateral scanning of the laser beam is performed using a two-dimensional galvanometer (Thorlabs GVS212), while the vertical focal spot position is controlled by a piezo-driven positioner (Edmund Optics at room temperture; attocube at cryogenic temperature). For gadolinium measurements, we put the DAC into a closed-cycle cryostat (attocube attoDRY 800) for temperature control from $35-320$~K. The AOM and the SPCM are gated by a programmable multi-channel pulse generator (SpinCore PulseBlasterESR-PRO 500) with $2$~ns temporal resolution.

% MW delivery
A microwave source (Stanford Research Systems SG384) in combination with a 16W amplifier (Mini-Circuits ZHL-16W-43+) serves to generate signals for NV spin state manipulation. The microwave field is delivered to DAC through a 4~$\mu$m thick platinum foil compressed between the gasket and anvil pavilion facets, followed by a 40~dB attenuator and a 50~$\Omega$ termination.
\subsection{Optically detected magnetic resonance (ODMR)}
In this work, we use continous-wave optically detected magnetic resonance (ODMR) spectroscopy to probe the NV spin resonances. The laser and microwave field are both on for the entire measurement, while the frequency of the microwave field is swept. When the microwave field is resonant with one of the NV spin transitions, it drives the spin from $|m_\textrm{s}=0 \rangle$ to $|m_\textrm{s}= \pm1 \rangle$, resulting in a decrease in NV fluorescence.

\section{Sensitivity and accuracy}\label{sec:sensitivity}
\subsection{Theoretical sensitivity}
The magnetic field sensitivity for continuous-wave ODMR \cite{Dreau:2011ec} is given by:
\begin{equation} \label{eq:etaB}
    \eta_\textrm{B}= \mathcal{P}_G\frac{1}{\gamma_{\textrm{B}}}\frac{\Delta\nu}{\mathcal{C}\sqrt{\mathcal{R}}},
\end{equation}
where $\gamma_{\textrm{B}}$ is the gyromagnetic ratio, $\mathcal{P}_G\approx 0.7$ is a unitless numerical factor for a Gaussian lineshape, $\Delta\nu=10\textrm{ MHz}$ is the resonance linewidth, $\mathcal{C}\approx 1.8\%$ is the resonance contrast, and $\mathcal{R}\approx 2.5\times10^{6}$~s$^{-1}$
is the photon collection rate. 
One can relate this to magnetic moment sensitivity by assuming that the field is generated by a point dipole located a distance $d$ from the NV center (pointing along the NV axis). Then the dipole moment sensitivity is given by
\begin{equation}
    \eta_\textrm{m}=\mathcal{P}_G\frac{1}{\gamma_{\textrm{B}}}\frac{\Delta\nu}{\mathcal{C}\sqrt{\mathcal{R}}}\frac{2\pi d^3}{\mu_0},\label{eq:etam}
\end{equation}
where $\mu_0$ is the vacuum permeability. 

Analogous to Eq.~\ref{eq:etaB}, the stress sensitivity for continuous-wave ODMR is given by 
\begin{equation}
    \eta_\textrm{S}= \mathcal{P}_G\frac{1}{\xi}\frac{\Delta\nu}{\mathcal{C}\sqrt{\mathcal{R}}},\label{eq:etas}
\end{equation}
where $\xi$ is the susceptibility for the relevant stress quantity. More specifically, $\xi$ is a tensor defined by:
\begin{equation} \label{eq:susc}
    \xi_{\alpha \beta} = \left | \frac {\delta f_\alpha} {\delta \sigma_\beta} \right |_{\sigma^{(0)}}
\end{equation}
where $f_\alpha$, $\alpha \in [1,8] $ are the resonance frequences associated with the 4 NV crytallographic orientations; $\sigma^{(0)}$ is an initial stress state; and $\delta \sigma_\beta$ is a small perturbation to a given stress component, e.g. $\beta \in \{XX,YY,ZZ,XY,XZ,YZ\}$. 
For optimal sensitivity, we consider perturbations about an unstressed state (i.e. $\sigma^{(0)} = \mathbf{0}$)
\footnote{Equivalently, one can begin from any hydrostatic stress, i.e. $\sigma^{(0)} \sim \mathbf{I}$. Non-hydrostatic stress, however, will generally reduce the stress susceptibilities, as will the presence of electric or magnetic fields.}. The resulting susceptibilities for stress components in a (111)-cut diamond frame\footnote{The $Z$ axis is normal to the diamond surface, and the $XZ$ plane contains two of the NV axes (the vertical axis and one of the three non-vertical axes).} are
\begin{center}
$\xi_{\alpha \beta} = (2\pi)\times$ 
$\begin{bmatrix}
    10.5 & 10.5 & 2.5 & 3.9 & 9.0 & 9.0 \\ 
    6.6 & 6.6 & 2.5 & 3.9 & 9.0 & 9.0 \\ 
    1.3 & 10.5 & 11.9 & 9.8 & 12.7 & 0.7 \\ 
    3.9 & 6.6 & 2.8 & 9.8 & 1.2 & 0.7 \\
    10.8 & 6.1 & 11.9 & 13.5 & 0.5 & 11.1 \\
    1.4 & 3.7 & 2.8 & 3.6 & 6.4 & 1.0 \\
    10.8 & 6.1 & 11.9 & 3.6 & 0.5 & 1.0 \\
    1.4 & 3.7 & 2.8 & 13.5 & 6.4 & 11.1
\end{bmatrix} \mathrm{[MHz/GPa].}$
% $\begin{bmatrix}
%     8.6 & 8.6 & 2.5 & 0 & 0 & 0 \\ 1.3 & 8.6 & 7.3 & 0 & 7.0 & 0\\ 6.1 & 1.2 & 7.3 & 8.5 & 3.5 & 6.0 \\ 
%     6.1 & 1.2 & 7.3 & 8.5 & 3.5 & 6.0 \\
%     2.0 & 2.0 & 0 & 3.9 & 9.0 & 9.0 \\
%     2.6 & 2.0 & 4.6 & 9.8 & 5.8 & 0.7 \\
%     4.7 & 4.9 & 4.6 & 4.9 & 2.9 & 5.0 \\
%     4.7 & 4.9 & 4.6 & 4.9 & 2.9 & 5.0
% \end{bmatrix} \mathrm{[MHz/GPa].}$
\end{center}
In the main text and in Table \ref{tab:sensitivity}, we compute the sensitivity using the maximum susceptibility for each stress component:
\begin{equation} \label{eq:max_susc}
\xi_\beta^{\textrm(max)} = \max_\alpha \xi_{\alpha \beta}
\end{equation}

% and compute the maximum frequency dependence for each stress component: $\xi_\beta = \max_\alpha \xi_{\alpha \beta}$. In Table \ref{tab:s_susc}, we present the resulting susceptibilities for stress components for a (111)-cut diamond \footnote{The $Z$ axis is normal to the diamond surface, and the $XZ$ plane contains two of the NV axes (the vertical axis and one of the three non-vertical axes).}. 

% \begin{table}
% \centering
% \begin{tabular}{l|l|l}
%   \hline \hline
%   \multirow{2}{*}{Stress component} &  Max. susceptibility, $\xi$ & Systematic accuracy \\ & $(2\pi)\times$ MHz/GPa & GPa \\ \hline\hline
%   $P = \frac 1 3 \textrm{Tr}~\sigma$  & $14.6$ & 0.0012 \\\hline \hline
%   $\sigma_{XX}$  & $8.6$ & 0.0033 \\\hline
%   $\sigma_{YY}$  & $8.6$ & 0.0034 \\\hline
%   $\sigma_{ZZ}$  & $7.3$ & 0.0028 \\\hline
%   \hline
%   $\sigma_{XY}$  & $9.8$ & 0.0027 \\\hline
%   $\sigma_{XZ}$  & $9.0$ & 0.0033 \\\hline
%   $\sigma_{YZ}$  & $9.0$ & 0.0033 \\\hline
%   \hline 
% \end{tabular}
% \caption{Maximum susceptibility and systematic accuracy for individual stress components in the lab frame of a (111)-cut diamond. The maximum susceptibility is given by the largest frequency response to perturbations in each stress component, e.g.~  Eq.~(\ref{eq:susc}-\ref{eq:max_susc}). The accuracies are based on the minimum fitting error for the resonance frequencies, i.e.~$(2\pi)\times 0.056$~MHz.}
% \label{tab:s_susc} 
% \end{table}

\subsection{Experimental sensitivity and accuracy}\label{sec:exp_sensitivity}
\begin{figure}
    \begin{center}
     \includegraphics[width=.4\linewidth]{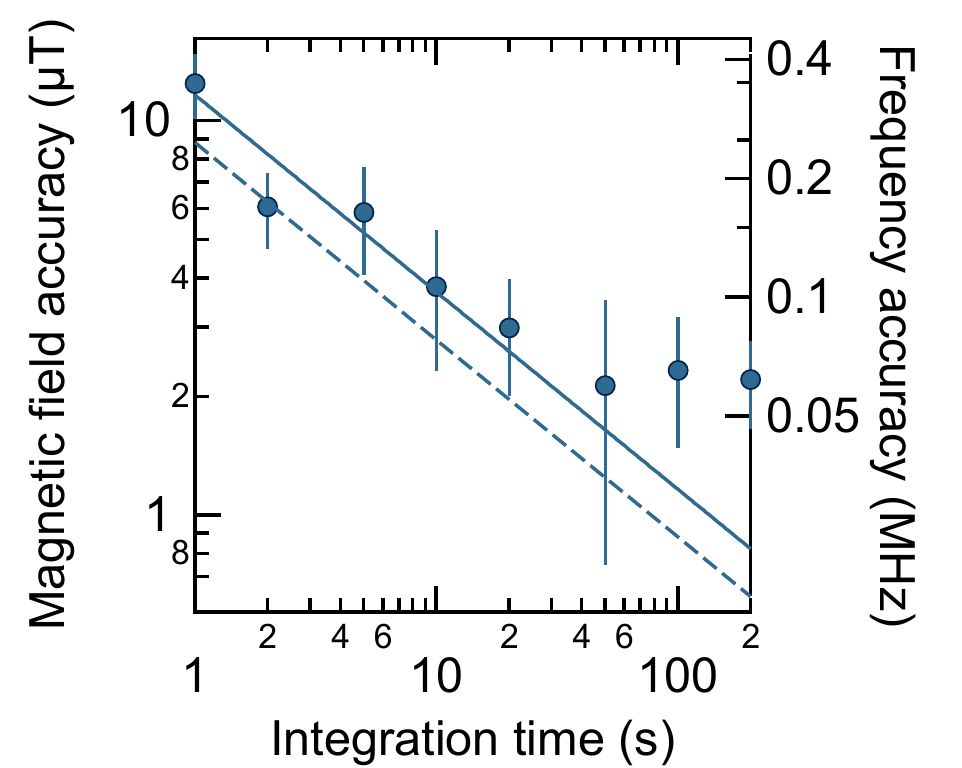}
    \caption{Scaling of magnetic field accuracy as a function of total integration time on a single resonance. Right axis corresponds to standard deviation of center frequency fitting. Solid line corresponds to a fit to $AT^{-1/2}$ where $A$ is the sensitivity reported in the main text and $T$ is the total integration time. Dashed line corresponds to the scaling predicted by Eq.~\ref{eq:etaB}. The experimental accuracy saturates for $T \gtrsim 100$~s due to systematic noise.}
    \label{fig:scaling}
    \end{center}
 \end{figure}

In order to characterize the sensitivity of our system, we perform ODMR spectroscopy on a single resonance. We fit a Gaussian lineshape to this resonance and observe the fitting error on the center frequency as a function of the total integration time, $T$ (Fig.~\ref{fig:scaling}). In particular, we fit the time scaling behavior of the fitting error to $AT^{-1/2}$, where $A$, divided by the susceptibility of interest, characterizes the experimental sensitivity for a given signal. For $T \gtrsim 100$~s, the experimental accuracy saturates due to systematic noise, which we define here as the ``systematic accuracy'' for each type of signal.

For scalar signals (e.g.~axial magnetic fields, temperature, etc.), the accuracy is directly proportional to the minimum fitting error. 
For stress components, however, determining the accuracy is more complicated as the relation between resonance frequencies and the full stress tensor is a multi-dimensional, nonlinear function (Section \ref{sec:stress_overview}). To this end, we quantify the accuracy of each stress component using a Monte Carlo procedure. 
We begin with an unstressed state, which corresponds to the initial set of frequencies $f^{(0)}_\alpha = D_{\textrm{gs}}$. 
We then apply noise to each of the freqencies based on the minimum fitting error determined above---i.e.~$f_\alpha^{(0)}+\delta f_\alpha$, where $\delta f_\alpha$ are sampled from a Gaussian distribution with a width of the fitting error---and calculate the corresponding stress tensor using a least-squared fit (Sec.~\ref{sec:stress_overview}).
Repeating this procedure over many noise realizations, we compute the standard deviation of each stress component. The results of this procedure are shown in Table \ref{tab:sensitivity}.

\begin{table}
\begin{tabular}{l|l|l|l}
  \hline \hline
  Signal (unit) & Theo Sensitivity  & Exp Sensitivity & Accuracy \\
  & (unit/$\sqrt{\textrm{Hz}}$) & (unit/$\sqrt{\textrm{Hz}}$)  &(unit) \\ \hline\hline
  Hydrostatic stress (GPa) &  $0.017$ & $0.023$ & $0.0012$ \\\hline
  Average normal stress (GPa) &  $0.022$ & $0.03$ & $0.0032$ \\\hline
  Average shear stress (GPa) &  $0.020$ & $0.027$ & $0.0031$ \\\hline\hline
  Magnetic field ($\mu$T) & $8.8$ & $12$ & $2.2$ \\\hline
  Magnetic dipole (emu), & $5.5\times10^{-12}$  & $7.5\times10^{-12}$& $1.4\times10^{-12}$  \\
  floating sample ($d=5$~$\mu$m)&&\\\hline
      \rowcolor{gray!25}
    Magnetic dipole (emu),  &  $1.7\times10^{-20}$   & $2.3\times10^{-20}$ & $4.3\times10^{-21}$  \\
    \rowcolor{gray!25}
    exfoliated sample ($d=5$~nm)$^{(*)}$&&&\\\hline
    \rowcolor{gray!25}
  Magnetic dipole (emu), &  $1.6\times10^{-21}$   & $2.2\times10^{-21}$ & $4.0\times10^{-22}$  \\
  \rowcolor{gray!25}
  exfoliated sample, &&&\\
  \rowcolor{gray!25}single NV ($d=5$~nm)$^{(\dagger)}$ &&& \\
  \hline \hline
  %Axial electric field (MV/cm) & $0.91$ & $0.16$ \\ \hline
  \rowcolor{gray!25} Electric field (kV/cm), & $1.8$ & $2.5$ & $0.45$ \\ \rowcolor{gray!25}single NV$^{(\dagger)}$ &&& \\\hline\hline
  \rowcolor{gray!25} Temperature (K), & $0.4$ & $0.55$ & $0.10$ \\\rowcolor{gray!25}single NV$^{(\dagger)}$ &&& \\\hline\hline
  
\end{tabular}
\caption{NV sensitivity and accuracy for various signals. Sensitivity is calculated using Eqs.~\ref{eq:etam}-\ref{eq:etas}. We also report the typical fitting error of the center frequency for the relevant experiments in the main text. Gray rows correspond to projected sensitivity given an exfoliated sample atop $^{(*)}$ an ensemble of $5$~nm depth NV centers or $^{(\dagger)}$ a single $5$~nm depth NV center with $\Delta\nu=1\textrm{ MHz},\mathcal{C}=0.1,\mathcal{R}=10^4\textrm{ s}^{-1}$. Magnetic dipoles are reported in units of emu, where $1$ emu $=10^{-3}$ A$\cdot$m$^2$. }
\label{tab:sensitivity}
\end{table}

\subsection{Comparison to other magnetometry techniques}
In this section, we discuss the comparison of magnetometry techniques presented in Fig.~1F of the main text.
%Lacking discussion on why we choose accuracy in main text.
%Sensitivity and accuracy serve as figures of merit for comparing the efficacy of competing sensing methods. 
%Dipole accuracy, an oft-reported figure of merit, describes the uncertainty in the dipole moment associated with a magnetometer; this approximately parameterizes the minimum dipole moment detectable in a given measurement realization.
%This is a somewhat different concept than dipole moment sensitivity, which describes the minimum moment one can distinguish in a given measurement time, specifically in the regime where statistical errors are the dominant source of moment uncertainty.
For each sensor, the corresponding dipole accuracy (as defined in Section~\ref{sec:exp_sensitivity}) is plotted against its relevant ``spatial resolution,'' roughly defined as the length scale within which one can localize the source of a magnetic signal. In the following discussion, we specify the length scale plotted for each method in Fig.~1F of the main text. We consider two broad categories of high pressure magnetometers.%All estimates are presented in Fig.~1e in the main text.

The first category encompasses inductive methods such as pickup coils \cite{Feng:2014coil2,Mito:2003coil3,Jackson:2003da} and superconducting quantum interference devices (SQUIDs) \cite{Alireza:2009squid2,Mito:2001squid3,Giriat:2010squid4,Takeda:2002squid6,Marizy:2017squid7}\footnote{Under the category of inductive methods, we also include the ``designer anvil'' which embeds a pickup coil directly into the diamond anvil.}.
Magnetic dipole measurement accuracies are readily reported in various studies employing inductive methods.
We estimate the relevant length scale of each implementation as the pickup coil or sample bore diameter.

The second class of magnetometers comprises high energy methods including M\"ossbauer spectroscopy \cite{Pasternak:1991mossbauer1,Pasternak:2001mossbauer2,Kantor:2004mossbauer3} and x-ray magnetic circular dichroism (XMCD) \cite{Mathon:2004it,Ishimatsu:2007eu,Watanabe:2011hk,Chen:2018}, which probe atomic scale magnetic environments.
%Rework structure so clause doesn't sound dangly:
For the M\"ossbauer studies considered in our analysis, we calculate magnetic dipole moment accuracies by converting $B$-field uncertainties into magnetic moments, assuming a distance to the dipole on order of the lattice spacing of the sample. 
We assess the length scale as either the size of the absorbing sample or the length scale associated with the sample chamber/culet area.
For XMCD studies, we accept the moment accuracies reported in the text. 
Length scales are reported as the square root of the spot size area.
Notably, we emphasize that both methods provide information about atomic scale dipole moments rather than a sample-integrated magnetic moment; these methods are thus not directly comparable to inductive methods.

We compare these methods alongside the NV center, whose accuracy is defined in Section~\ref{sec:exp_sensitivity} and shown in Table~\ref{tab:sensitivity}. For the current work, we estimate a length scale $\sim5$ $\mu$m, corresponding to the approximate distance between a sample (suspended in a pressure-transmitting medium) and the anvil culet. By exfoliating a sample onto the diamond surface, the diffraction-limit $\sim600$ nm bounds the transverse imaging resolution for ensemble NV centers; this limit can be further improved for single NV centers via super-resolution techniques \cite{rittweger2009resolution}.

\section{Stress tensor}
% \subsection{Stress reconstruction}\label{sec:stress_theory}
\subsection{Overview} \label{sec:stress_overview}
In this section, we describe our procedure for reconstructing the full stress tensor using NV spectroscopy. 
This technique relies on the fact that the four NV crystallographic orientations experience different projections of the stress tensor within their local reference frames. 
% In particular, each NV orientation is described by two effects. 
% First, the $\ket {m_s \pm 1}$ states are shifted from the zero-field energy (i.e. $D_{gs} = (2\pi) \times $2.87 GHz) by:
% \begin{equation}
% \Pi_{z,i} = \alpha_1 \left[\sigma^{(i)}_{xx} + \sigma^{(i)}_{yy}\right] + \beta_1 \sigma^{(i)}_{zz} 
% \end{equation}
% where $\sigma^{(i)}$ is the stress tensor in the local frame of the $i$-th NV orientation. Second, the states are split in energy by $2$
% In particular, each NV orientation is described by the amount that their $\ket {m_s \pm 1}$ states shift and split from their 
% the $\ket m_s \pm 1$ states of each NV oriention are shifted from their zero-field energy (i.e. $D_{gs} = (2\pi)$2.87 GHz) by 
% In particular, the $\ket m_s \pm 1$ states of each NV oriention are shifted from their zero-field energy (i.e. $D_{gs} = (2\pi)$2.87 GHz) by:
In particular, the full Hamiltonian describing the stress interaction is given by: 
\begin{equation}\label{eq:Hs}
% H_\textrm{S} = \sum_i \left [ \alpha_1 (\sigma_{xx} + \sigma_{yy} ) + \beta_1 \sigma_{zz}   \right ]  S_z^2 +   \left [  \alpha_2 (\sigma_{yy} - \sigma_{xx} ) +\beta_2 \sigma_{xz}   \right ] (S_y^2-S_x^2) 
H_\textrm{S} = \sum_i \Pi_{z,i}  S_{z,i}^2 + \Pi_{x,i}\left(S_{y,i}^2-S_{x,i}^2\right) + \Pi_{y,i}
\left(S_{x,i}S_{y,i}+S_{y,i}S_{x,i}\right)
\end{equation}
where
\begin{align} \label{eq:pi_s}
\Pi_{z,i} &= \alpha_1 \left(\sigma^{(i)}_{xx} + \sigma^{(i)}_{yy}\right) + \beta_1 \sigma^{(i)}_{zz} \\
\Pi_{x,i} &= \alpha_2 \left(\sigma^{(i)}_{yy} - \sigma^{(i)}_{xx}\right) +\beta_2 \left(2\sigma^{(i)}_{xz}\right)  \\
\Pi_{y,i} &=  \alpha_2 \left(2\sigma^{(i)}_{xy}\right) + \beta_2 \left(2\sigma^{(i)}_{yz}\right)
\end{align}
$\sigma^{(i)}$ is the stress tensor in the local frame of each of NV orientations labeled by $\{i = 1,2,3,4\}$, and $\{\alpha_{1,2},\beta_{1,2}\}$ are stress susceptibility parameters (Section \ref{sec:susceptibilities}).
% which is related to the stress in lab frame by a rotation, $\sigma^{(i)} = R^T_i \sigma^{(\textrm{lab})} R_i$.
Diagonalizing this Hamiltonian, one finds that the energy levels of each NV orientation exhibit two distinct effects: the $\ket {m_s =  \pm 1}$ states are \emph{shifted} in energy by $\Pi_{z,i}$ and \emph{split} by $2\Pi_{\perp,i} = 2\sqrt{\Pi_{x,i}^2+\Pi_{y,i}^2}$. 
Thus, the Hamiltonian can be thought of as a function that maps the stress tensor in the lab frame to eight observables: $H_\textrm{S}(\sigma^{\textrm(\textrm{lab})})=\{\Pi_{z,1},\Pi_{\perp,1},\Pi_{z,2},\Pi_{\perp,2},...\}$. Obtaining these observables through spectroscopy, one can numerically invert this function and solve for all six components of the corresponding stress tensor. 

In practice, resolving the resonances of the four NV orientation groups is not straightforward because the ensemble spectra can exhibit near degeneracies. When performing ensemble NV magnetometry, a common approach is to spectroscopically separate the resonances using an external bias magnetic field.
% \begin{equation}\label{eq:Hs}
% % H_\textrm{S} = \sum_i \left [ \alpha_1 (\sigma_{xx} + \sigma_{yy} ) + \beta_1 \sigma_{zz}   \right ]  S_z^2 +   \left [  \alpha_2 (\sigma_{yy} - \sigma_{xx} ) +\beta_2 \sigma_{xz}   \right ] (S_y^2-S_x^2) 
% H_\textrm{B} = \sum_i \vec B_i\cdot \vec S_i
% \end{equation}
% Here $\vec B_i$ is the magnetic field in the local frame of the $i$-th orientation, which can be tuned to spectrally separate the four orientations.
However, unlike magnetic contributions to the Hamiltonian, stress that couples via $\Pi_\perp$ is suppressed by an axial magnetic field. 
Therefore, a generic magnetic field provides only stress information via the shifting parameters, $\Pi_{z,i}$, which is insufficient for reconstructing the full tensor.

To address this issue, we demonstrate a novel technique that consists of applying a well-controlled external magnetic field perpendicular to each of the NV orientations. 
This technique leverages the symmetry of the NV center, which suppresses its sensitivity to transverse magnetic fields. 
In particular, for each perpendicular field choice, three of the four NV orientations exhibit a strong Zeeman splitting proportional to the projection of the external magnetic field along their symmetry axes, while the fourth (perpendicular) orientation is essentially unperturbed
\footnote{A transverse magnetic field leads to shifting and splitting at second order in field strength. We account for the former through a correction described in Section \ref{sec:stress_analysis}, while the latter effect is small enough to be neglected. More specifically, the effective splitting caused by magnetic fields is $(\gamma_{\textrm{B}}B_\perp)^2/D_{gs} \approx 5-10$~MHz, which is smaller than the typical splitting observed at zero field.}.
This enables one to resolve $\Pi_{z,i}$ for all four orientations and $\Pi_{\perp,i}$ for the orientation that is perpendicular to the field. Repeating this procedure for each NV orientation, one can obtain the remaining splitting parameters and thus reconstruct the full stress tensor.

In the following sections, we provide additional details regarding our experimental procedure and analysis. In Section \ref{sec:magnet}, we describe how to use the four NV orientations to calibrate three-dimensional magnetic coils and to determine the crystal frame relative to the lab frame. In Section \ref{sec:stress_analysis}, we discuss our fitting procedure, the role of the NV's local charge environment, and the origin of the stress susceptibility parameters. In Section \ref{sec:stress_results}, we present the results of our stress reconstruction procedure for both (111)- and (110)-cut diamond. In Section \ref{sec:fem}, we compare our experimental results to finite element simulations.
\subsection{Experimental details}
\label{sec:magnet}
\subsubsection{Electromagnet calibration procedure}
To apply carefully aligned magnetic fields, we utilize a set of three electromagnets that are approximately spatially orthogonal with one another and can be controlled independently via the application of current. 
Each coil is placed $>$10~cm away from the sample to reduce the magnetic gradient across the $(200~\mu$m$)^2$ culet area \footnote{We note that the pressure cell, pressure medium and gasket are nonmagnetic.}. 

To calibrate the magnetic field at the location of the sample, we assume that the field produced by each coil is linearly porportional to the applied current, $I$. Our goal is then to find the set of coefficients, $a_{mn}$ such that
\begin{equation}
    B_m=\sum_{m}a_{mn}I_n,
\end{equation}
where $B_m=\{B_{\mathcal{X}}, B_{\mathcal{Y}}, B_{\mathcal{Z}}\}$ is the magnetic field in the crystal frame and $n=\{1, 2, 3\}$ indexes the three electromagnets. We note that this construction does not require the electromagnets to be spatially orthogonal. 

To determine the nine coefficients, we apply arbitrary currents and measure the Zeeman splitting of the four NV orientations via ODMR spectroscopy. 
Notably, this requires the ability to accurately assign each pair of resonances to their NV crystallographic orientation. 
We achieve this by considering the amplitudes of the four pairs of resonances, which are proportional to the relative angles between the polarization of the excitation laser and the four crystallagraphic orientations. 
In particular, the $\ket{ m_s =  0 }\leftrightarrow \ket {m_s = \pm 1}$ transition is driven by the perpendicular component of the laser field polarization with respect to the NV's symmetry axis. 
Therefore, tuning the laser polarization allows us to assign each pair of resonances to a particular NV orientation.

In order to minimize the number of fitting variables, we choose magnetic fields whose projection along each NV orientation is sufficient to suppress their transverse stress-induced energy splitting, i.e. $\gamma_{\textrm{B}} B\gg\Pi_\perp$. As a result, the spectrum measured at each magnetic field is determined by (a) the stress-induced shift $\Pi_{z,i}$ for each NV orientation, which is constant for all applied fields, and (b) the applied vector magnetic field $\{B_{\mathcal{X}}, B_{\mathcal{Y}}, B_{\mathcal{Z}}\}$. Sequentially applying different currents to the electromagnet coils and determining the subsequent vector magnetic field at the sample three times, we obtain sufficient information to determine the matrix $a_{mn}$ as well as the shift $\Pi_z$ for all NV orientations. We find that the calibration technique is precise to within 2$\%$.

\subsubsection{Calibration of crystal and laboratory frames}
To determine the orientation of the crystal frame (i.e. the [100] diamond axis) with respect to the lab frame, we apply an arbitrary magnetic field and measure its angle (a) in the lab frame via a handheld magnetometer, and (b) in the crystal frame via the Zeeman splittings (see \ref{sec:magnet}). Together with the known diamond cut, this provides a system of equations for the rotation matrix, $R_c$, that relates the lab frame and the crystal frame:
\begin{equation}\label{eq:R}
R_\textrm{c} \hat B^{(\textrm{lab})} = \hat B^{(\textrm{crystal})}\; , \quad R_\textrm{c} \hat Z = \hat e^{(\textrm{crystal})}\; 
\end{equation}
where $\hat Z = (0,0,1)^\top$ is the longitudinal axis in the lab frame, and $\hat e^{(\textrm{crystal})}$ is the unit vector perpendicular to the diamond cut surface in crystal frame, e.g. $\hat e^{(\textrm{crystal})} \propto (1,1,1)^\top$ for the (111)-cut diamond. We solve for $R_c$ by numerically minimizing the least-squared residue of these two equations.

However, we note that the magnetic field determined by the Zeeman splittings contains an overall sign ambiguity. To account for this, we numerically solve Eq.~\eqref{eq:R} using both signs for $\hat B^{(\textrm{crystal})}$ and select the solution for $R_\textrm{c}$ with the smaller residue. Based on this residue, we estimate that our calibration is precise to within a few degrees. 
% We can estimate our calibration accuracy from this residue and typically find that it corresponds to an error of a few degrees. 
% This residue typically corresponds to calibration accuracy of a few degrees. 
% in our calibration, i.e. the magnetic field is a few degrees away from a 

% a calibration accuracy two within  
% In practice, we find that one of the signs leads to a significantly smaller residue than the other, which implies that our solution is unique. %For example, the absolute orientations of the four NV groups in the lab frame are shown for the (111)-cut diamond in Fig.~\ref{fig:111-fits}.

\subsection{Analysis}\label{sec:stress_analysis}
\subsubsection{Extracting splitting and shifting information}
\begin{figure}
    \begin{center}
     \includegraphics[width=\linewidth]{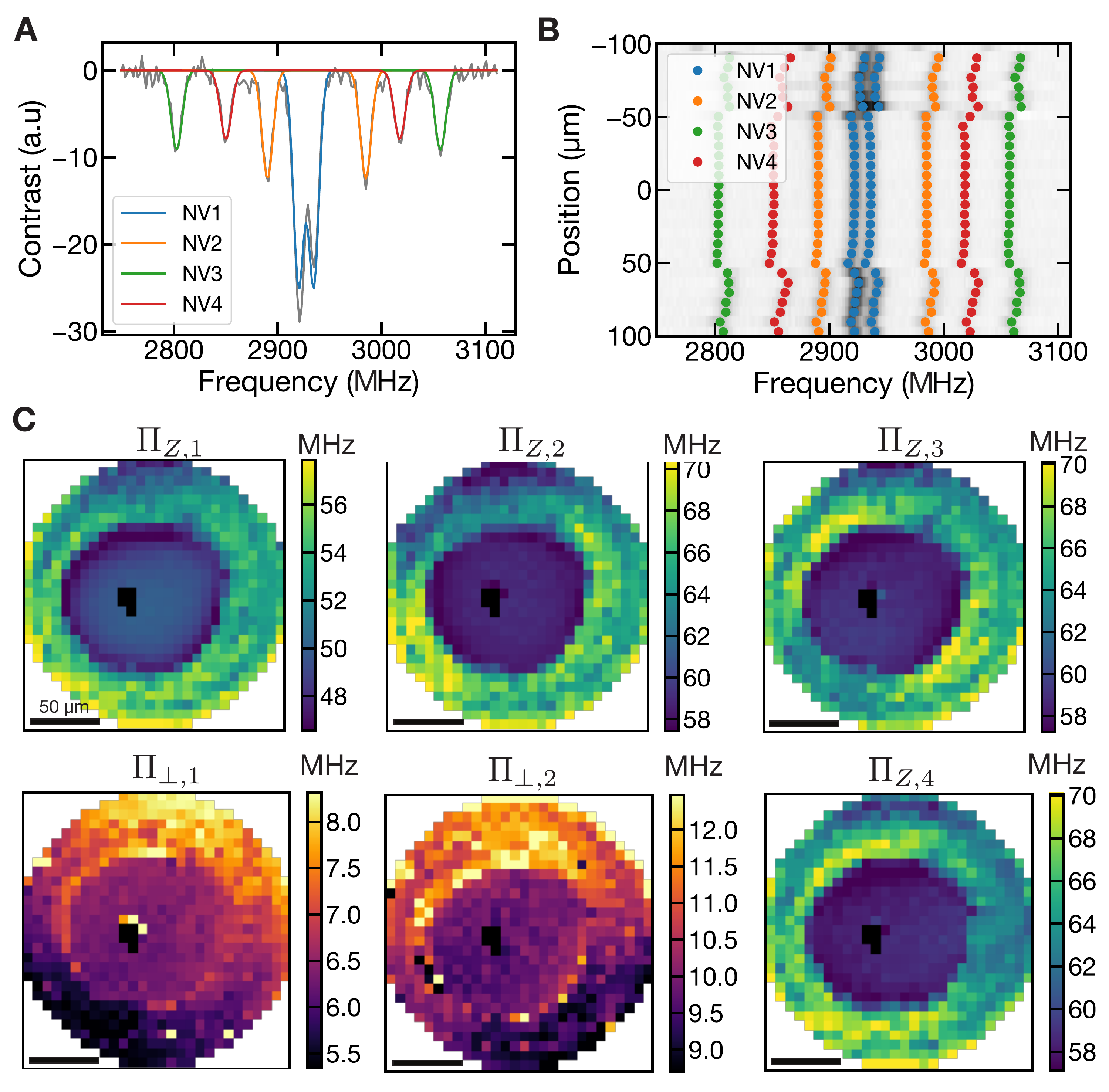}
    \caption{Stress reconstruction procedure applied to the (111)-cut diamond at 4.9 GPa. (\textbf{A}) A typical ODMR spectrum with the resonances corresponding to each NV orientation fit a pair of Lorentzian lineshapes. (\textbf{B}) A linecut indicating the fitted resonance energies (colored points) superimposed on the measured spectra (grey colormap). (\textbf{C}) 2D maps of the shifting ($\Pi_{z,i}$) and splitting parameters ($\Pi_{\perp,i}$) for each NV orientation across the entire culet.}
    \label{fig:111-fits}    
    \end{center}
 \end{figure}

Having developed a technique to spectrally resolve the resonances, we fit the resulting spectra to four pairs of Lorentzian lineshapes. 
Each pair of Lorentzians is defined by a center frequency, a splitting, and a common amplitude and width.
To sweep across the two-dimensional layer of implanted NV centers, we sequentially fit the spectrum at each point by seeding with the best-fit parameters of nearby points.
We ensure the accuracy of the fits by inspecting the frequencies of each resonance across linecuts of the 2D data (Fig.~\ref{fig:111-fits}B). 

Converting the fitted energies to shifting ($\Pi_{z,i}$) and splitting parameters  ($\Pi_{\perp,i}$) requires us to take into account two additional effects.
First, in the case of the shifting parameter, we subtract off the second-order shifting induced by transverse magnetic fields. 
In particular, the effective shifting is given by $\Pi_{z,\textrm{B}} \approx (\gamma_{\textrm{B}}B_\perp)^2/D_{\textrm{gs}}$, which, under our experimental conditions, corresponds to $\Pi_{z,\textrm{B}} \approx 5-10$ MHz. 
To characterize this shift, one can measure each of the NV orientations with a magnetic field aligned parallel to its principal axis, such that the transverse magnetic shift vanishes.
In practice, we obtain the zero-field shifting for each of the NV orientations without the need for additional measurements, as part of our electromagnet calibration scheme (Section \ref{sec:magnet}).
We perform this calibration at a single point in the two-dimensional map and use this point to characterize and subtract off the magnetic-induced shift in subsequent measurements with arbitrary applied field. 
Second, in the case of the splitting parameter, we correct for an effect arising from the NV's charge environment. We discuss this effect in the following section. The final results for the shifting ($\Pi_{z,i}$) and splitting ($\Pi_{\perp,i}$) parameters for the (111)-cut diamond at 4.9 GPa are shown in Fig.~\ref{fig:111-fits}C. 

\subsubsection{Effect of local charge environment}
It is routinely observed that ensemble spectra of high-density samples (i.e.~Type 1b) exhibit a large ($5-10$~MHz) splitting even under ambient conditions. While commonly attributed to instrinsic stresses in the diamond, it has since been suggested that the splitting is, in fact, due to electric fields originating from nearby charges \cite{mittiga2018imaging}. This effect should be subtracted from the total splitting to determine the stress-induced splitting. 

To this end, let us first recall the NV interaction with transverse electric fields:
\begin{equation}
H_{E} = d_\perp \left[ \mathcal{E}_x (S_y^2-S_x^2) + \mathcal{E}_x
(S_xS_y+S_yS_x) \right]
\end{equation}
where $d_\perp = 17$~Hz cm/V. Observing the similarity with Eq.~\eqref{eq:Hs}, we can define 
\begin{align}
\tilde \Pi_x &= \Pi_{\mathrm{s},x} + \Pi_{\mathrm{E},x}\\
\tilde \Pi_y &=  \Pi_{\mathrm{s},y} + \Pi_{\mathrm{E},y}
\end{align}
where $\Pi_{\mathrm{S},\{x,y\}}$ are defined in Eq.~\eqref{eq:pi_s} and $\Pi_{\mathrm{E},\{x,y\}} = d_\perp \mathrm{E}_{\{x,y\}}$. The combined splitting for electric fields and stress is then given by
\begin{equation}
2\tilde \Pi_\perp = 2\left(\left(\Pi_{\mathrm{s},x}+\Pi_{\mathrm{E},x} \right)^2+\left(\Pi_{\mathrm{s},y}+\Pi_{\mathrm{E},y} \right)^2\right)^{1/2} \: .
\end{equation}
We note that the NV center also couples to longitudinal fields, but its susceptibility is $\sim50$ times weaker and is thus negligible in the present context. 

To model the charge environment, we consider a distribution of transverse electric fields. For simplicity, we assume that the electric field strength is given by a single value $\mathcal{E}_0$, and its angle is randomly sampled in the perpendicular plane. 
% This approximation leads to consistent results with the full distribution of electric fields produced by random charges, up to .... 
Adding the contributions from stress and electric fields and averaging over angles, the total splitting becomes
% \begin{equation}
\begin{align} 
\tilde \Pi_{\perp,\textrm{avg}}  &= \int d\theta (\Pi_{\textrm{S},\perp}^2+\Pi_{\textrm{E},\perp}^2+2\Pi_{\textrm{S},\perp}\Pi_{\textrm{E},\perp}\cos{\theta})^{1/2}\notag \\ 
&= \frac 1 \pi \left [\sqrt{\Pi_{\textrm{s},\perp}^2-\Pi_{\textrm{E},\perp}^2}\textrm{EllipticE}\left( - \frac {4 \Pi_{\textrm{S},\perp} \Pi_{\textrm{E},\perp}}{\sqrt{\Pi_{\textrm{S},\perp}^2-\Pi_{\textrm{E},\perp}^2}}\right)\right. \notag \\ 
&\quad \quad+ \left.\sqrt{\Pi_{\textrm{S},\perp}^2+\Pi_{\textrm{E},\perp}^2}\textrm{EllipticE}\left( - \frac {4 \Pi_{\textrm{s},\perp} \Pi_{\textrm{E},\perp}}{\sqrt{\Pi_{\textrm{S},\perp}^2+\Pi_{\textrm{E},\perp}^2}}\right)\right] \label{eq:split_avg}
\end{align}
% \end{equation}
where $\textrm{EllipticE}(z)$ is the elliptic integral of the second kind.
This function is plotted in Fig.~\ref{fig:charge}A, and we note its qualitative similarity to a quadrature sum.

To characterize the intrinsic charge splitting ($\Pi_{\textrm{E},\perp}$), we first aquire an ODMR spectrum for each diamond sample under ambient conditions. For example, for the (111)-cut diamond, we measured $\Pi_{\mathrm{E},\perp} \approx 4.5$ MHz. For subsequent measures under pressure, we then subtract off the charge contribution from the observed splitting by numerically from inverting Eq.~\eqref{eq:split_avg} and solving for $\Pi_{\mathrm{s},\perp}$. 

\subsubsection{Susceptibility parameters} \label{sec:susceptibilities}
A recent calibration experiment established the four stress susceptibilities relevant to this work \cite{Barson:2017ba}. In this section, we discuss the conversion of their susceptibilities to our choice of basis (the local NV frame), and we reinterpret their results for the splitting parameters taking into account the effect of charge. 

In their paper, Barson et. al. define the stress susceptilities with respect diamond crystal frame: 
\begin{align}
\Pi_{z} &= a_1 (\sigma_{\mathcal{XX}}+\sigma_{\mathcal{YY}}+\sigma_{\mathcal{ZZ}})+2a_2(\sigma_{\mathcal{YZ}}+\sigma_{\mathcal{ZX}}+\sigma_{\mathcal{XY}}) \\ 
\Pi_{x}  &= b(2 \sigma_{\mathcal{ZZ}}-\sigma_{\mathcal{XX}}-\sigma_{\mathcal{YY}})+c(2\sigma_{\mathcal{XY}}-\sigma_{\mathcal{YZ}}-\sigma_{\mathcal{ZX}}) \\
\Pi_{y} &= \sqrt {3} \left [b(\sigma_{\mathcal{XX}}-\sigma_{\mathcal{YY}})+c(\sigma_{\mathcal{YZ}}-\sigma_{\mathcal{ZX}})\right]  
\end{align}
where $\mathcal{XYZ}$ are the principal axes of the crystal frame. Their reported results are $\{a_1,a_2,b,c\}=(2\pi)\times\{4.86(2),-3.7(2),−2.3(3), 3.5(3)\}$~MHz/GPa. 

To convert these susceptibilities to our notation (Eq.~\ref{eq:Hs}), one must rotate the stress tensor from the crystal frame to the NV frame, i.e. $\sigma_{xyz} = R \sigma_{\mathcal{XYZ}}R^\top $.
The rotation matrix that accomplishes this is:
\begin{equation}
R = \left(
\begin{array}{ccc}
 -\frac{1}{\sqrt{6}} & -\frac{1}{\sqrt{6}} & \sqrt{\frac{2}{3}} \\
 \frac{1}{\sqrt{2}} & -\frac{1}{\sqrt{2}} & 0 \\
 \frac{1}{\sqrt{3}} & \frac{1}{\sqrt{3}} & \frac{1}{\sqrt{3}} \\
\end{array}
\right).
\end{equation}
Applying this rotation, one finds that the above equations become (in the NV frame)
\begin{align}
% H_\textrm{S} = \left [ \alpha_1 (\sigma_{xx} + \sigma_{yy} ) + \beta_1 \sigma_{zz}   \right ]  S_z^2 +   \left [  \alpha_2 (\sigma_{yy} - \sigma_{xx} ) +\beta_2 \sigma_{xz}   \right ] (S_y^2-S_x^2) 
\Pi_z &= (a_1-a_2) (\sigma_{xx} + \sigma_{yy} ) + (a_1+2a_2) \sigma_{zz} \\
\Pi_x &= (-b-c) (\sigma_{yy} - \sigma_{xx} ) +(\sqrt 2 b-\frac{\sqrt 2}{2}c) (2\sigma_{xz} ) \\
\Pi_x &=  (-b-c) (2\sigma_{xy}) + (\sqrt 2 b-\frac{\sqrt 2}{2}c)(2\sigma_{yz})
\end{align}
Thus, the conversion between the two notations is
\begin{equation}
\begin{aligned} \label{eq:conversion}
\begin{pmatrix} \alpha_1 \\ \beta_1 \end{pmatrix} &= \begin{pmatrix} 1 & -1 \\ 1 & 2 \end{pmatrix} \begin{pmatrix} a_1 \\ a_2 \end{pmatrix} \\ 
\begin{pmatrix} \alpha_2 \\ \beta_2 \end{pmatrix} &= \begin{pmatrix} -1 & -1 \\ \sqrt{2} & -\frac{\sqrt{2}}{2} \end{pmatrix} \begin{pmatrix} b \\ c \end{pmatrix}  
\end{aligned}
\end{equation}

\begin{figure}
    \begin{center}
         \includegraphics[width=\textwidth]{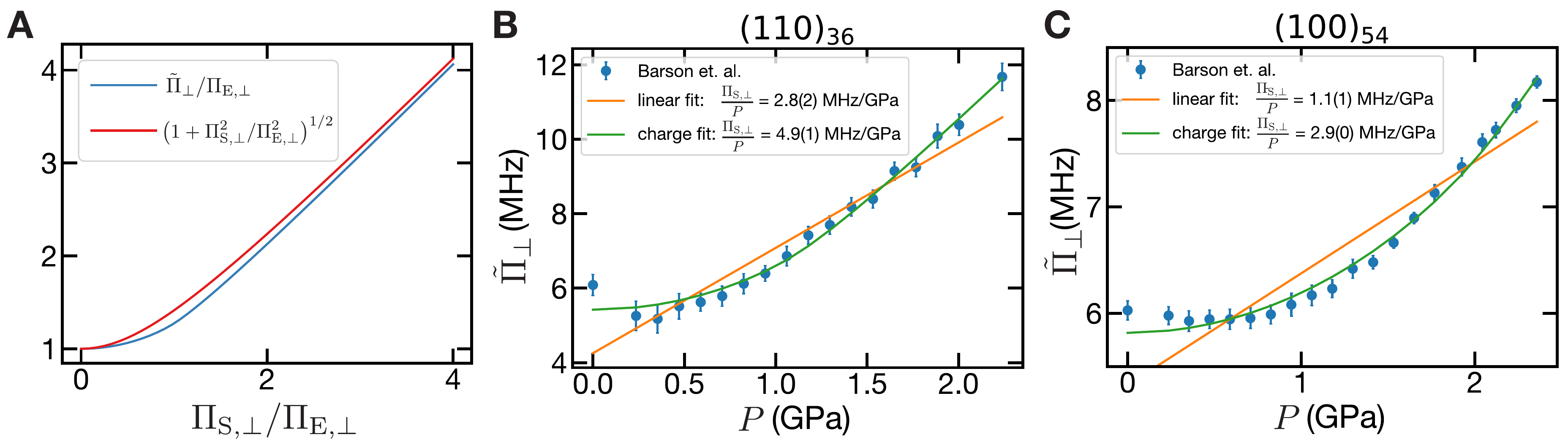}
    \caption{Interplay between stress and random electric fields. (\textbf{A}) Theoretical curve (blue) for the total splitting in the presence of stress and electric fields, Eq.~(\ref{eq:split_avg}). We compare this to a quadratic sum (red). (\textbf{B}-\textbf{C}) Measured splitting parameter (blue) for uniaxial pressure applied to a (110)-cut and (100)-cut diamond, reprinted with permission from \cite{Barson:2017ba}. We fit the data using (a) a linear function (orange), $\tilde \Pi_\perp = \Pi_{\textrm{E},\perp} + \Pi_{\textrm{S},\perp}$, and (b) the aforementioned theoretical curve, Eq.~\eqref{eq:split_avg} (green). Both fits include two free parameters: $\Pi_{\textrm{E},\perp}$ and $a = \Pi_{\textrm{S},\perp}/P$. We report the best-fit value for the latter parameter in the inset.}
    \label{fig:charge}    
    \end{center}
 \end{figure}

In characterizing the splitting parameters ($b$ and $c$), Barson et. al. assumed a linear dependence between the observed splitting and $\Pi_{\textrm{S},\perp}$. 
However, our charge model suggests that for $\Pi_{\textrm{S},\perp} \lesssim \Pi_{\textrm{E},\perp}$ the dependence should be nonlinear. 
To account for this, we re-analyze their data using Eq.~\ref{eq:split_avg} as our fitting form, rather than a linear function as in the original work. 
The results are shown in Fig.~\ref{fig:charge} for two NV orientation groups measured in the experiment: $(110)_{36}$ and $(100)_{54}$, where $(\cdots)$ denotes the crystal cut and the subscript is the angle of the NV group with respect to the crystal surface. 
From the fits, we extract the linear response, $\Pi_{\textrm{s},\perp}/P$, for the two groups. 
These are related to the stress parameters by $b-c$ and $2b$, respectively. 
Using these relations and the results of the fits, one finds $\{b,c\}=(2\pi)\times\{-1.47(2), 3.42(7)\}$ ~MHz/GPa \footnote{Note that the overall sign of these parameters cannot be determined through these methods, as the energy splitting is related to the quadrature sum of $\Pi_x$ and $\Pi_y$. To determine the sign, one would need to measure the phase of the perturbed states \cite{mittiga2018imaging}.}. Finally, we convert these and the original reported for $\{a_1,a_2\}$ to the NV frame using Eq.~\ref{eq:conversion}. This leads to the susceptibilites that we use for our analysis:
\begin{equation}
    \{\alpha_1,\beta_1,\alpha_2,\beta_2\}=(2\pi)\times\{8.6(2),-2.5(4),-1.95(9),-4.50(8)\} \: \: \text{MHz/GPa}.
\end{equation}

\subsection{Results}\label{sec:stress_results}
\begin{figure}
    \begin{center}
     \includegraphics[width=0.85\linewidth]{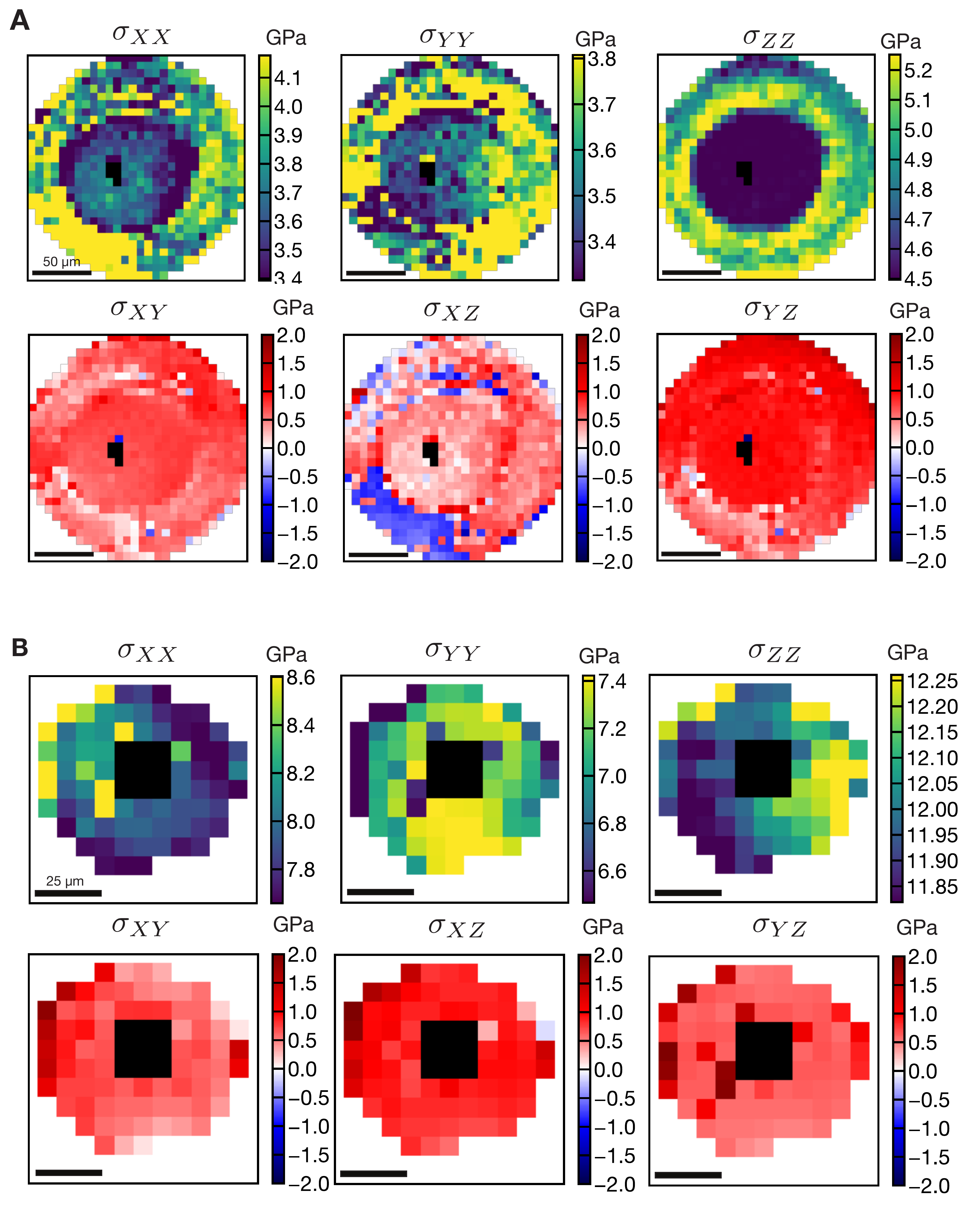}
    \caption{Stress tensor reconstruction of (111)-cut diamond at (\textbf{A}) 4.9 GPa and (\textbf{B}) 13.6 GPa. In the former case, we reconstruct both the inner region in contact with the fluid-transmitting medium, and the outer region in contact with the gasket. In the latter case, we reconstruct only the inner region owing to the large stress gradients at the contact with the gasket; note that the black pixels in the center indicates where the spectra is obscured by the ruby flourescence. As described in the main text, both pressures exhibit inward concentration of the normal lateral stress ($\sigma_{XX}$ and $\sigma_{YY}$). In contrast, the normal loading stress is uniform for the lower pressure and spatially varying at the higher pressure, indicating that the pressure medium has solidified.}
    \label{fig:111-stress}    
    \end{center}
 \end{figure}

\begin{figure}
    \begin{center}
     \includegraphics[width=0.85\linewidth]{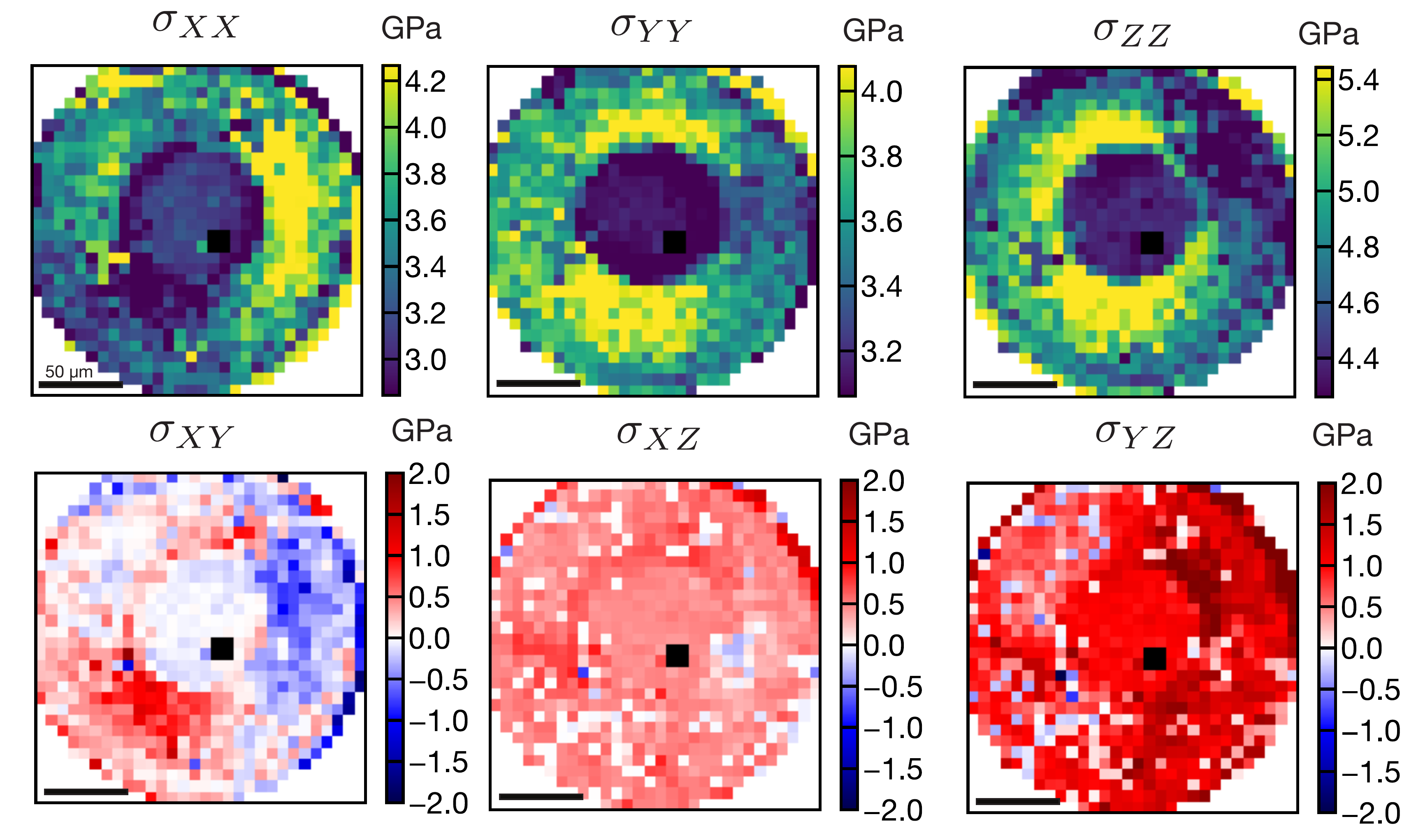}
    \caption{Stress tensor reconstruction of (110)-cut diamond at 4.8~GPa pressure. Analogous to the (111)-cut at low pressure, we observe an inward concentration of lateral stress and a uniform loading stress in the fluid-contact region.}
    \label{fig:110-stress}    
    \end{center}
 \end{figure}

In this section, we discuss our stress reconstruction results for (a)  the (111)-cut diamond at 4.9 GPa and 13.6 GPa (Fig.~\ref{fig:111-stress}), and (b) the (110)-cut diamond at 4.8 GPa (Fig.~\ref{fig:110-stress}). The stress tensors were obtained by numerically minimizing the least-squared residue with respect to the measured shifting and splitting parameters (i.e. $\Pi_{z,i},\Pi_{\perp,i}$). 
While ideally we would measure all eight observables, in this experiment we measured only six: all four shifting parameters and two splitting parameters.
We find that this information allows for the robust characterization of $\sigma_{ZZ}$ and $\sigma_{\perp} = \frac 1 2 (\sigma_{XX}+\sigma_{YY})$, i.e. the two azimuthally symmetric normal components.

We can estimate the accuracy of the reconstructed tensors from the spatial variations of $\sigma_{ZZ}$ at 4.9 GPa. Assuming the medium is an ideal fluid, one would expect that $\sigma_{ZZ}$ to be flat in the region above the gasket hole. 
In practice, we observe spatial fluctuations characterized by a standard deviation $\approx 0.01$~GPa; this is consistent with the expected accuracy based on frequency noise (Table \ref{tab:sensitivity}). The errorbars in the reconstructed stress tensor are estimated using the aforementioned experimental accuracy.

Interestingly, the measured values for $\sigma_{ZZ}$ differs from the ruby pressure scale by $\sim 10\%$. 
This discrepancy is likely explained by inaccuracies in the susceptibility parameters; in particular, the reported susceptibility to axial strain (i.e. $\beta_1$) contains an error bound that is also $\sim 10\%$. 
Other potential sources of systematic error include inaccuracies in our calibration scheme or the presence of plastic deformation. 

Finally, we note that, in many cases, our reconstruction procedure yielded two degenerate solutions for the non-symmetric stress components;
that is, while $\sigma_{ZZ}$ and $\sigma_{\perp}$ have a unique solution, we find two different distributions for $\sigma_{XX},\sigma_{XY},$ etc.
This degeneracy arises from the squared term in the splitting parameter, $\Pi_{\perp,i} = 2\sqrt{\Pi_{x,i}^2+\Pi_{y,i}^2}$, and the fact we measure only six of the eight observables. 
In Fig.~\ref{fig:111-stress} and Fig.~\ref{fig:110-stress} (and Fig.~2B of the main text), we show the solution for the stress tensor that is more azymuthally symmetric, as physically motivated by our geometry.

%applying perp B fields
\subsection{Finite element simulations of the stress tensor}
\label{sec:fem}
Using equations from elasticity theory under the finite element approach, a numerical simulation was coded in ABAQUS for the stress and strain tensor fields in the diamond anvil cell.
%
% Diamond crystal has a cubic structure with three independent elastic constants. %why is this here?
The diamond anvil cell is approximately axially symmetric about the diamond loading axis, in this case the crytallographic (111) axis (i.e.~the $Z$ axis).
This permits us to improve simulation efficiency by reducing the initially 3D tensor of elastic moduli to the 2D axisymmetric cylindrical frame of the diamond as follows.
Initially, the tensor can be written in 3D with cubic axes $c_{11}=1076$ GPa, $c_{12}=125$ GPa, $c_{44}=577$ GPa.
Next, we rotate cubic axes such that the (111) direction is along the $Z$ axis of the cylindrical coordinate system. 
Finally, the coordinate system is rotated by angle $\theta$ around the $Z$ axis and the elastic constants are averaged over $360^{\circ}$ rotation. 
The resulting elasticity tensor in the cylindrical  coordinate  system is
\begin{center}
$\begin{bmatrix}
    1177.5 & 57.4 & 91 & 0 \\ 57.4 & 1211.6 & 57.4 & 0 \\ 91 & 57.4 & 1177.5 & 0 \\ 0 & 0 & 0 & 509.2
\end{bmatrix} \mathrm{[GPa].}$
\end{center}

\begin{figure}
    \begin{center}
     \includegraphics[scale=0.9]{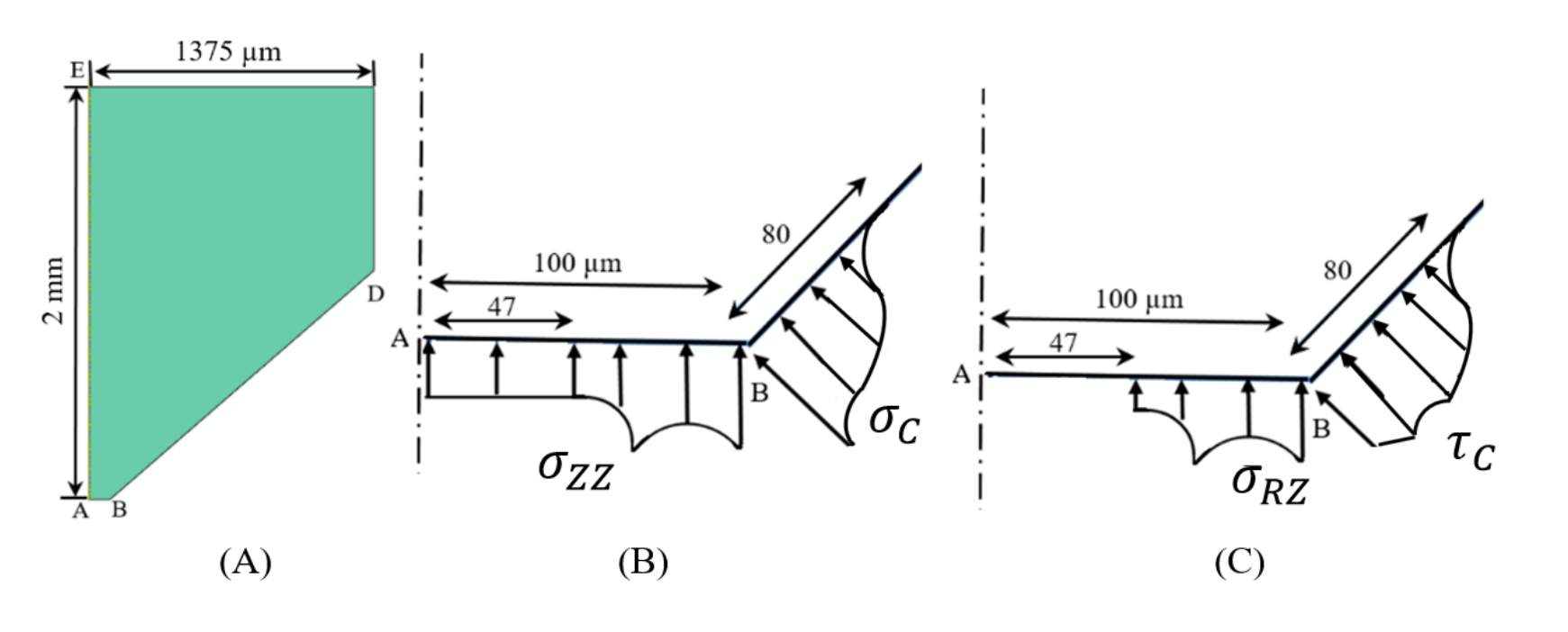}
    \caption{({\bf A}) Diamond geometry, ({\bf B}) anvil tip with distribution of the applied normal stress, ({\bf C}) distribution of the applied shear stress. Normal stress $\sigma_{\textrm{ZZ}}$ at the culet and zero shear stress $\sigma_{\textrm{RZ}}$ along the pressure-transmitting medium/anvil boundary ($r\leq 47~\mu m$) are taken from experiment. Normal and shear contact stresses along all other contact surfaces are determined from the best fit of the mean in-plane stress distribution $\sigma_{\bot}=0.5 (\sigma_{RR}+\sigma_{\Theta\Theta})$ to experiment (main text Fig. 2A and Fig.~\ref{FigFES_S2})}
    \label{FigFES_S1}
    \end{center}
 \end{figure}

The geometry of the anvil and boundary conditions (Fig.~\ref{FigFES_S1}) are as follows:
\begin{enumerate}
    \item The top surface of the anvil is assumed to be fixed. %Fixed to what? Space?
    The distribution of stresses or displacements along this surface does not affect our solution close to the diamond culet line AB.
    \item The normal stress $\sigma_{ZZ}$ along the line AB is taken from the experimental measurements (main text Fig. 2A and ~\ref{FigFES_S2}). The pressure-transmitting medium/gasket boundary runs along the innermost 47 $\mu$m of this radius.
    \item  Along the pressure-transmitting medium/anvil boundary ($r\leq 47~\mu m$) and also at the symmetry axis $r=0$ (line AE) shear stress $\sigma_{\textrm{RZ}}$ is zero. Horizontal displacements at the symmetry axis are also zero.
    \item Normal and shear contact stresses along all other contact surfaces are determined from the best fit to the  mean in-plane stress distribution $\sigma_{\bot}=0.5 (\sigma_{RR}+\sigma_{\Theta\Theta})$ measured in the experiment (main text Fig. 2A and Fig.~\ref{FigFES_S2} ). We chose to fit to $\sigma_{\bot}$ rather than to other measured stresses is because it has the smallest noise in experiment. With this, the normal stress on the line BD with the origin at point B is found to be
\begin{equation}
    \sigma_c = 3.3\times10^5 x^4 - 7.5\times10^4 x^3 +4.5\times10^3 x^2 - 10^2 x + 4.1,
\end{equation}
where $\sigma_c$ is in units of GPa, and the position $x$ along the lateral side is in units of mm. The distribution of the normal stresses is shown in Fig.~\ref{FigFES_S1}B and Fig.~\ref{FigFES_S3}.
\begin{figure}[!t]
 \begin{center}
 \includegraphics[scale=0.9]{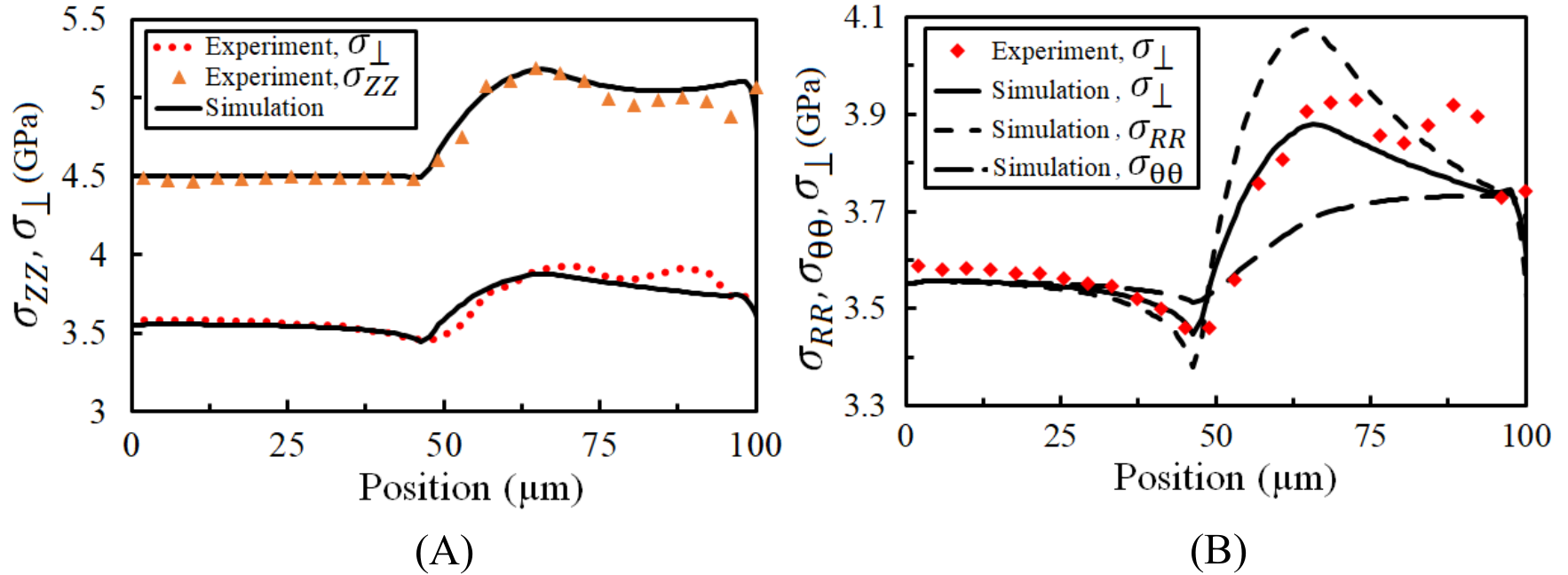}
 \caption{({\bf A}) Distribution of applied normal stress  $\sigma_{ZZ}$ and the  mean in-plane stress  $\sigma_{\bot}$ along the culet  surface of the diamond
 from the experiment and FEM simulations.  
 ({\bf B}) Distribution of  the  mean in-plane stress  $\sigma_{\bot}$ (experimental and simulated) as well as the simulated radial  $\sigma_{RR}$ and circumferential $\sigma_{\Theta\Theta}$ stresses along the culet  surface of the diamond.}
  \label{FigFES_S2}
 \end{center}
\end{figure}
\item At the contact surface between the gasket and the anvil, a Coulomb friction model is applied. The friction coefficient on the culet is found to be 0.02 and along the inclined surface of the anvil (line BD) is found to vary  from 0.15 at point B to 0.3 at 80 $\mathrm{\mu}$m from the culet. The distribution of shear stresses is shown in Fig.~\ref{FigFES_S1}C and Fig.~\ref{FigFES_S3}.
\item Other surfaces not mentioned above are stress-free.
\end{enumerate}

\begin{figure}[!t]
 \begin{center}
 \includegraphics[scale=0.9]{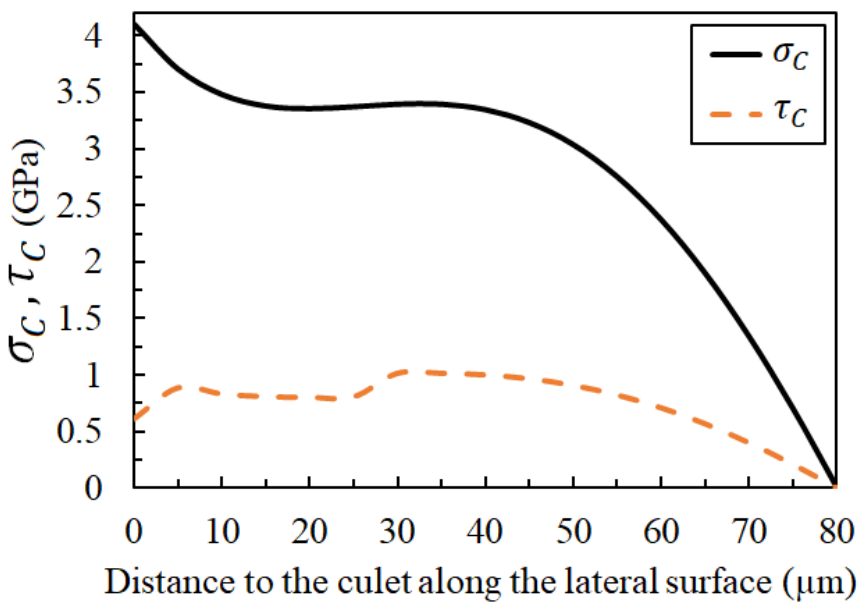}
 \caption{Distribution of applied normal and shear stress along the lateral surface of the diamond determined from the best fit of the  mean in-plane stress distribution $\sigma_{\bot}$ to experiment (main text Fig. 2A  and Fig.~\ref{FigFES_S2}).}
  \label{FigFES_S3}
 \end{center}
\end{figure}

\begin{figure}[!t]
 \begin{center}
 \includegraphics[scale=0.9]{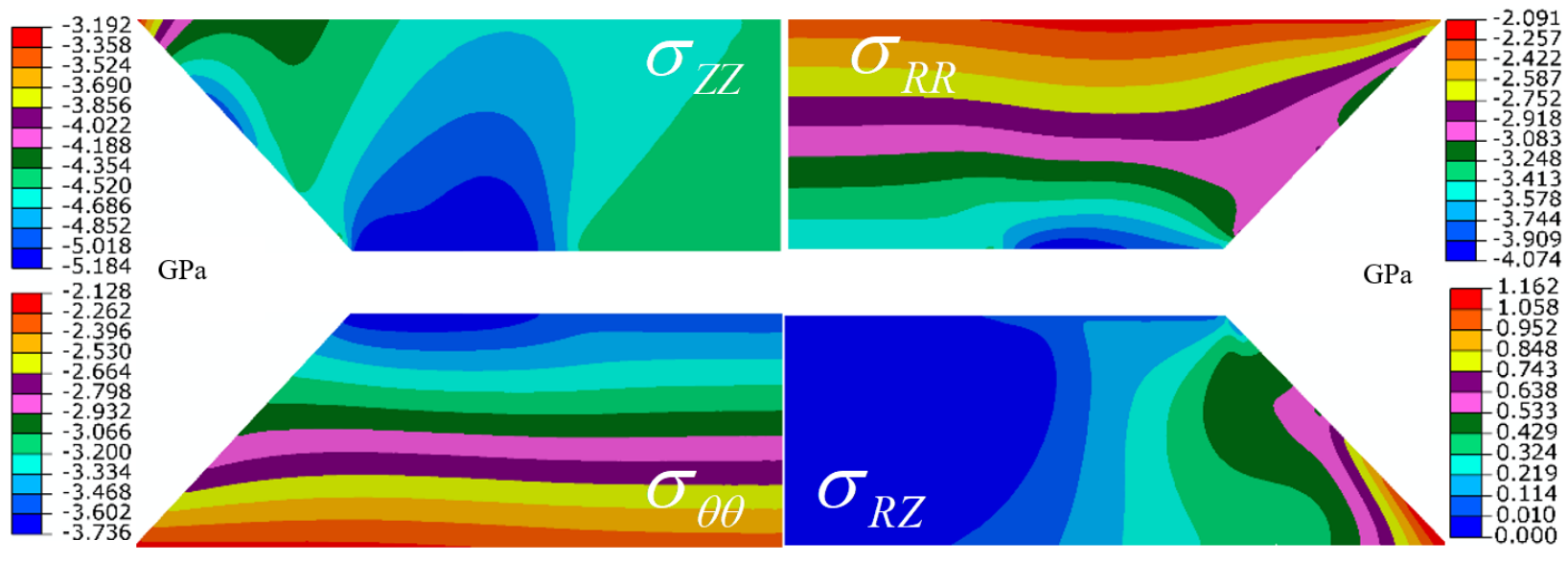}
 \caption{Calculated distributions of the components of stress tensor in the anvil for $r<150$ and $z<475$ $\mu m$.}
 \label{FigFES_S4}
 \end{center}
\end{figure}

The calculated distributions of the stress tensor components near the tip of the anvil are shown in Fig.~\ref{FigFES_S4}.

% Note that the distribution of the normal stress $\sigma_{ZZ}$ and the  mean in-plane stress  $\sigma_{\bot}=0.5 (\sigma_{RR}+\sigma_{\Theta\Theta})$ along the culet  surface of the diamond from the experiment allowed us to reconstruct all normal and friction stresses required for the formulation of the elastic problem and determine the distribution of all components of the stress tensor in the entire anvil. %I want to rewrite this to say the experimental data provided the previously-unknown boundary conditions on the elastic problem. Is this accurate?

\section{Iron dipole reconstruction}
% THe title is more specific than the first couple of sentences suggests

% (Satcher)
%In this section, we describe the details pertaining to the iron $\alpha \leftrightarrow \epsilon$ pressure induced transition.
In this section, we discuss the study of the pressure-induced $\alpha \leftrightarrow \epsilon$ transition in iron. In particular, we provide the experimental details, describe the model used for fitting the data, and outline the procedure to ascertain the transition pressure.
%We describe the experimental details, the model and its fitting procedure, and the extraction of the transition pressure.

For this experiment, the DAC is prepared with a rhenium gasket preindented to $60~\mu$m thickness and laser drilled with a $100~\mu$m diameter hole. 
We load a $\sim 10~\mu$m iron pellet, extracted from a powder (Alfa Aesar Stock No. 00737-30), and a ruby microsphere for pressure calibration.
A solution of methanol, ethanol and water (16:3:1 by volume) is used as the pressure-transmitting medium.

The focused laser is sequentially scanned across a 10$\times$10 grid corresponding to a $\sim30\times30~\mu$m area of the NV layer in the vicinity of the iron pellet, taking an ODMR spectrum at each point. 
As discussed in the main text, the energy levels of the NV are determined by both the magnetic field and the stress in the diamond.
Owing to their different crystallographic orientations, the four NV orientations in general respond differently to these two local parameters. 
As a result, for each location in the scan, eight resonances are observed.

A large bias magnetic field ($\sim 180$~G), not perpendicular to any of the axes, is used to suppress the effect of the transverse stress in the splitting for each NV orientation.
However, the longitudinal stress still induces an orientation-dependent shift of the resonances which is nearly constant across the imaging area, as measured independently (Fig~\ref{fig:111-fits}C).

By analyzing the splittings of the NV resonances across the culet, we can determine the local magnetic field and thereby reconstruct the dipole moment of the iron pellet.

To estimate the error in pressure, a ruby fluorescence spectrum was measured before and after the ODMR mapping, from which the pressure could be obtained \cite{MezouarRubyPressure}. 
The pressure was taken to be the mean value, while the error was estimated using both the pressure range and the uncertainty associated with each pressure point.

\subsection{Extracting Splitting Information}

\begin{figure}
    \centering
    \includegraphics[width=0.9\textwidth]{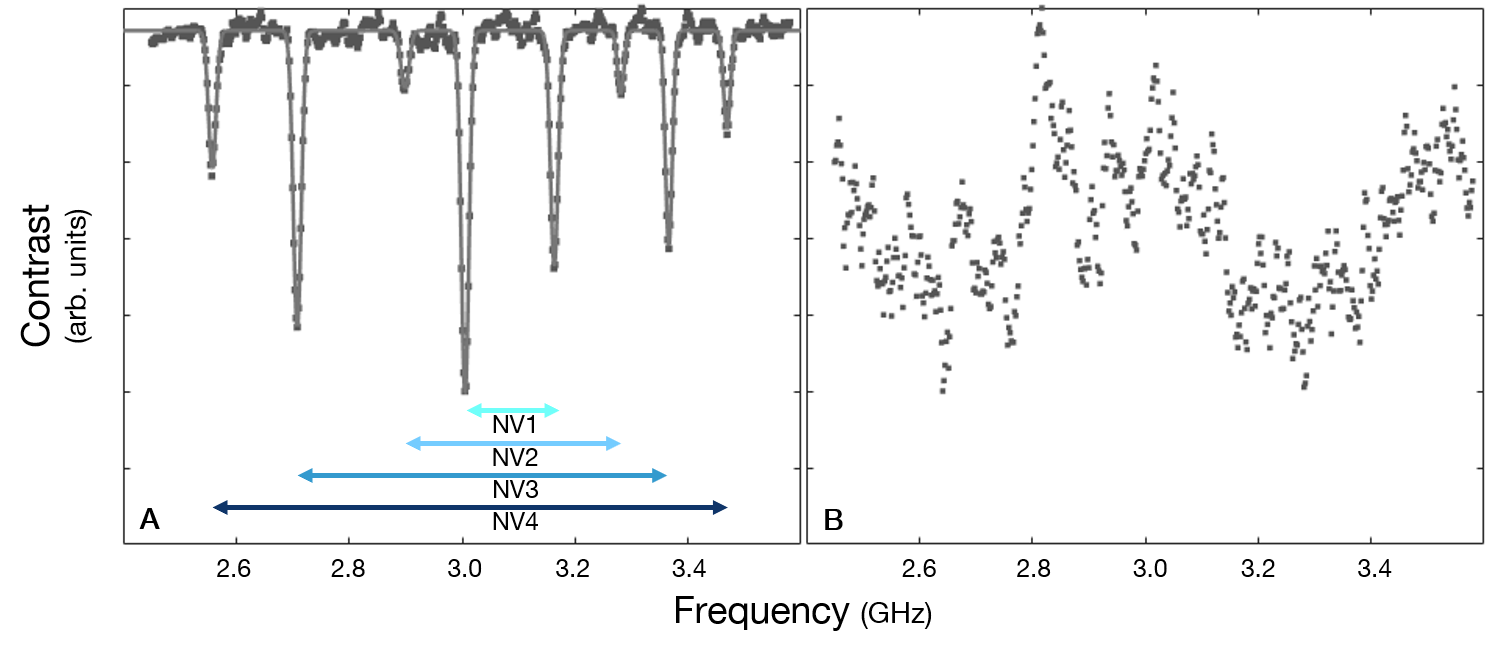}
    \caption{\textbf{(A)} Example of a typical spectrum with a fit to eight free Gaussians. Resonance pairs are identified as in Fig.~1D of the main text: NV4 has the strongest magnetic field projection and NV1 has the weakest. \textbf{(B)} Example spectrum for which resonances are broadened and shifted. In this case we cannot correlate any resonances in the spectrum to specific NV orientations.}
    \label{fig:IronSpectra}
\end{figure}

The eight resonances in a typical ODMR spectrum are fit to Gaussian lineshapes to extract the resonance frequency (Fig~\ref{fig:IronSpectra}A). 
Resonances are paired as in Fig.~1D of the main text: from outermost resonances to innermost, corresponding to NV orientations with the strongest magnetic field projection to the weakest, respectively.
Once identified, we calculate the splitting and magnetic field projection for each NV orientation.

We note that there are two regimes where our spectra cannot confidently resolve and identify all the eight resonances.
First, at high pressure, the resonance contrast for some NV orientations is diminished, possibly due to 
%\red{stress-induced changes to the NV excited states which modify its dipole moment matrix elements} or 
a modification of the frequency response of the microwave delivery system.
Second, close to or on top of the iron pellet, the resonances are broadened; we attribute this to the large magnetic field gradients (relative to the imaging resolution) caused by the sample. 
The resulting overlap in spectral features obfuscates the identity of each resonance (Fig.~\ref{fig:IronSpectra}B). 
In both cases, we fit and extract splittings only for the orientations we could identify with certainty.

% Discussion of the experimental data:
% - What we exctracted
% - Why that is enough
% How we calibrated the NV axes
\subsection{Point Dipole Model}
\label{sec:point_dipole}

We model the magnetization of our pellet sample as a point dipole at some location within the sample chamber.
The total magnetic field is then characterized by the external applied field, $\bm{B}_0$, the dipole of the sample, $\bm{d}$, and the position of the dipole, $\bm{r}$.
Because of the presence of a large applied field, we observe that the magnetization of the sample aligns with $\bm{B}_0$, and thus, we require only the strength of the dipole to characterize its moment, $\bm{d} = D \hat{B}_0$.
We expect the external magnetic field and the depth of the particle to remain nearly constant at different pressures.
This is indeed borne out by the data, see Sec.~\ref{sec:fit_all}.
As a result, we consider the external magnetic field  $\bm{B}_0 = (-23(7), -160(1), 92(2))$~G and depth of the iron pellet $r_Z = -5(1)~\mathrm{\mu m}$ to be fixed.
% We discuss the robustness of our results in Sec.~\ref{sec:fit_all}, by comparing the results when these parameters are not fixed \blue{(\textbf{the word 'result' used twice in this sentence})}.
%\red{Francisco Update Values}

Due to the dipole of the iron pellet, the magnetic field across the NV layer at position $\bm{x}$ is given by:
\begin{equation}
    \bm{B}(\bm{x}) = \bm{B}_0 + \frac{\mu_0}{4\pi} \frac{1}{|\bm{x}|^3}\left( 3 \hat{x}(\bm{d}\cdot \hat{x}) - \bm{d}\right),
\end{equation}
where hats represent unit vectors.
At each point, the local field induces a different splitting, $\Delta^{(i)}$, to the 4 NV crytallographic orientations $i \in \{1,2,3,4\} $, measured by diagonalizing the Hamiltonian $H = D_{gs} S_z^2 + B_z^{(i)} S_z + B_\perp^{(i)} S_x$,
% :
% \begin{equation}
% \begin{equation}
%     \Delta^{(i)} \approx 2\gamma_\mathrm{B}\sqrt{ (B_z^{(i)})^2 + \left[\frac{ (B_\perp^{(i)})^2}{D_{\mathrm{gs}}}\right]^2 } ~,
% \end{equation}
where $B_z = |\bm{B}\cdot \hat{z}^{(i)}|$ is the projection of $\bm{B}$ onto the axis of the NV, and $B^{(i)}_\perp = \sqrt{|\bm{B}|^2 - (B^{(i)}_z)^2}$, its transverse component. $D_{gs}$ is the zero field splitting of the NV.
For each choice of $D$, $r_X$ and $r_Y$, we obtain a two dimensional map of $\{\Delta^{(i)}\}$.
Performing a least squares fit of this map against the experimental splittings determines the best parameters for each pressure point. %\blue{(\textbf{Should we consider changing "pressure point" to "2D scan"?})}
% This enables the measurement of the dipole moment of the iron pellet as a function of pressure.
% Since the external magnetic field is set constant, it provides a direct probe of the magnetic susceptibility of the sample and thus its magnetic phase, Fig.~3 of the main text. \blue{(\textbf{The first sentence is superfluous. With regards to the second, we previously mention that we assume paramagnetic response. Perhasp we should change the wording around to reflect that.})}
The error in the fitting procedure is taken as the error in the dipole strength $D$.

\subsection{Determining Transition Pressure}

Although the $\alpha \leftrightarrow \epsilon$ structural phase transition in iron is a first order phase transition, we do not observe a sharp change in the dipole moment of the sample, observing instead a cross-over between the two magnetic behaviors.
We attribute this to the non-hydrostatic behavior of the sample chamber at high pressures.
As a result, different parts of the iron pellet can experience different amounts of pressure and, thus, undergo a phase transition at different applied pressures.
The measured dipole moment should scale with the proportion of the sample that has undergone the phase transition.
This proportion, $p(P)$, should plateau at either $0$ or $1$ on different sides of the phase transition, and vary smoothly across it.
To model this behavior we use a logistic function:
\begin{equation}
    p(P) = \frac{1}{e^{B(P-P_c)}+1}~.
\end{equation}
The dipole strength is then given by:
\begin{equation}
    D = p(P) D_{\alpha} + [1-p(P)] D_{\epsilon}~,
\end{equation}
where $D_{\alpha}~(D_\epsilon)$ is the dipole moment of the sample in the $\alpha$~($\epsilon$) structural phase and $1/B$ corresponds to the width of the transition, thus its uncertainty.

% Details, why we expect it to work, simplifications
\subsubsection{Large error bar in the 11~GPa decompression point}

\begin{figure}
    \centering
    \includegraphics[width=0.6\textwidth]{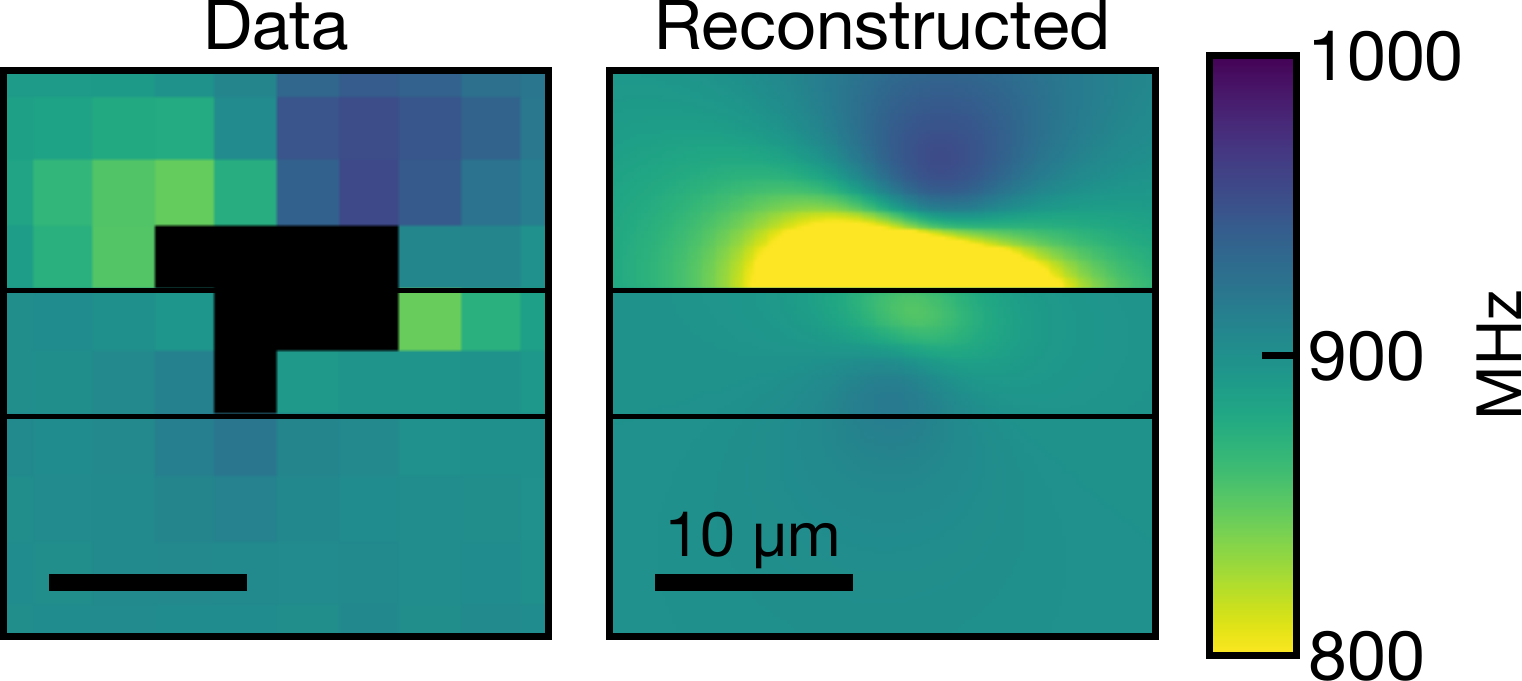}
    \caption{
    Measured map of the splittings of one of the NV orientations (left).
    Near the top of the plot we observe a much stronger splitting compared to the bottom of the plot.
    Throughout the measurement, the shift in the pressure induced a shift in the dipole moment of the sample.
    We consider 3 different regions (seperated by horizontal lines) corresponding to 3 different dipole strengths.
    The reconstructed map of the splittings is shown on the right in agreement with the data.
    From the center and the spread of dipole strengths, we extract the dipole moment and its error.
    Black bar corresponds to $10~\mathrm{\mu m}$.
    }
    \label{fig:11GPa}
\end{figure}

During the decompression, around $11$~GPa, we observed a significant drift of the pressure during measurement of the ODMR spectra.
Unfortunately, the starting pressure was close to the transition pressure, and the drift in pressure led to a very large change in the pellet's dipole moment throughout the scanning measurement.
This is clear in the measured data, Fig.~\ref{fig:11GPa}, with the top-half of the map displaying a significantly larger shift with respect to the bottom-half.

To extract the drift in the dipole moment, we divide the two-dimensional map into three different regions, each assumed to arise from a constant value of the dipole moment of the pellet.
By fitting to three different dipole moments (given a fixed position, $r_X$ and $r_Y$) we obtain an estimate of the drift of the dipole moment that allows us to compute an errorbar of that measurement.
The estimated dipole moment at this pressure point is taken as the midpoint of the three extracted values, $\dfrac{D_{max}+D_{min}}{2}$, while the error is estimated by the range, $\dfrac{D_{max}-D_{min}}{2}$.

\subsection{Fitting to external magnetic field and depth}

\begin{figure}
    \centering
    \includegraphics[width=0.8\textwidth]{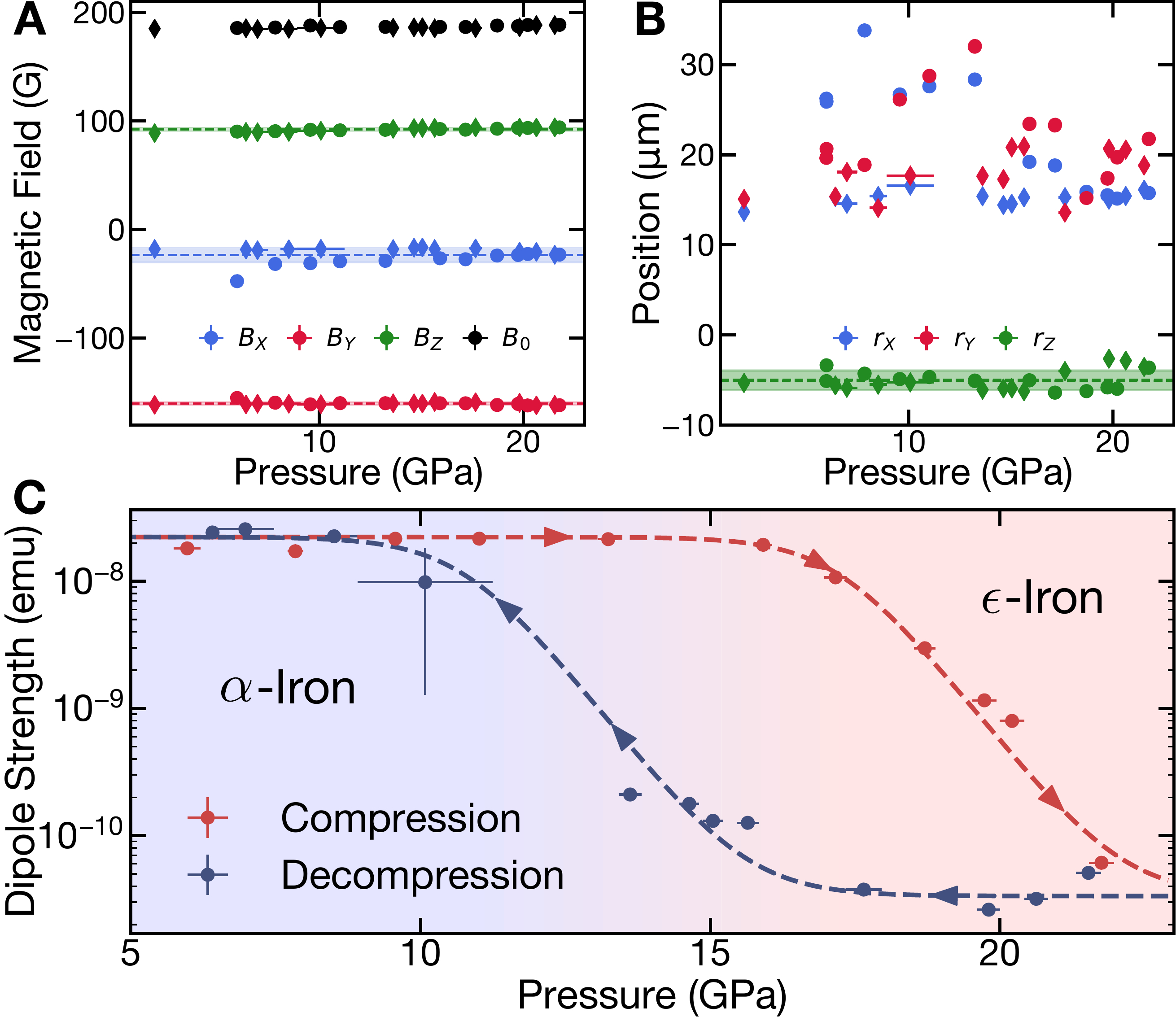}
    \caption{Result of fitting procedure when the external magnetic field and the depth of the iron pellet is allowed to vary at each pressure.
    ({\bf A})[({\bf B})] External magnetic field [position of the pellet] extracted as a function of pressure (circles correspond to compression while diamonds correspond to decompression).
    Across the entire range of pressures, the extracted external magnetic field and the depth of the iron pellet is approximately constant.
    %The fluctuactions are most likely associated with the fitting rather than an alteration of the physical parameters.
    In the final fitting procedure, these values are fixed to their extracted mean (dashed lines).
    Shaded regions correspond to a standard deviation above and below the mean value.
    ({\bf C}) Dipole strength of the iron pellet, extracted when all seven parameters ($B_X, B_Y, B_Z, D, r_X,r_Y,r_Z$) are fitted. 
    The resulting transitions occur at $17.2$~GPa and $10.8$~GPa for compression and decompression, respectively.
    Comparing with the width of the transition ($1.3$~GPa), these values are in excellent agreement with those presented in the main text.
    }
    \label{fig:fit_all}
\end{figure}

In this section we present additional data where we have allowed both the external magnetic field and the depth of the iron pellet to vary in the fitting procedure.
The result of the fitting procedure is summarized in Fig.~\ref{fig:fit_all}.

In particular, we expect the external magnetic field and the depth of the pellet to remain constant at different pressures.
Indeed, we observe this trend in the extracted parameters, Fig.~\ref{fig:fit_all}(A,B).
Using the mean and standard deviation, we estimate these values and their errors, quoted in Sec.~\ref{sec:point_dipole}.
The final fitting procedure with these values fixed is presented in the main text.

\label{sec:fit_all}
%draft figure SMfigIronExample

% (Francisco, Thomas)
% how to reconstruct dipole from field lines/what was fixed, why broadening occurs near iron, 
% Landau theory does not apply, use Sigmoid to fit
\section{Gadolinium}
\subsection{Experimental detail}
We use a custom-built closed cycle cryostat (Attocube attoDRY800) to study the $P$-$T$ phase diagram of Gd. The DAC is placed on the sample mount of the cryostat, which is incorporated with a heater and a temperature sensor for temperature control and readout. 
% what is the best way to write this sentence? 
% FM: I don't think we need to given that we don't make statements about the placement of the heater and temp sensor
%We neglect temperature gradients between the sample chamber of the DAC and the sample mount of the cryostat because of the DAC's small thermal mass and high thermal conductivity. 

For this experiment, we use beryllium copper gaskets. The Gd sample is cut from a 25 $\mu$m thick Gd foil (Alfa Aesar Stock No. 12397-FF) to a size of $\sim 30 \, \mu$m $\times$ $30 \, \mu$m and loaded with cesium iodide (CsI) as the pressure-transmitting medium. A single ruby microsphere loaded into the chamber is used as a pressure scale.

\begin{figure}
    \centering
    \includegraphics[width=0.8\textwidth]{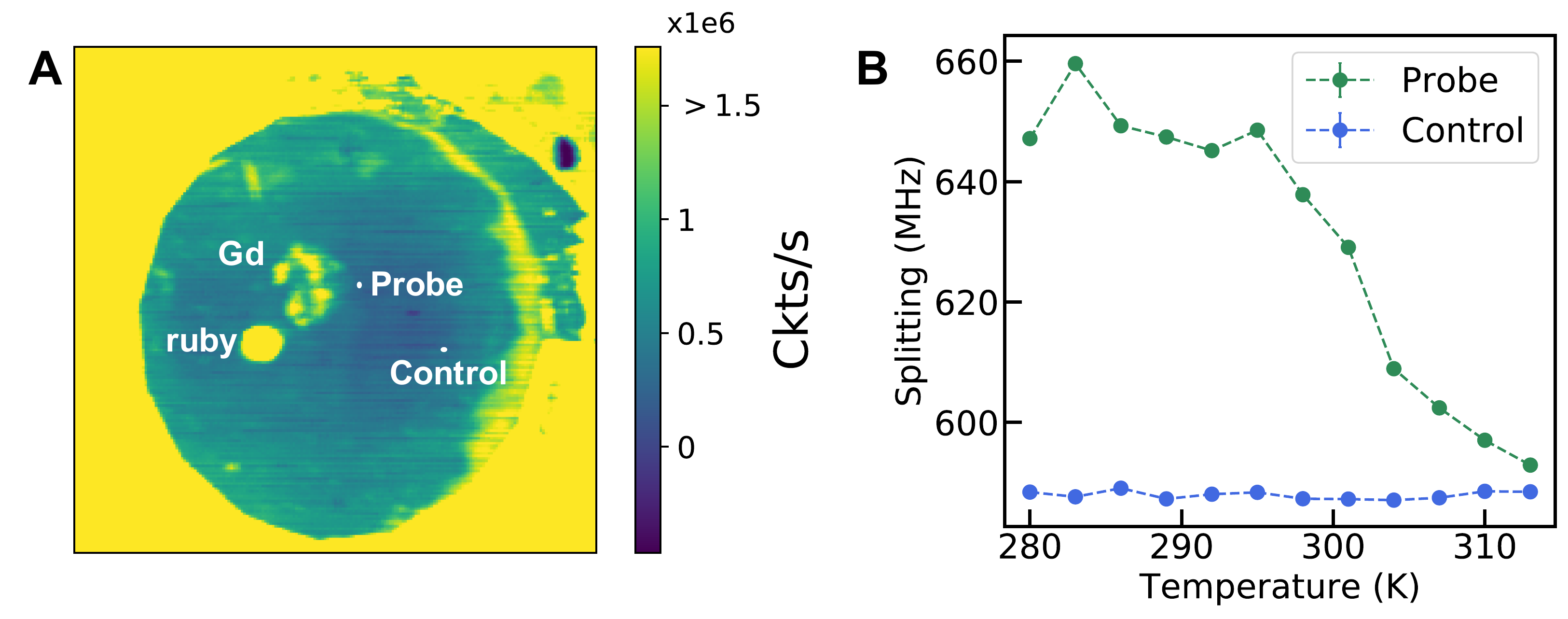}
    \caption{({\bf A}) The protocol for obtaining $P$-$T$ phase map of Gd relies on monitoring the ODMR spectrum versus temperature and pressure at a point of interest (probe) near the sample. To verify that the observed signal is from the Gd flake, one can perform the same measurement on a control point further away from the sample. 
    ({\bf B}) The difference in the splitting between the probe and control points isolates the magnetic field generated by the Gd sample, allowing us to monitor the magnetic behavior of the sample.
    %The comparison of the data at these two points shows a clear magnetic signal from the sample that manifests as a change in splitting of the NV centers.
    }
    \label{fig:Gd-probe-control}
\end{figure}

For each experimental run, we start with an initial pressure (applied at room temperature $~300$~K) and cool the cell in the cryostat. Due to contraction of the DAC components with decreasing temperature, each run of the experiment traverses a non-isobaric path in $P$-$T$ phase space, Fig. \ref{fig:PM_to_FM}A. 
Using fiducial markers in the confocal scans of the sample chamber, we track points near and far from the Gd sample throughout the measurement. 
By performing ODMR spectroscopy at these points for each temperature, we monitor the magnetic behavior of the sample.
More specifically, comparing the spectra between the close point (probe) against the far away one (control), Fig. \ref{fig:Gd-probe-control}, enables us to isolate the induced field from the Gd sample.
%\red{Control against what? Laser pointing error from thermal contraction of sample, ...}

\subsection{Fitting phase transition}
% (Francisco)
There are three different transitions we which to locate in the study of the Gd's $P$-$T$ phase diagram: a magnetic transition from PM dhcp to FM dchp; structural phase transitions, either hcp $\to$ dhcp or Sm-type $\leftrightarrow$ dhcp; and a magnetic phase transition from PM Sm-type to AFM Sm-type.

In order to extract the transition temperature of the paramagnet to ferromagnet transition from our data, we model the magnetization of our sample near the magnetic phase transition using a regularized mean field theory.
%combination of mean field theory and a regularization far away from the transition.

The magnetism of gadolinium is well-described by a three dimensional Heisenberg magnet of core electrons \cite{oroszlany2015}.
In the presence of an external magnetic field, the free energy near the critical point is expanded in even powers of the magnetization with a linear term that couples to the external magnetic field:
\begin{equation} \label{eq:LandauTheory}
    f = - B m + \frac{\alpha}{2} (T-T_\textrm{C}) m^2 + \frac{\beta}{4} m^4,
\end{equation}
where $m$ is the magnetization, $B$ is the external magnetic field, $\alpha$ and $\beta$ the expansion coefficients, $T$ the temperature, and $T_\textrm{C}$ the transition temperature.
In this treatment, we implicitly assume that $\alpha$ and $\beta$ do not vary significantly with pressure and thus can be taken to be constant across paths in $P$-$T$ phase space.
The magnetization $m_{\mathrm{min}}$ is then obtained by minimizing the free energy. 

Because our observation region extends far away from the transition, we observe a plateauing of the splittings that emerges from the microscopics of Gd.
Using $R$ as the regularization scale and $\tilde{A}$ as the maximum magnetization of the sample we propose the simple regularization scheme:
\begin{equation}
    m(T,P) = \tilde{A} \frac{m_{\mathrm{min}}}{m_{\mathrm{min}} + R}.
\end{equation}

The splitting of the NV group, up to some offset, is proportional to the magnetization of the sample.
This proportionality constant, $A$, captures he relation between magnetization and induced magnetic field, the geometry of sample relative to the measurement spot, as well as the susceptibility of the NV to the magnetic field.
The splitting of the NV is then given by:
\begin{equation}
    \Delta = A\frac{m_{\mathrm{min}}}{m_{\mathrm{min}} + R} + c
\end{equation}
where we incorporated $\tilde{A}$ into $A$ as well. Normalizing $\alpha$ and $\beta$ with respect to $B$, we obtain six parameters that describe the magnetization profile, directly extracting $T_\textrm{C}$.

In the case of the first order structural phase transitions, similar to that of iron, we take the susceptibility to follow a logistic distribution. We model the observed splitting as:
\begin{equation}
    \Delta = \frac{A}{e^{B(T-T_\textrm{C})} + 1} + c
\end{equation}
Fitting to the functional form provides the transition temperature $T_\textrm{C}$.
Error bar is taken as largest between $1/B$ and the fitting error.

In the case of the paramagnetic to antiferromagnetic transition, we use the mean field susceptibility across the phase transition of the system.
The susceptibility across such transition is peaked at the transition temperature:
\begin{equation}
    \chi(T) \propto \begin{cases}
        \dfrac{1}{T- \theta_p} & T > T_c\\
        \;\\
        C \dfrac{3L'(H/T)}{T - \theta_p 3L'(H/T)} & T < T_c
    \end{cases}
\end{equation}
where $C$ is chosen to ensure continuity of $\chi$, $L'(x)$ is the derivative of the Langevin function $L(x)$at, $H$ is a meaasure of the applied field, and $\theta_p$ is the assymptotic Curie point.
Finally, we fit the observed splitting to:
\begin{equation}
    \Delta = A\chi(T; T_c, H, \theta_p) + c
\end{equation}
where, as before, $A$ captures both the geometric effects, as well as the response of the chosen NV group to the magnetic field.

\subsection{Additional data}

\begin{figure}
    \centering
    \includegraphics[width = 0.6\textwidth]{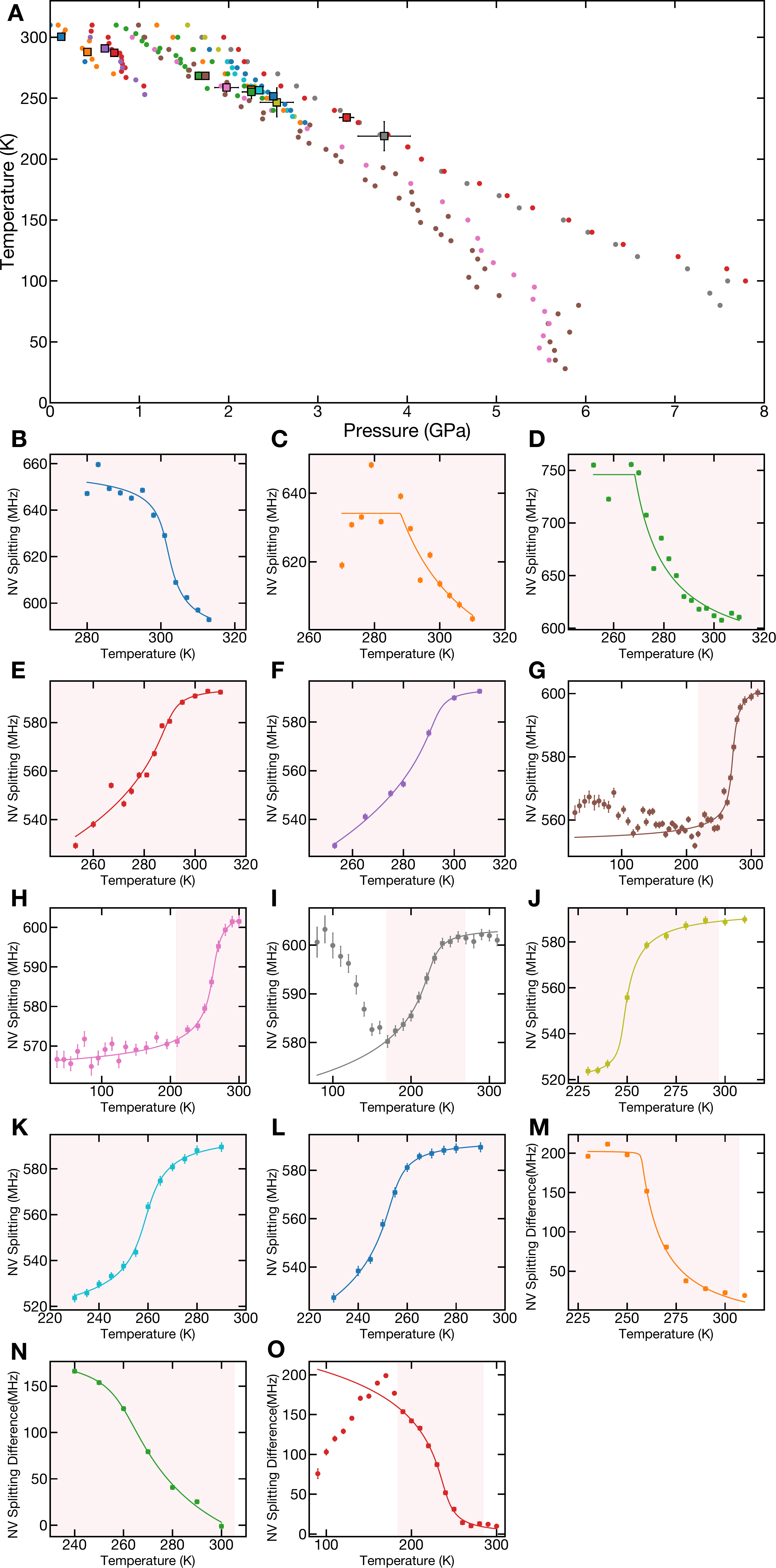}
    \caption{
    {\bf (A)} Paths in the $P$-$T$ phase space that inform about the hcp PM phase to the hcp FM phase.
    {\bf (B-O)} Measured NV splitting and corresponding fit.
    The resulting transition temperatures are highlighted in (A) with squares.
    Shaded region corresponds to the part of the spectrum fitted.
    }
    \label{fig:PM_to_FM}
\end{figure}

\begin{figure}
    \centering
    \includegraphics[width = 0.8\textwidth]{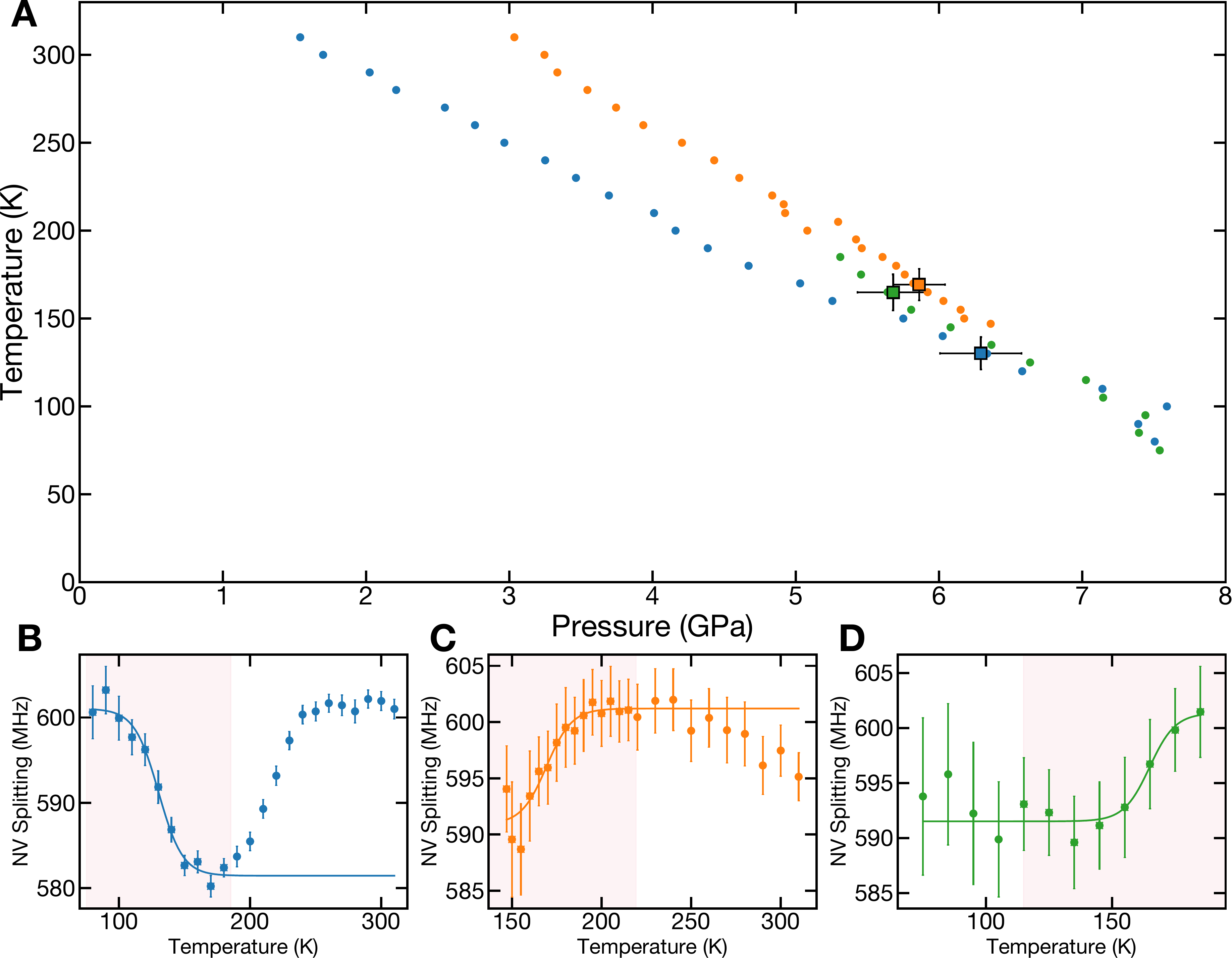}
    \caption{
    {\bf (A)} Paths in the $P$-$T$ phase space that inform about the transition to the PM dhcp phase.
    {\bf (B-D)} Measured NV splitting and corresponding fit.
    The resulting transition temperatures are highlighted in (A) with squares.
    We interpret (B) as a transition from FM hcp to PM dhcp, while (C),(D) as a transition from PM Sm-type to PM dhcp.
    Shaded region corresponds to the part of the spectrum fitted.
    }
    \label{fig:to_dhcp}
\end{figure}

\begin{figure}
    \centering
    \includegraphics[width = 0.8\textwidth]{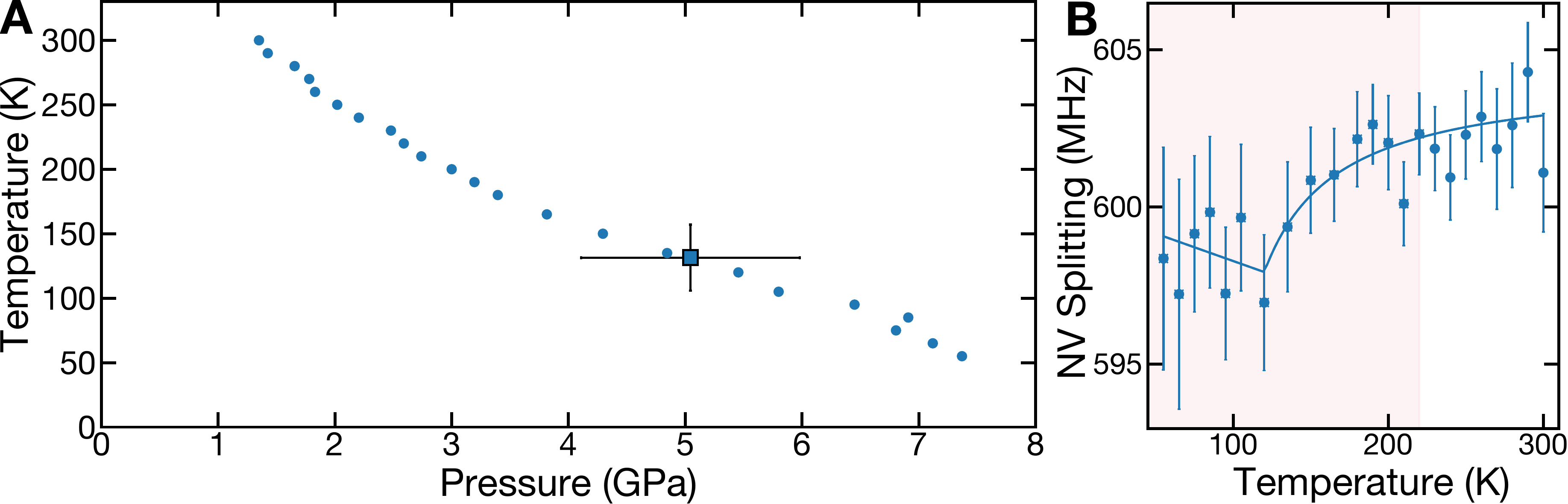}
    \caption{
    {\bf (A)} Path in the $P$-$T$ phase space where a signal consistent with the purported AFM transition in Sm-type Gd is seen {\bf (B)}.
    Shaded region corresponds to the part of the spectrum fitted.
    }
    \label{fig:afm_suggestion}
\end{figure}

In this section we present the data for the different paths taken in $P$-$T$ phase and the resulting fits.
Table \ref{PTruns} summarizes the observations for all experimental runs.
Fig.~\ref{fig:PM_to_FM} contains the data used in determining the linear pressure dependence of the hcp phase.
Fig.~\ref{fig:to_dhcp} comprises the data used in determining the transition to the dhcp phase, either via the FM hcp to PM dhcp transition, Fig.~\ref{fig:to_dhcp}B, or via the difference in susceptibilities between PM Sm-type and PM dhcp  of Gd, Fig.~\ref{fig:to_dhcp}C and D. 
We emphasize that in the blue path, we begin the experiment below $2$~GPa and thus in the hcp structure, while for the orange and green, we begin above $2$~GPa, so we expect the system to be in Sm-type.
Finally, Fig.~\ref{fig:afm_suggestion} contains the data where we observe a change in the susceptibility of Gd that occurs at the purported Sm-type PM to AFM transition.

\begin{table}
\centering
\begin{tabular}{l|l|l|l}
\hline \hline
  Run & Direction  & Phase transition & Remarks, visible in Fig. \\ \hline
  1 & Heat-up & hcp (FM) $\longrightarrow$ hcp (PM) & New sample, Fig. \ref{fig:PM_to_FM}B \\ \hline
  2 & Cool-down & hcp (PM) $\longrightarrow$ hcp (FM)  & Fig. \ref{fig:PM_to_FM}C \\ \hline
  3 & Cool-down & hcp (PM) $\longrightarrow$ hcp (FM)  & Fig. \ref{fig:PM_to_FM}D  \\ \hline
  4 & Cool-down & No observation & Probably starting in Sm due \\
  & & & to large initial pressure \\ \hline \hline
  %5 & Cool-down & Inconclusive & Technical issues  \\ \hline
  %6 & Heat-up & No observation &  \\ \hline
  %7 & Cool-down & Weak evidence for hcp (PM) $\longrightarrow$ hcp (FM) &   \\ \hline
  5 & Cool-down & hcp (PM) $\longrightarrow$ hcp (FM) & New sample, Fig. \ref{fig:PM_to_FM}E  \\ \hline
  6 & Heat-up & hcp (FM) $\longrightarrow$ hcp (PM) & Fig. \ref{fig:PM_to_FM}F  \\ \hline
  7 & Cool-down & hcp (PM) $\longrightarrow$ hcp (FM)  & Fig. \ref{fig:PM_to_FM}G  \\ \hline
  8 & Heat-up & hcp (FM) $\longrightarrow$ hcp (PM) & Fig. \ref{fig:PM_to_FM}H  \\ \hline
  9 & Cool-down & hcp (PM) $\longrightarrow$ hcp (FM) & Fig. \ref{fig:PM_to_FM}I, \ref{fig:to_dhcp}B  \\
  & & $\longrightarrow$ dhcp (PM) & \\ \hline
  %13 & Cool-down & Weak evidence for Sm (PM) $\longrightarrow$ dhcp (PM) &   \\ \hline
  10 & Cool-down & Weak evidence for &  Probably starting in Sm due \\
  & & Sm (PM) $\longrightarrow$ Sm (AFM) & to metastability, Fig. \ref{fig:afm_suggestion}B \\ \hline \hline
  %15 & Cool-down & No observation & Possibly still in Sm phase   \\ \hline
  11 & Cool-down & hcp (PM) $\longrightarrow$ hcp (FM) & New sample, Fig. \ref{fig:PM_to_FM}J  \\ \hline
  12 & Heat-up & hcp (FM) $\longrightarrow$ hcp (PM) & Fig. \ref{fig:PM_to_FM}K  \\ \hline
  13 & Cool-down & hcp (PM) $\longrightarrow$ hcp (FM) & Fig. \ref{fig:PM_to_FM}L  \\ \hline
  14 & Cool-down & Weak evidence for & Probably starting in Sm due \\
  & & Sm (PM) $\longrightarrow$ dhcp (PM) & to large initial pressure \\ \hline
  %20 & Heat-up & Weak evidence for dhcp (PM) $ \longrightarrow$ Sm (PM) &  \\ \hline
  15 & Cool-down & Weak evidence for & Probably starting in Sm due  \\
  & & Sm (PM) $ \longrightarrow$ dhcp (PM) & to metastability, Fig. \ref{fig:to_dhcp}C \\ \hline
  16 & Heat-up & Weak evidence for & Fig. \ref{fig:to_dhcp}D \\
  & & dhcp (PM) $ \longrightarrow$ Sm (PM) & \\ \hline \hline
  %23 & Cool-down & No observation & \\ \hline
  %24 & Cool-down & hcp (PM) $\longrightarrow$ hcp (FM) & Fig. \ref{fig:PM_to_FM}M \\ \hline
  %25 & Heat-up & No observation &  \\  \hline
  17 & Cool-down & hcp (PM) $\longrightarrow$ hcp (FM) & New sample, Fig. \ref{fig:PM_to_FM}M  \\  \hline
  18 & Heat-up & hcp (FM) $\longrightarrow$ hcp (PM) & Fig. \ref{fig:PM_to_FM}N \\  \hline
%   29 & Cool-down & hcp (PM) $\longrightarrow$ hcp (FM) &  \\ \hline
  19 & Cool-down & hcp (PM) $\longrightarrow$ hcp (FM)  & Fig. \ref{fig:PM_to_FM}O\\
     &  & and start of transition to dhcp (PM) &  \\
\hline \hline
\end{tabular}
\caption{Summary of all experimental runs in the $P$-$T$ phase diagram, indexing either a decrease or increase in temperature during this path, and the observed phase transitions.
Each group of runs, between double lines in the table, corresponds to a different sample.
    }
\label{PTruns}
\end{table}

\subsection{Recreating the $P$-$T$ phase diagram of Gd}
% 1. Summary of literature review of Gd
% 2. Our experimental runs are summarized in Table 2. Based on these runs, our hypothesis is summarized in Fig. 11.
%
%
The rich magnetic behavior of Gd is partially dependent on its structural phases, captured in the sequence: hexagonal closed packed (hcp) to Samarium (Sm) type at $\sim 2$~GPa, and then to  double hexagonal closed packed (dhcp) at $\sim6$~GPa.
In particular, while the paramagnetic (PM) phase of hcp orders to a ferromagnet (FM), the PM phase of Sm-type orders to an antiferromagnet (AFM) \cite{JAYARAMAN1978707}.
Similarly, dhcp undergoes a PM to magnetically ordered phase transition.
%Notably, the hcp to Sm-type transition is "sluggish" \cite{JAYARAMAN1978707, PhysRev.139.A682}, leading to a mixture of the two phases over the timescale of the experimental run.

For experimental runs with initial pressures $< 2$~GPa (runs 1-3, 5-9, 11-13, 17-19), we observe a PM $\leftrightarrow$ FM phase transition in hcp Gd. In agreement with previous studies, we see a linear decrease of the Curie temperature with increasing pressure up to $\sim4$~GPa \cite{PhysRevB.71.184416, PhysRev.139.A682,IWAMOTO2003667}. Notably, prior studies have shown a structural transition from hcp to Sm-type at $2$~GPa \cite{PhysRev.139.A682,doi:10.1080/08957959.2014.977277,AKELLA1988573}, which is believed to be ``sluggish'' \cite{JAYARAMAN1978707,PhysRev.139.A682}. This is indeed consistent with our observation that the linear dependence of the Curie temperature persists well into the Sm-type region, suggesting the existence of both structural phases over our experimental timescales.

Furthermore, in run 9 (Table \ref{PTruns} and Fig. \ref{fig:afm_suggestion}A,B), we observe a complete loss of FM signal when pressures exceed $\sim6$ GPa at $\sim 150$~K, in good agreement with the previously reported phase transition from hcp (FM) to dhcp (PM) structure \cite{doi:10.1080/08957959.2014.977277,PhysRev.139.A682}. 
Upon performing a similar path in $P$-$T$ space (run 19), we observe the same behavior. % refer to run 14 (Fig 8)
In contrast to the previous slow hcp to Sm-type transition, we believe that the equilibrium timescale for the hcp (FM) to dhcp (PM) transition is much faster at this temperature.
%Notably, in run 19, our system attains $6.3$~GPa at $130$~K, yet we don't observe this hcp (FM) to dhcp (PM) transition, suggesting either that at lower temperature the transition occurs at higher pressure, or that it has slower timescale.
%

After entering the dhcp structure (run 9), we no longer observe a clear FM signal from the sample even after heating to $315$~K and depressurizing $<0.1$~GPa.
This can be explained by the retention of dhcp or Sm-type structure in the sample. 
Previous studies, suggesting that the Sm-type phase in Gd is metastable up to ambient pressure and temperature \cite{JAYARAMAN1978707}, corroborate that our sample is likely still in the Sm-type structural phase. 
It is not too surprising, that by continuing to cool down and walking along a slightly different $P$-$T$ path, we observe only a small change in the NV splitting at $\sim150$~K and $\sim5$~GPa as we cross the purported Sm-type PM to AFM phase boundary (run 10 in Table \ref{PTruns})\cite{doi:10.1080/08957959.2014.977277, JAYARAMAN1978707,PhysRev.139.A682}.

Moreover, the metastable dynamics of hcp to Sm-type transitions are strongly pressure and temperature dependent, suggesting that different starting points (in the $P$-$T$ phase diagram) can lead to dramatically different behaviors.
Indeed, by preparing the sample above 2~GPa at room temperature (run 4), we no longer detect evidence for a ferromagnetic Curie transition, hinting the transition to the Sm-type structure. Instead, we only observe a small change in the NV splitting at $\sim 6$~GPa and $\sim170$~K, which could be related to the presence of different paramagnetic susceptibilities of the Sm-type and dhcp structural phases. Interestingly, by cycling temperature across the transition (run 14-16 in Table \ref{PTruns}), we observe negligible hysteresis, suggesting fast equilibration of this structural transition.

% \begin{figure}
% \begin{center}
% \includegraphics[scale=0.5]{gd-phase-map.png}
% \caption{Characterization of the equilibration %timescales of gadolinium's structural phase %transitions from $150$-$250$~K.}
%  \label{fig:gd-phase-boundary}
% \end{center}
% \end{figure}

% Use Landau theory
\subsection{Noise spectroscopy}
% (Tim)
In order to perform magnetic noise spectroscopy of Gd at temperatures ranging from 273~K to 340~K, we attach a small chunk of Gd foil (100~$\mu$m $\times$ 100~$\mu$m $\times$ 25~$\mu$m) close to a microwave wire on a Peltier element with which we tune the temperature.
Instead of mm-scale diamonds as before, we use nano-diamonds (\emph{Adamas}, $\sim$ 140~nm average diameter) drop-cast onto the Gd foil to minimize the distance to the surface of our sample.

With no external field applied, all eight resonances of the NVs inside the nano-diamonds are found within our resolution to be at the zero-field splitting $D_{gs}$ for either para- and ferromagnetic phase of Gd, leading to a larger resonance contrast since we can drive all NVs with the same microwave frequency.
Measuring the NV's spin relaxation time $T_1$ under these circumstances is equivalent to ascertaining the AC magnetic noise at $\sim$ 2.87~GHz.

\begin{figure}
 \centering
 \includegraphics[scale=0.38]{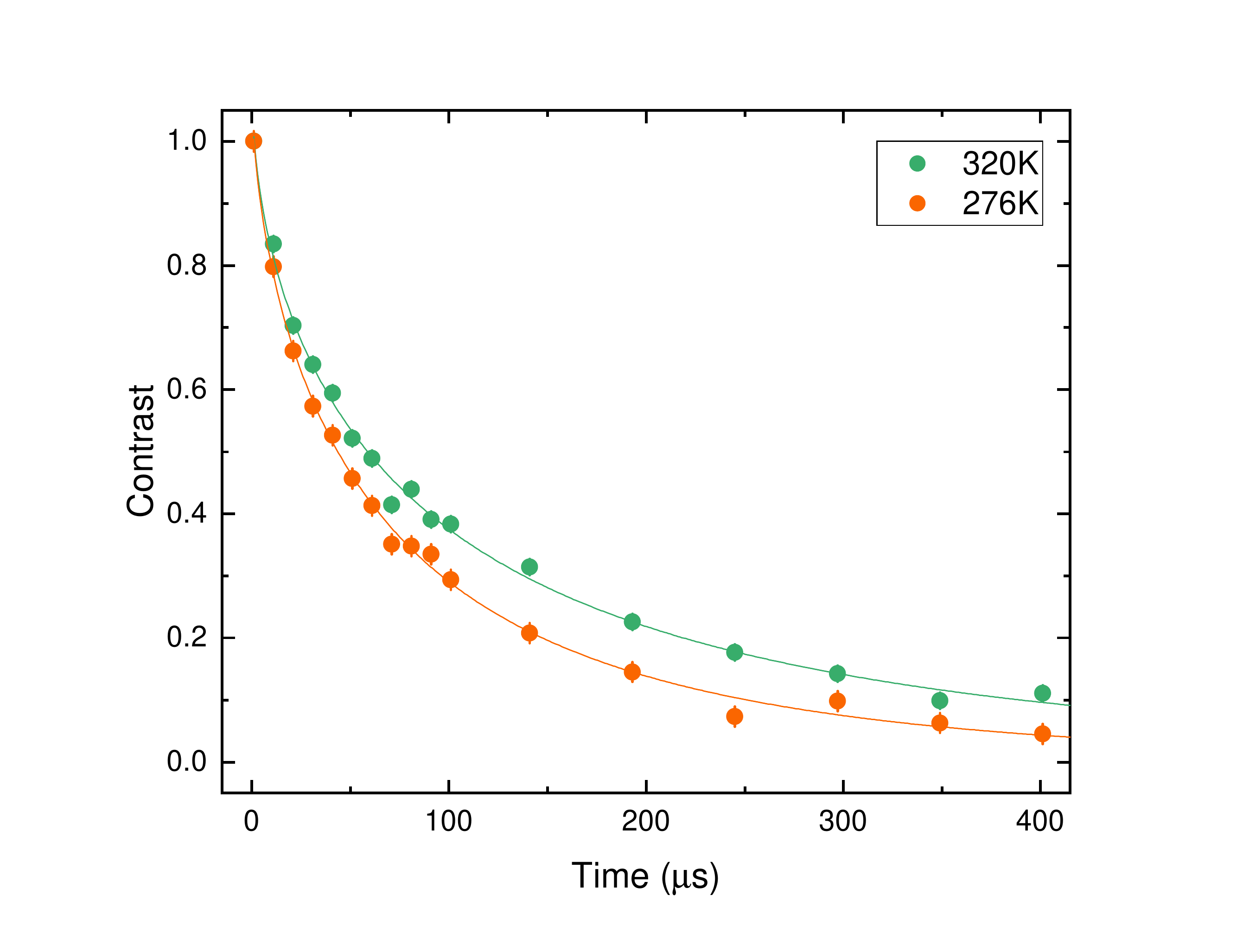}
 \caption{Plots of $T_1$ measurements below and above the magnetic phase transition in Gd. The green (orange) curve was measured at 320~K (276~K) and yields $T_1$ = 91$\pm$4~$\mu$s (66$\pm$3~$\mu$s), indicating a clear reduction of the spin polarization lifetime in the ferromagnetic phase. A stretched exponential function with exponent $\alpha$ = 0.6 (0.65) was used for fitting.}
  \label{fig:T1}
 \end{figure}

For this purpose, we utilize the following pulse sequence to measure $T_1$. First, we apply a 10~$\mu$s laser pulse to intialize the spin into the $|m_s = 0\rangle$ state. After laser pumping, we let the spin state relax for a variable time $\tau$, before turning on a second laser pulse to detect the spin state (signal bright). We repeat the exact same sequence once more, but right before spin detection, an additional NV $\pi$-pulse is applied to swap the $|m_s = 0\rangle$ and $|m_s = \pm 1\rangle$ populations (signal dark). The difference between signal bright and dark gives us a reliable measurement of the NV polarization (Fig.~4D top inset in main text) after time $\tau$. The resulting $T_1$ curve exhibits a stretched exponential decay $\propto e^{{-(\tau/T_1)}^\alpha}$, with $\alpha \sim 0.65$ (Fig.~\ref{fig:T1}).
 
 \begin{figure}
 \begin{center}
 \includegraphics[scale=0.38]{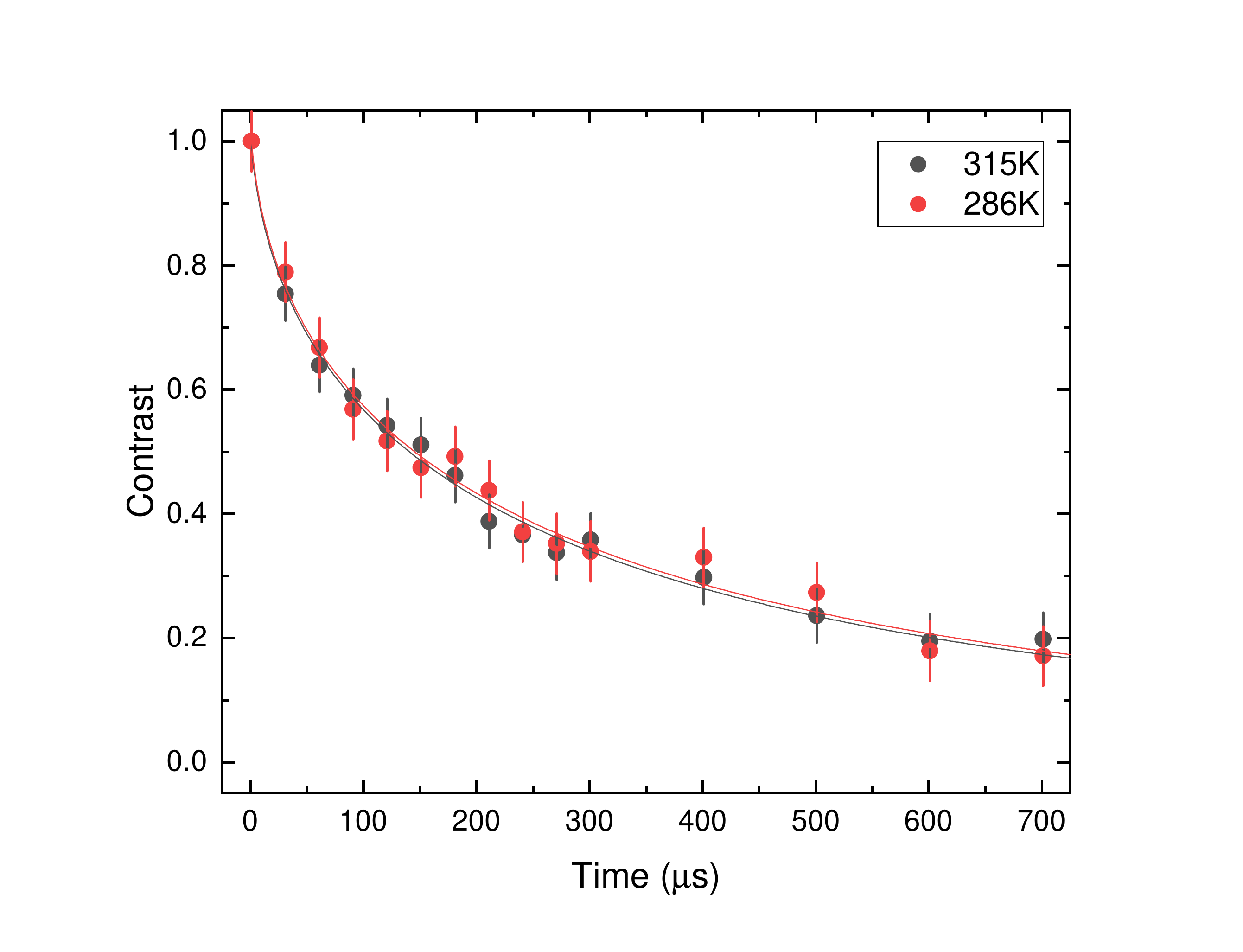}
 \caption{Plots of $T_1$ measurements away from the Gd flake at 315~K (grey curve) and 286~K (red curve). The resulting spin polarization lifetimes $T_1$ = 243$\pm$14~$\mu$s (315~K) and 247$\pm$20~$\mu$s (286~K) are the identical within the errorbar.}
  \label{fig:T1_away}
 \end{center}
 \end{figure}

By sweeping the Peltier current over a range of $\sim$ 3.5~A, we adjust the temperature of the sample from $273$~K to $340$~K, therefore determining the temperature dependence of $T_1$.

This procedure is performed on two different nano-diamonds on top of the Gd flake to confirm that the signal is not an artifact.
% and the temperature is not changed linearly in one direction, but adjusted randomly such that a temporal decay of the intrinsic $T_1$ can be excluded.
%
Furthermore, this is contrasted with an additional measurement at a nano-diamond far away from the Gd foil, exhibiting no temperature dependence of $T_1$. 
%
% Also, the intrinsic depolarization lifetime of the nano-diamonds is hereby available and determined to be about twice the value of the previously measured NVs on Gd.
%
% A measurement of the decoherence time $T_2$ shows no signal of a significant change related to the phase transition.

% describe the experiment: dropcasting nanodiamond (adamas, ~140nm average diameter) on to Gd foil, T1 measurement in absence of external magnetic field
%

% (Shubhayu)
\subsection{Theoretical analysis of $T_1$}
The depolarization time $T_1$ of NV centers shows a distinct drop when we decrease the temperature $T$ to across the ferromagnetic phase transition of Gd, Fig.~4D of the main text. 
Assuming that Johnson noise is the main contribution, because we are working at a fixed small transition frequency ($\omega\sim 2.87$~GHz) and in the thermal limit ($\hbar \omega \ll k_B T$), we can consider the DC limit.
In this case, we have $T_1 \propto \rho(T)/T$, where $\rho(T) = 1/\sigma(T)$ is the DC resistivity \cite{Agarwal2017}. 
Importantly, previously measurements of the resistivity curve for Gd show a kink at $T_\textrm{C}$, with a sharper temperature dependence below $T_\textrm{C}$ \cite{Nigh63,Jacobsson89}. 
However, this sudden change in slope is insufficient to explain our observations of $T_1$; in particular, given the magnitude of the resistivity, the change in temperature dominates the $T_1$ behavior.
This implies that $T_1$ should increase in the ferromagnetic phase if the sole contribution is bulk Johnson noise, whereas observations indicate otherwise.

A hint to the resolution of this puzzle comes from two observations. First, NV centers drop-cast onto Gd samples are very close to the sample, and hence far more sensitive to the surface than the bulk.
Second, the surface of Gd is well known to show a higher ferromagnetic transition temperature than the bulk; the drop in $T_1$ starts at a larger temperature ($\approx 300$~K) compared to the bulk $T_{\textrm{C}} \approx 292$~K.
These observations strongly suggest that the NV is detecting a large drop of surface resisitivity as we lower $T$ across the surface critical temperature, and this dominates over the small drop of bulk resistivity in the observed behavior.

In order to quantitatively estimate the relative contribution of the surface to the bulk, we write down, following Ref.~\cite{Agarwal2017}, the contribution to the noise for a single two-dimensional layer at a distance $z$ from the probe for a sample with conductivity $\sigma(T)$ 
\begin{equation}
\frac{1}{T_1} \propto  N(\omega) = \frac{k_B T \mu_0^2 \sigma(T)}{16 \pi z^2} ~ .
\end{equation}
Here we have assumed that the optical conductivity has a smooth dc limit (true for typical metals) and taken the extreme thermal limit to neglect the small frequency dependence of $\sigma$. Gd has a hcp structure with $c \approx 2a$, so we approximate the sample as being composed of decoupled two-dimensional layers and add their individual contributions to the noise. If the distance from the surface to the probe is $d$, the surface thickness is $D$ (infinite bulk thickness), and the surface and bulk conductivity are denoted by $\sigma_s$ and $\sigma_b$ respectively, then we have:
\begin{equation}
\label{eq:Rel}
    \frac{1}{T_1} \propto T \left[ \int_d^{d+D} dz \; \frac{ \sigma_s(T)}{z^2} +  \int_{d+D}^\infty dz \; \frac{ \sigma_b(T)}{z^2}  \right]  = T \sigma_s(T) \left( \frac{1}{d} - \frac{1}{d+D}\right) + \frac{T \sigma_b(T) }{d+D} ~.
\end{equation}
Eq.~(\ref{eq:Rel}) makes it explicit that when $D/d$ is an $\mathcal{O}(1)$ number (i.e. the surface thickness is of the order of sample-probe distance) the surface and bulk contributions are comparable. 
On the other hand, if $D/d \ll 1$, the bulk noise dominates. 
For our drop-cast nano-diamonds on the surface of Gd, we can estimate $D \approx 10$~nm, given the distinct surface signatures in the density of states even 6 layers deep \cite{oroszlany2015}. We also estimate the average distance as approximately half the radius of a nano-diamond, $d\approx50$~nm.
Therefore, we see that, for our samples, a large rise in surface conductivity can cause a significant increase in magnetic noise, even if the bulk conductivity remains roughly constant across the transition to the ferromagnetic phase. 
Hence, we conjecture that an enhanced surface conductivity below the surface critical temperature $T_{c,s}$ is responsible for the observed drop in $T_1$. 

% Contrast this with the net conductivity of such a sample, given by the weighted average of surface and bulk conductivity.
% \begin{equation}
% \sigma = \frac{D \sigma_s + (\Lambda - D)\sigma_b}{\Lambda} = \sigma_b + \frac{D}{\Lambda}(\sigma_s - \sigma_b) \approx \sigma_b, \; \text{ for } \frac{D}{\Lambda} \ll 1 
% \end{equation}
% We see that the corrections to the bulk conductivity are of the order $D/\Lambda$, where $D$ is the surface depth and $\Lambda$ is the sample thickness which is pratically infinite for a bulk sample. Therefore, unless we have a thin film where $D \approx \Lambda$, the surface correction to the conductivity is negligible and only the bulk value has been measured in experiments. 

The sharp drop of surface resisitivity below the surface ordering temperature can be due to several reasons.
It can be caused by the critical behavior of surface magnetism, or a different electron-magnon coupling on the surface because the surface electrons have more localized wave-functions. 
Here, we provide one consistent picture for the drop in surface resisitivity in terms of a distinct surface criticality relative to the bulk.

\begin{figure}
 \begin{center}
 \includegraphics[scale=0.75]{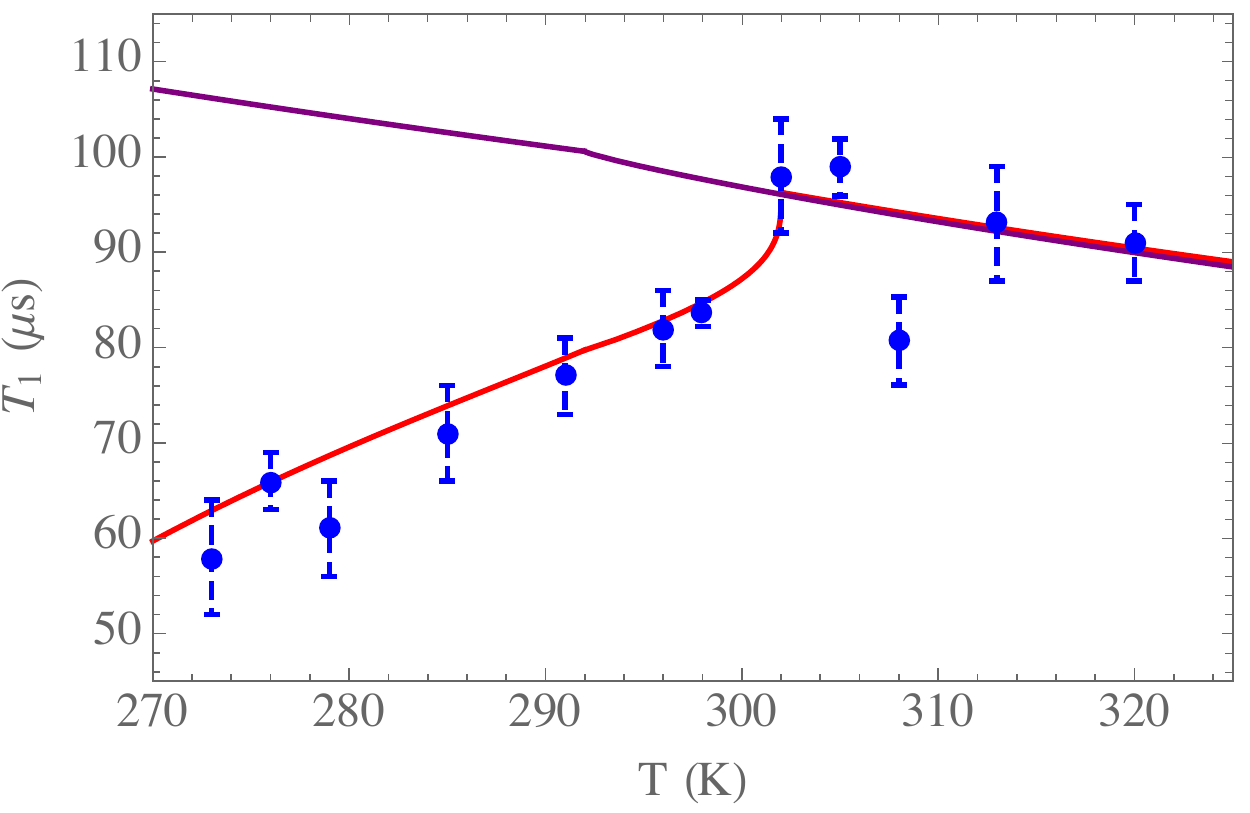}
 \caption{The purple curve shows $T_1$ taking only the bulk contribution to Johnson noise into account. The red curve shows $T_1$ taking both surface and bulks contribution into account, with $T_\textrm{C} = 292 $ K and $T_{c,s} = 302$ K. The blue dots are experimental data.}
  \label{fig:T1Fit}
 \end{center}
 \end{figure}

From Ref.~\cite{Colvin60,Jacobsson89,Nigh63} we know that both the bulk residual resistivity and the phonon contribution to the resistivity is quite small, and electron scattering below the bulk $T_\textrm{C}$ is dominated by magnetic excitations. 
Since $T_\textrm{C} = 292$~K is much larger than the Debye temperature $\Theta_D \approx 170$~K \cite{Bodryakov99,Jacobsson89}, the phonon contribution to scattering is expected to be linear in $T$ near $T_\textrm{C}$. 
Above $T_\textrm{C}$, the slope $d\rho/dT$ for Gd is very small. Hence the majority of scattering below $T_\textrm{C}$ takes place due to magnetic correlations, which, below $T_\textrm{C}$, changes resistivity by $d\rho/dT \propto t^{2\beta -1}$ where $t = |T_\textrm{C} - T|/T_\textrm{C}$ \cite{FisherLanger}. $\beta$ can be significantly different from 1, leading to a cusp in $\rho(T)$ at $T_\textrm{C}$. For the bulk, we can write:
\begin{equation}
\rho_b(T) = \rho_b(T_\textrm{C}) - \alpha_{ph} \left( \frac{T_\textrm{C}-T}{T_\textrm{C}}\right)  - \alpha_{mag}  \left( \frac{T_\textrm{C}-T}{T_\textrm{C}}\right)^{2\beta} \Theta(T_\textrm{C} - T)
\end{equation}
Above $T_\textrm{C}$, the singularity in $d\rho/dT$ is of the form $t^{-\alpha}$. 
However, for both Heisenberg and Ising universality classes of ferromagnetic transitions, $\alpha$ is close to zero ($\alpha \approx -0.1$), and the surface enhancement of the surface density of states is negligible.
Therefore, for $T > T_\textrm{C}$ we assume that the surface conductivity is identical to the bulk conductivity. 
Moreover, the scattering from uncorrelated core-spins should be constant at high temperatures away from $T_\textrm{C}$, so the slope $d\rho/dT$ is entirely from phonons for $T \gg T_\textrm{C}$. 
Using this relation, we can estimate $\alpha_{ph} \approx 27 ~\mathrm{ \mu \Omega cm}$ using the data for T between $350$ and $400$~K \cite{Jacobsson89}. 
Using the data for $\rho$ at $T = 280$~K in Ref.~\cite{Nigh63} to extract $\alpha_{mag}$ and $\beta \approx 0.35$ for the three dimensional Heisenberg model, which is believed to describe quite well the ordering of local moments in Gd \cite{oroszlany2015}, we obtain $\alpha_{mag}$:
\begin{equation}
\rho_b(T) - \rho_b(T_\textrm{C}) = - 4~\mathrm{ \mu \Omega cm} = - \alpha_{ph} \left(\frac{12}{292}\right) - \alpha_{mag} \left(\frac{12}{292}\right)^{0.7} \implies \alpha_{mag} \approx 27~\mathrm{ \mu \Omega cm} 
\end{equation}
This gives the bulk resistivity as a function of temperature, but it does not replicate the experimental observations, purple line in Fig.~\ref{fig:T1Fit}.
We now postulate a similar critical behavior at the surface but with surface critical exponent $\beta_s$ for the magnetization. 
On a two-dimensional surface, the Mermin-Wagner theorem forbids the spontaneous breaking of a continuous spin-rotation symmetry at a non-zero temperature \cite{KogutRMP}. 
For a surface ferromagnetic phase transition, we must have theory with reduced symmetry. 
Given the easy axis anisotropy in Gd \cite{Nigh63,oroszlany2015}, the surface magnetic phase transition is plausibly in the Ising universality class, with $\beta_s = 0.125$ \cite{KogutRMP}. Therefore, on the surface, we have:
\begin{equation}
\rho_s(T) = \rho_s(T_{c,s}) -  \alpha_{ph,s} \left(\frac{T_{c,s}-T}{T_{c,s}}\right) - \alpha_{mag,s} \left(\frac{T_{c,s}-T}{T_{c,s}}\right)^{0.25} \Theta(T_{c,s}-T) 
\end{equation}

In absence of evidence otherwise, we take $\alpha_{ph,s}  = \alpha_{ph}$ (same value as in the bulk). However, $\alpha_{mag,s}$ can be significantly enhanced relative to the bulk value. This can be due to several reasons. The surface electrons can be more localized than the bulk, therefore increasing the electron core-spin coupling. Further, the surface local moments can have a larger net spin $S$ relative to the bulk which orders more slowly. Since the electron-spin scattering cross-section is proportional to $S(S+1)$ \cite{FisherLanger}, a fully polarized core 4f state with $S = 7/2$ will have a larger scattering rate with an itinerant electron compared to a partially polarized state with $S < 7/2$. The exact value of $\alpha_{mag,s}$ thus depends on delicate surface physics; here we treat it as a free parameter. Fig.~\ref{fig:T1Fit} shows a good fit to our data with the estimates $\alpha_{mag,s} = 7 \alpha_{mag} \approx 189 ~\mathrm{ \mu \Omega cm}$, surface thickness $D = 10$~nm $\approx 17 c$, and sample-probe distance $d = 50$~nm (we have used an overall proportionality factor for the fit). 

We note that spin-fluctuations in Gd can also cause cause the NV polarization to relax. 
Although such fluctuations are negligible in the paramagnetic phase as our sample-probe distance is much larger than the lattice spacing \cite{Agarwal2017}, gapless critical fluctuations and spin-wave modes can indeed have a larger contribution to magnetic noise. 
However, the magnon contribution is related to magnon occupancies and decreases with decreasing temperature \cite{CRD2018}, implying that $T_1$ should increase as one lowers temperature in the ferromagnetic phase. 
This is inconsistent with the behavior we observe. 
Bulk critical spin-fluctuations should make the largest contribution at $T_\textrm{C}$, which is also not observed. 
An even more involved theoretical analysis is required to rule out critical surface spin-fluctuations.
This analysis is left for future work.

\bibliography{suppbib.bib}